\documentclass[useAMS, usenatbib]{mn2e}

\usepackage{amssymb} \usepackage{graphicx}
\usepackage{subcaption} \usepackage{float}
\usepackage{booktabs} \usepackage{multirow} \usepackage{amsfonts}
\usepackage{lmodern}
\usepackage{soul,color}

\let\oldbullet\bullet \renewcommand{\bullet}[1][0pt]{%
\mathrel{\raisebox{#1}{$\oldbullet$}}%
}

\title{The environment of bright QSOs at $z \sim 6$: Star forming galaxies
  and X-ray emission} \author[Costa et al.]{Tiago Costa$^1$\footnotemark[1], Debora
  Sijacki$^{1,2}$, Michele Trenti$^1$ and Martin G. Haehnelt$^1$ \\
  $^1$ Institute of Astronomy and Kavli Institute for Cosmology,
  University of Cambridge, Madingley Road, Cambridge CB$3$ $0$HA, UK
  \\ $^2$ Harvard-Smithsonian Center for Astrophysics, 60 Garden
  Street, Cambridge, MA, 02138, USA}

\begin{document}

\maketitle

\begin{abstract}
We employ cosmological hydrodynamical simulations to investigate
models in which the supermassive black holes powering luminous  $z
\sim 6$ quasars (QSOs) grow from massive seeds. We simulate 
$18$ regions with densities ranging from
the mean cosmic density to the highest $\sigma$ peaks in
the Millennium simulation volume. 
Only in the most massive halos 
situated in the most overdense regions, can black holes
 grow to masses up to
$\approx 10^{9}\, \mathrm{M_{\odot}}$ by $z \sim 6$ without
invoking super-Eddington accretion. Accretion onto the most massive
black holes becomes limited by thermal AGN feedback by $z \sim 9 \--
8$ with further growth proceeding in short Eddington
limited bursts. Our modelling suggests that current flux-limited
surveys of QSOs at high redshift preferentially detect objects
at their peak luminosity and therefore miss a substantial
population of QSOs powered by similarly massive black holes but with
low accretion rates.  
To test whether the required host halo masses are consistent with
the observed galaxy environments of $z \sim 6$ QSOs, we produce
realistic rest-frame UV images of our simulated galaxies. Without strong stellar feedback, our
simulations predict numbers of bright galaxies larger than observed
by a factor ten or more. Supernova-driven galactic winds reduce the
predicted numbers to a level consistent with observations indicating
that stellar feedback was already very efficient at high redshifts.    
We further investigate the effect of thermal AGN feedback on the
surrounding gas. AGN outflows are highly anisotropic and mostly energy-driven, pushing gas at
$\gtrsim 1000 \,  \mathrm{km \,s^{-1}}$ out to tens of 
$\mathrm{kpc}$ consistently with observations. The spatially extended thermal X-ray emission around bright QSOs is powered by these outflows 
and is an important
diagnostic of the mechanism whereby AGN feedback energy couples to surrounding gas.
\end{abstract}


\section{Introduction}
\renewcommand{\thefootnote}{\fnsymbol{footnote}}
\footnotetext[1]{E-mail: taf34@ast.cam.ac.uk}

There is solid observational evidence  that most galaxies host a
supermassive black hole in  their centre with properties that
correlate with those of the galactic bulge \citep{Dressler:89,
  Kormendy:95, Magorrian:98, Gebhardt:00, Ferrarese:00}.  Currently
inactive supermassive black holes are thought to have undergone a
quasar (QSO) phase in the past, when they grew by accretion to their
current sizes, emitting  enormous amounts of radiation into their
surroundings \citep{Schmidt:63, Salpeter:64,
 Lynden-Bell:69}. Important clues  to the formation of these
enigmatic objects may thereby be provided   by the very luminous QSOs
at redshifts $\gtrsim 6$, revealed by deep observations such as the
\emph{Sloan Digital Sky Survey} \citep{Fan:01, Fan:03, Fan:04,
  Fan:06}. These bright high-redshift QSOs  are believed to be powered
by supermassive black holes with  masses of  $\sim 10^9 \,
\mathrm{M_{\odot}}$. Some of these supermassive black holes appear to
be even in place as early as  $z \approx 7$  \citep{Mortlock:11}.  The
existence of such massive black holes less than a Gyr after the Big Bang
poses strong constraints on  models in which supermassive black holes
grow out of stellar mass seeds via Eddington limited accretion
\citep{Volonteri:03, Volonteri:05, Lodato:06, Haiman:06, Trenti:07}.   Models in
which black holes grow from initially massive seeds are therefore
perhaps the most promising in this regard.  In this class of models,
seed black holes of mass $\sim 10^{4 \-- 6} \, \mathrm{M_{\odot}}$
form via direct collapse of metal free gas onto protogalactic halos at
$z > 10$ \citep{Haehnelt:93, Umemura:93, Kauffmann:00, Oh:02,
  Bromm:03, Koushiappas:04, Begelman:06, Lodato:06, Volonteri:08,
  Tanaka:09}.  Numerical simulations focusing on protogalactic halos
with virial temperatures just above the atomic cooling threshold at $z
\approx 15$ suggest that this scenario is  indeed viable
\citep{Wise:08, Regan:09, Regan:09b, Wolcott-Green:11, Choi:13,
  Prieto:13, Latif:13}.  Growth of supermassive black holes starting
from massive seeds has also been modelled with full cosmological
hydrodynamical simulations, where black hole sink particles of mass
$10^5 \-- 10^6 h^{-1}\mathrm{M_{\odot}}$ placed at the centre of halos
above a threshold mass of $10^{9} \-- 10^{11} h^{-1}
\mathrm{M_{\odot}}$ evolve into a population of supermassive black
holes with masses and accretion luminosities in line with
observational estimates by $z \sim 6$ \citep{Sijacki:09, Khandai:12}.

The clustering analysis of bright QSOs at intermediate redshifts, i.e. $z
\sim 2\-- 5$, suggests that the black holes with mass $\ga 10^{9} \,
\mathrm{M_{\odot}}$ are located at the centre  of dark matter halos
with masses of $\sim 3 \times 10^{12} \-- 10^{13} \,
\mathrm{M_{\odot}}$  or higher (\citet{Porciani:04, Croom:05, Coil:07,
  Myers:07, Shen:07, daAngela:08, Padmanabhan:09, Ross:09}, but see 
also \citet{Trainor:12}). At $z
\sim 6$ halos that massive are still very rare and thus highly biased
with respect to the overall matter distribution. Unless the fraction
of the baryons finding their way into supermassive black holes is much
higher than at lower redshifts, black holes with masses of  $\sim
10^{9} \, \mathrm{M_{\odot}}$ observed at $z \sim 6$ are expected to
be surrounded by a wealth of other massive structures and should
appear strongly clustered when compared to more `average regions' of
the cosmic density field. How  such overdensities of massive halos
translate into an overdensity of bright galaxies surrounding the
luminous QSOs at $z \gtrsim 6$ is, however, less clear
\citep{Kauffmann:02, Munoz:08, Overzier:09, Romano-Diaz:11,
  Fanidakis:13}.

Accurate direct mass measurements of the host galaxy/halo have so far
proven difficult if not impossible. Some measurements of the line
widths of molecular gas have returned surprisingly low values
\citep{Walter:04, Wang:10, Willott:13, Wang:13}.  It is, however, not
clear to what extent these are actually representative of the full
depth of the gravitational potential of the host halo and whether low
velocity widths should be interpreted as evidence  that these
supermassive black holes are located in significantly less massive
halos \citep{Willott:05}.

Studies of the environment of the observed high-redshift QSOs are
therefore probably our best bet to find out what is the typical mass
of the dark matter host halos. None the less, observational studies aimed at
verifying whether the number of galaxies surrounding high redshift
QSOs are consistent with the presence of a strong overdensity have
been inconclusive so far. Significant overdensities were detected
around two different SDSS QSOs at $z \approx 6$ \citep{Stiavelli:05,
  Zheng:06}, taking the GOODS field \citep{Giavalisco:04} as a
reference average field.   \citet{Kim:09} analysed a sample of
$i_{775}$-dropout galaxy candidates identified in five \emph{Hubble
  Space Telescope} (HST) \emph{Advanced Camera for Surveys} (ACS) QSO
fields which includes \citet{Stiavelli:05} sample, revealing two overdense and  two underdense fields and one
field with similar density compared to the mean density of the GOODS
survey.  Note, however, that with an angular  surface area of $202
\times 202 \, \mathrm{arcsec^2}$ \citet{Kim:09} only probed a region
of $\approx 6 \times 6 \, \rm comoving \, Mpc^2$ around the QSOs.
Using the \emph{Subaru `Suprime-Cam'}, with the larger field of view
of $34 \times 27 \, \mathrm{arcmin^2}$, \citet{Utsumi:10} 
\citep[see also recent studies by][]{Banados:13, Husband:13} reported an
overdensity of Lyman-break galaxies (LBGs) around another QSO at $z
\approx 6.4$. The quoted excess of LBGs occurs at a projected distance
greater than $2 \, \mathrm{Mpc}$, suggesting the QSO that appears
to be located in an underdensity on small scales might still reside in
an overdensity, when larger scales are considered. Taken at face
value, all of these observational results appear at odds with
expectations from the $\Lambda$CDM paradigm of galaxy formation, where
pronounced overdensities are predicted around the massive halos
supposed to host the bright $z \sim 6$ QSOs \citep{Romano-Diaz:11}.
Using a mock survey of LBGs at $z \approx 6$ constructed using the
semi-analytic prescriptions applied to the Millennium simulation
\citep{Springel:05}, \citet{Overzier:09} concluded that the $10^{12}
\-- 10^{13} \, \mathrm{M_{\odot}}$ QSO host halo hypothesis for the
$z \sim 6$ QSOs is marginally consistent, but poorly constrained by
available observational data due to the expected large scatter in the
number of neighbouring LBGs.

Deeper data reaching to fainter magnitudes as well as improved
modelling including relevant baryonic/feedback physics are clearly
required for further progress. One of the main aims of this paper is
to investigate this question for the first time with realistic mock
images based on hydrodynamical cosmological simulations  which include
detailed supernovae and AGN feedback physics.

Observations of the environment of bright QSOs are also an important
diagnostic tool for the study of the large scale  outflows driven by
the momentum and energy released by  the central engine. Strong
coupling between these outflows  and the interstellar medium of the
host galaxy is the basis  for most of the explanations of the
correlations identified between  the mass of supermassive black holes
and the bulge properties  \citep{Silk:98, Haehnelt:98, Fabian:99b,
  King:03, Fabian:12}.    By expelling baryons and suppressing star
formation, AGN feedback may also help explain the properties of local
massive galaxies, such as their dearth in young stars and low gas
content.  Observational evidence for AGN feedback includes the
detection of massive outflows in ULIRGS \citep{Sturm:11} and QSOs
\citep{Fischer:10, Feruglio:10, Rupke:11, Greene:12,
  Cicone:12}. AGN-driven outflows have also been detected in high
redshift QSOs \citep{Alexander:10, Nesvadba:11, Maiolino:12}, for
which there is some evidence of quenched star formation
\citep{Cano-Diaz:12}.  Typical outflow rates range from hundreds to
thousands solar masses per year with wind velocities of the order of
$1000 \, \mathrm{km/s}$.  In the case of the $z \approx 6.4$ SDSS QSO
J$1148+5251$ \citep{Maiolino:12}, the inferred outflow rates are $>
3500 \, \mathrm{M_{\odot} \, \mathrm{yr}^{-1}}$, even higher than the
already large  star formation rate ($\approx 3000 \, \mathrm{M_{\odot} \,
  \mathrm{yr}^{-1}}$) of the QSO host. The observed wind speeds are of
the order of $1300 \, \mathrm{km} \, \mathrm{s}^{-1}$ and are observed
at a distance from the supermassive black hole as large as  $\approx
16 \, \mathrm{kpc}$. A theoretical understanding of the driving
mechanisms and properties of AGN-driven winds is starting to build
up. A number of analytical and numerical works predicts AGN winds with
speeds and outflow rates comparable to the observational findings
\citep{King:11, Debuhr:12, Giguere:12, Roth:12, Sharma:12}. The
spatial scale at which the energy of the wind is radiated away should
depend strongly on the physical details of the driving of the wind.
As the wind is launched from the AGN it will collide with the
surrounding medium  driving both a shock trough it and a reverse shock
into the wind itself. Depending on the local gas properties,  the
reverse shock might cool on a time-scale which is shorter that the
characteristic outflow time-scale,  thus leading to a
`momentum-driven' flow. Alternatively, if the wind energy is not lost
to the radiation but  rather transferred to the surrounding medium, an
`energy-driven' flow develops. The presence and size of an extended
cooling emission is therefore expected to be tightly linked to the
detailed physics of AGN feedback  as well as to the properties of gas
surrounding the black hole. An additional aim of this paper is to
investigate  the predicted extended cooling emission around bright
QSOs at $z \sim 6$.

This paper is organized as follows. In Section \ref{secmethod}, we
describe the setup of our numerical simulations and lay down our
method for constructing realistic mock galaxy catalogues.  We provide
the results of the analysis of our simulations in Section
\ref{secresults}.  Finally, we present our conclusions in Section
\ref{secconclusions}.

\section{Method}\label{secmethod}

\subsection{Numerical setup}\label{secmethod1}

In order to study the cosmological environments expected to host
bright high-redshift QSOs, we selected the six most massive halos at
$z \,=\, 6.2$ from the dark matter-only Millennium simulation
\citep{Springel:05}. This simulation follows cosmological structure
formation in a box of $500 h^{-1} \, \mathrm{comoving \,Mpc}$ on a
side. This is about a factor hundred smaller volume than probed  by
high-redshift QSO surveys at $z \sim 6$ and  should thus be just
sufficiently large to contain a few bright QSOs given their estimated
spatial density \citep{Fan:04}. The descendants of these halos all lie
in rich clusters at $z \,=\, 0$ with masses in the range $10^{14} \,
\mathrm{M_{\odot}} \-- 10^{15}\, \mathrm{M_{\odot}}$ \citep[see][for a
  detailed discussion of the descendants of massive high-redshift
  halos]{Trenti:08, Angulo:12}.  For comparison we also selected six halos at $z
\,=\, 6.2$ that reside in large-scale regions with a matter density
very close to the average cosmic density as well  as six halos in
regions with a more moderate overdensity.   As our average regions we
randomly selected six regions for which the density smoothed with a
Gaussian filter with radius $5 h^{-1}$ and $10 h^{-1} \,
\mathrm{comoving \, Mpc}$ was within $\sigma/2$ at $z \,=\, 6.2$,
where $\sigma$ is the standard deviation of the respective Gaussian
distribution. For our intermediate regions we  selected six halos with mass
in the range $4 \-- 8 \times 10^{11} \, \mathrm{M_{\odot}}$ at $z \,=\, 6.2$ located in regions which are intermediate in overdensity
between our overdense and average regions.

We then generated high resolution initial conditions at $z \,=\, 127$
centred on each of these $18$ halos and populated them with dark
matter as well as gas. Beyond the high resolution region, mass
resolution was gradually degraded with distance, in order to maintain
computational efficiency, while modelling large-scale tidal fields
acting on the regions of interest correctly \citep[for further details
  on resimulation technique adopted see][]{Sijacki:09}. Numerical
parameters of the simulations are summarized in Table~\ref{table1}. 

\begin{table*}
\centering
\begin{tabular}{lccccccc}
\toprule Simulation Resolution & $N_{\rm tot}$ & $N_{\rm HR}$ &
$m_{\rm DM} (h^{-1} \, \mathrm{M_{\odot}})$  & $m_{\rm gas} (h^{-1} \,
\mathrm{M_{\odot}})$ & $m_{\rm star} (h^{-1} \mathrm{M_{\odot}})$ &
$\epsilon (h^{-1} \, \mathrm{kpc})$ & $R_{\rm HR}(h^{-1} \,
\mathrm{Mpc})$\\ \midrule 
zoom$1$  & $3.6 \times 10^6$ & $5.6 \times 10^5$  & $8.44 \times 10^8$ & $1.66 \times 10^8$  & $8.28 \times 10^7$ & $5$ & $12 h^{-1} \, \mathrm{Mpc}$ \\ 
zoom$5$  & $7.3 \times 10^7$ & $7.0 \times 10^7$ & $6.75 \times 10^6$ & $1.32 \times 10^6$  & $6.62 \times 10^5$ & $1$ & $12 h^{-1} \, \mathrm{Mpc}$ \\ 
\bottomrule
\end{tabular}
\caption{Basic numerical parameters  of the simulations presented in
  this paper. In the first column, we give the factor by which
  the spatial resolution is increased compared the parent Millennium simulation. Most of
  our simulations have five times higher spatial resolution and
  dark matter particle masses a factor $5^3$ lower than the
  Millennium simulation. We also compare some of our results from high
  resolution simulations with those of simulations run at the equivalent
  resolution as the Millennium simulation (while for the Millennium simulation total particle number is $2160^3$, here we use mass resolution corresponding to the closest power of two i.e. $2048^3$). The total number of
  particles and the number of high-resolution particles in the entire
  cosmological box are given in the second and third columns,
  respectively. These numbers can vary over a factor of two from
  simulation to simulation due to variations in mass from region to
  region and we therefore provide typical values here. The fourth,
  fifth and sixth columns give the mass of high-resolution dark
  matter, gas and stellar particles, respectively. The gravitational
  softening length is given in comoving units in the seventh
  column. Note that at $z \,=\, 6.2$, we therefore resolve
  spatial scales $\approx 400 h^{-1} \, \mathrm{pc}$ ($\sim 3 \times \epsilon$) in our
  highest-resolution simulations. Finally, we provide an estimate for
  the typical radius of the approximately spherical high resolution
  regions in comoving units in the eighth column.}
\label{table1} 
\end{table*}

The numerical setup of our simulations is similar to that of
\citet{Sijacki:09}, so we refer the reader to that paper for a
detailed description. Here we present a brief overview of the code and
physical modules adopted and highlight the differences with respect to
the \citet{Sijacki:09} setup. For most of our simulations, we employed
the massively parallel TreePM-SPH code GADGET-$3$
\citep{Springel:05a}. Dark matter is treated as a purely collisionless
fluid and its evolution is completely determined by gravitational
interactions, which are followed with a TreePM algorithm. The code
also tracks gravity, hydrodynamics and radiative cooling of a gaseous
component, assumed to behave like an ideal optically thin mixture of
hydrogen and helium. The gas is subjected to a spatially constant but
time-dependent UV background with reionisation occurring at $z \approx
6$. Tabulated UV background intensities have been taken from
\citet{Haardt:96} as well as from \citet{Faucher-Giguere:09} (updated
version from December 2011). Additionally to the GADGET simulations we
have performed a number of simulations with the moving mesh code AREPO
\citep{Springel:10}. AREPO employs the same gravity solver as the
GADGET code and can be started from identical initial
conditions. However, in order to evolve gas hydrodynamics, AREPO uses
a second-order accurate finite-volume method on an unstructured
Voronoi mesh that is allowed to move together with the fluid. GADGET
and AREPO share the same implementation for gas cooling and sub-grid
star formation. The comparison of simulation results performed with
these two codes therefore allows us to identify possible numerical
inaccuracies stemming from the hydro solver \citep[see
  e.g.][]{Sijacki:12, Vogelsberger:12, Keres:12} and to gauge their
magnitude with respect to the uncertainties in the relevant physics
that we are trying to simulate (for further details see Section
\ref{seccodecomparison}).  
 
Due to the inevitable lack of resolution, our simulations cannot follow the internal structure 
of star forming regions and the formation of molecular hydrogen.
 We are thus forced to adopt a
sub-grid prescription to account for the relevant physics of star
formation and associated feedback. We here adopt the sub-grid model of
\citet{Springel:03}. In this model, stellar particles interact purely
gravitationally with other particles in the simulation and are spawned
stochastically out of gas particles with densities exceeding a density
threshold $\rho_0$. The dense star forming gas follows an effective
equation of state that accounts for the balance between supernova
heating and associated cloud evaporation, gas cooling and cloud
formation due to thermal instabilities. With respect to the
\citet{Springel:03} implementation we slightly modify the star
formation module in two ways: i) for gas particles/cells which are hot
and have densities above $\rho_0$ we estimate their new temperature as
reduced due to radiative cooling losses; if such estimated temperature
would fall below the effective temperature of the multiphase medium we
set it equal to the effective temperature; ii) we prevent star
formation from gas that is hot (i.e. $T > 10^5$ K) but above the
density threshold $\rho_0$; this is done to prevent artificial
spawning of star particles out of gas which e.g. was part of the
multiphase medium but then heated to high temperature by AGN
feedback. In some of the  simulations we additionally incorporate
starburst-driven galactic outflows \citep[as in][]{Springel:03}, and
we vary the mass loading of the winds in our various numerical
experiments.  

Black holes are represented as collisionless, sink particles in the
simulation. The unresolved physics of accretion onto black holes is
also modelled in a sub-grid fashion \citep{DiMatteo:05,
  Springel:05}. The accretion is assumed to be of a
Bondi-Hoyle-Lyttleton type and is capped at the Eddington rate for
each black hole. Hence, the black hole accretion rate $\dot{M}_{\rm
  BH}$ is given by
\begin{equation}
\dot{M}_{\rm BH} \ = \ \min{\left[ \frac{4 \pi \alpha G^2 M_{\rm BH}^2
      \rho_{\rm gas}}{\left( c_{\rm s}^2 + v^2\right)^{3/2}},\frac{4
      \pi G M_{\rm BH} m_{\rm p}}{\epsilon_{\rm r} \sigma_{\rm T}
      c}\right]} \, ,
\end{equation}

where $\alpha \, = \, 100$ is a dimensionless parameter that is
introduced to recover a volume average of the Bondi rates for the cold
and hot interstellar medium phases which are not resolved. $\rho_{\rm
  gas}$ and $c_{\rm s}$ are the density and sound speed of gas
respectively, $v$ is the speed of the black hole particle relative to
the local gas speed, $m_{\rm p}$ is the proton mass, $\sigma_{\rm T}$
the cross-section for Thomson scattering, $c$ the speed of light and
$\epsilon_{\rm r}$ the radiative efficiency. We set
$\epsilon_{\rm r}$ to the standard value of $0.1$, as expected for accretion
onto moderately spinning black holes.

A fraction $\epsilon_{\rm f}$ of the AGN bolometric luminosity is
assumed to couple thermally with the surrounding gas,

\begin{equation}
\dot{E}_{\rm feed} \ = \ \epsilon_{\rm f} L_{\rm BOL} \ =
\ \epsilon_{\rm f} \epsilon_{\rm r} \dot{M}_{\rm BH} c^2 \, ,
\end{equation}

where $\epsilon_{\rm f}$ is the feedback efficiency and is set to
$0.05$ so that the normalisation of the locally inferred $M_{\rm BH} -
\sigma$ relation is reproduced
\citep{DiMatteo:05,Sijacki:07,DiMatteo:08}. The energy is deposited
into the thermal budget of all gas particles within the smoothing
length of the black hole particle, calculated by taking into account
the SPH kernel of the $64$ nearest gas particle neighbours. Black hole
mergers are assumed to be efficient and occur every time two black
holes come within each others' smoothing lengths and have relative
velocities lower or comparable to the local sound speed.  Seed black
holes are assumed to be massive, with $M_{\rm seed} \,=\, 10^5
h^{-1}\, \mathrm{M_{\odot}}$ and are placed at the centre of halos
with masses above a threshold value of $M_{200} \,=\, 10^9 h^{-1} \,
\mathrm{M_{\odot}}$, identified by frequently running a
Friends-of-Friends (FoF) algorithm on the fly. 
Note that we do not resolve halos below the atomic cooling threshold where H$2$ cooling becomes dominant. 
Thus, neglecting H$2$ cooling does not significantly impact on our results.

For our cosmological parameters, we assume  a $\Lambda$CDM cosmology
with parameters: $h \,=\, 0.73$, $\Omega_{\rm m} \,=\, 0.25$,
$\Omega_{\Lambda} \,=\, 0.75$, $\Omega_{\rm b} \,=\, 0.041$, identical
to the parent Millennium simulation, but use a lower $\sigma_8$ value
of $0.8$, in better accordance with more recent results
\citep{Planck:13}. Note that this parameter sensitively affects the
mass of the high-$\sigma$ peaks where the first QSOs are thought to
form and therefore plays an important role in this work. Since
emphasis is placed on observational predictions or direct
comparison with observations, all quantities are given in physical,
rather than comoving, units throughout this paper, unless when
explicitly stated otherwise.

\subsection{Producing mock UV catalogues}\label{secmethod2}

One of the main goals of this work is to assess the likelihood of
observing enhanced number counts of star-forming galaxies around bright 
$z\sim 6$ QSOs compared to average regions of the Universe. We thus want to
realistically ``observe'' our simulation snapshots and compile galaxy
catalogues that can be directly compared to HST observations. For this
purpose we construct simulated Hubble-like images, rather
than relying on identifying dark-matter halos and subhalos and
subsequently investigating the properties of the stellar populations
residing within these subhalos. This direct procedure to construct
projected luminosity maps has the advantage of being independent of
the details of the halo and subhalo finding algorithms, galaxy radius
definitions, as well as to account for line-of-sight superposition.

We start from the snapshots of our simulations, which store
information on the positions, masses, ages and metallicities of all
the stellar particles. Using the \citet{Bruzual:03} models, we compute
the rest-frame UV luminosity (at $1450~\mathrm{\AA}$, corresponding to
observed $1~\mathrm{\mu m}$ at $z\sim 6$) for each stellar particle
that has been created, assuming a Salpeter IMF for consistency with
the chemical enrichment and supernova feedback implementation in our
simulations. We then project the innermost cube of our high-resolution
region (with a volume of $10^3 h^{-3} \mathrm{comoving \, Mpc^3}$) along the
z-axis. At $z \sim 6$, the resulting two-dimensional luminosity map
corresponds to an observed field of view of approximately $6\times 6$
arcmin$^2$ with a redshift depth of $\Delta z = 0.032$, which we
construct on a grid with a cell-size of $0.1 \, \mathrm{arcsec}$. To
convert the absolute luminosity into observed magnitude in the z-band
(for comparison with the \citet{Kim:09} HST observations), we adopt a
distance modulus $D \,=\, 48.9$ and a K-correction $K \,=\, -2.1$
(flat $f(\nu)$ spectrum), in accordance with the cosmology we use
throughout the paper. Finally, we add uncorrelated Gaussian noise in
each pixel of the synthetic images corresponding to  a $5\sigma$
sensitivity limit $m_{\rm UV} = 26.5$ in an aperture of radius
$r=0.2''$ (for comparison with \citealt{Kim:09}). We also
consider higher S/N images which reach one magnitude deeper. Note that
all magnitudes are given in the AB magnitude system \citep{Oke:83}.

To mimic galaxy detection in actual HST data, we then run the source
finder code SExtractor \citep{Bertin:96} on our synthetic images,
using detection threshold and deblending parameters typical for LBGs surveys at high redshifts \citep{Bouwens:11, Trenti:11}. As in \citet{Kim:09}, we define the magnitude of a source
as SExtractor's $\mathrm{MAG\textunderscore{AUTO}}$ value, which
represents the code's best estimate of the total luminosity and
automatically includes aperture corrections.

\begin{figure}
\centering \includegraphics[scale = 0.5]{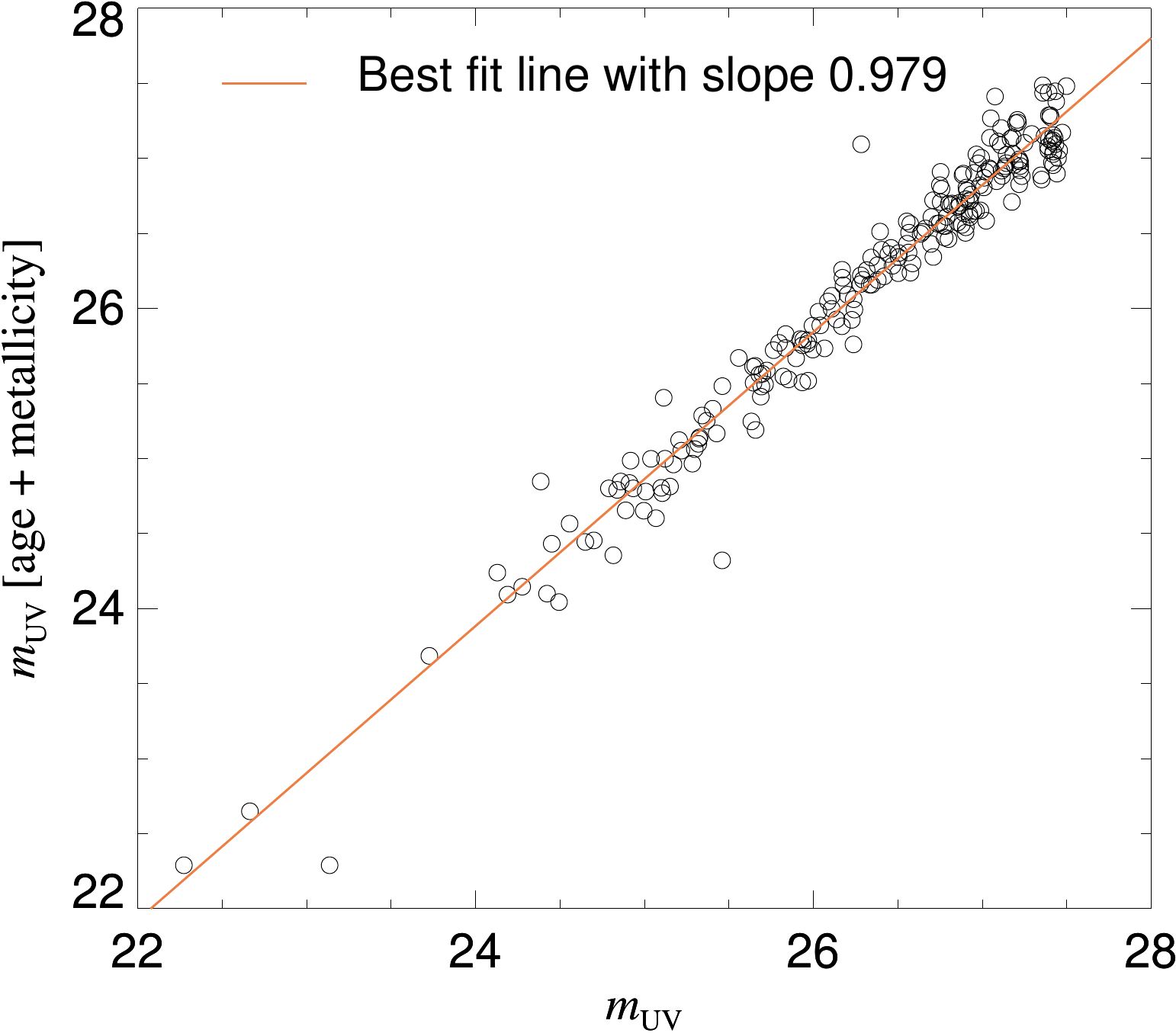}
\caption{Apparent UV magnitude (AB magnitude system) obtained using
  the stellar population synthesis catalogues of \citet{Bruzual:03}
  (taking into account stellar age and metallicity) shown as a function of
  the apparent UV magnitude derived from a typical $SFR - m_{\rm UV}$
  relation \citep[here as in][]{Labbe:10} for galaxies in one of our
  simulations of an overdense region at $z \,=\,
  6.2$. The best fit line shows the agreement between the two to be
  excellent.}
\label{uvcompare}
\end{figure}

This procedure closely resembles the construction of observational
catalogues and provides us with a realistic sample of galaxies within
our high-resolution computational volume. Minor discrepancies, which
are, however, unlikely to significantly affect any conclusion drawn in
this paper, are the use of Gaussian uncorrelated noise, while real
data may include an extended noise tail (for example because of
detector defects or residual cosmic rays), or correlated noise induced
by the drizzling procedure \citep{Casertano:00} commonly used to
combine different HST exposures. Another difference with actual data
is the fact that our projection only includes galaxies in a narrow
redshift interval with $\Delta z \sim 0.032$, so we are effectively
neglecting crowding from foreground sources (which typically impacts
$\sim 10 \%$ of the area in medium-depth observations like
\citealt{Kim:09}, but which is accounted for in actual observations by
means of artificial source recovery simulations).

In Figure~\ref{uvcompare}, we show how predictions for the apparent
magnitudes of galaxies in one of our overdense regions compare with
typical $SFR \-- m_{\rm UV}$ relations found in the literature.
Assuming the $SFR \-- m_{\rm UV}$ relation found in \citet{Labbe:10},
we construct mock images where we convert the star
formation rates of each particle into a UV luminosity. We match
galaxies between maps constructed in this way as well as according to
our default method presented above. We then directly compare
predictions of both methods. As shown in Figure~\ref{uvcompare}, the 
agreement is excellent, indicating that predictions for the UV luminosity of different galaxies
should be robust as long as their star formation rates are numerically converged (see Section~\ref{numconv}). 
Note also that UV bright $z \sim 6$ galaxies are thought not to be significantly affected by dust 
attenuation \citep{Bouwens:07, Labbe:10}, which we therefore do not model in our mock 
catalogues.

\begin{table*}
  \centering
\begin{tabular}{lcccccccccc}
\toprule Simulation    & $M_{\rm BH}$ & $M_{\rm FoF}$ & $M_{\rm 200}$
& $R_{\rm 200} \dagger$ & $V_{\mathrm{c},200}$ & $T_{\rm 200}$ &
$\rho_5/\bar{\rho_0}$ & $\rho_8/\bar{\rho_0}$ & $\delta_5/\sigma_5 (z
\,=\, 6.2)$ & $\delta_8/\sigma_8 (z \,=\, 6.2)$\\ & ($\times 10^{8} \,
\mathrm{M_{\odot}}$) & \multicolumn{2}{c}{($\times 10^{12} \,
  \mathrm{M_{\odot}}$)} & ($\mathrm{kpc}$) & $(\times\mathrm{km \,
  s^{-1}})$ & $(\times 10^6 \mathrm{K})$ & \\ \midrule 
A$1$      & 0.15	& 0.14   & 0.16    & 23.88	  & 169.7   & 1.62    & 1.07     & 1.05   & 0.3 & 0.3 \\
A$2$      & 0.01	& 0.04   & 0.04    & 14.90	  & 107.4   & 0.00$\ddagger$    & 0.98     & 0.97   & 0.1 & 0.2 \\
A$3$      & 0.07	& 0.06   & 0.06    & 17.38	  & 121.8   & 0.64    & 1.10     & 0.98   & 0.5 & 0.1 \\
A$4$      & 0.21	& 0.14   & 0.14    & 23.17	  & 161.2   & 1.13    & 0.98     & 0.97   & 0.1 & 0.2 \\
A$5$      & 0.07	& 0.10   & 0.11    & 20.97	  & 150.2   & 0.98    & 1.03     & 1.00   & 0.1 & 0 \\
A$6$      & 0.17	& 0.11   & 0.12    & 21.41	  & 155.2   & 1.22    & 1.04     & 1.06   & 0.2 & 0.4 \\
\midrule
I$1$      & 0.47    & 0.77   & 0.38    & 31.96	  & 226.1   & 2.15    & 1.52     & 1.23   & 2.6 & 1.5 \\
I$2$      & 0.98    & 0.63   & 0.65    & 38.35	  & 270.0   & 2.91    & 1.24     & 1.02   & 1.2 & 0.1 \\
I$3$      & 0.40    & 0.68   & 0.66    & 38.49	  & 271.5   & 2.81    & 1.54     & 1.34   & 2.7 & 2.3 \\
I$4$      & 0.41    & 0.39   & 0.27    & 28.71	  & 201.1   & 1.64    & 1.56     & 1.29   & 2.8 & 1.9 \\
I$5$      & 0.36    & 0.45   & 0.38    & 32.07 	  & 225.7   & 1.71    & 1.58     & 1.35   & 2.9 & 2.3 \\
I$6$      & 0.39    & 0.63   & 0.35    & 31.13    & 219.9   & 1.86    & 1.60     & 1.37   & 3.0 & 2.5 \\
\midrule
O$1$      & 5.65	& 4.13   & 3.84    & 69.26	  & 488.3   & 7.87    & 2.01     & 1.55   & 5.0 & 3.7 \\
O$2$      & 6.50	& 4.11   & 2.60    & 60.80	  & 428.8   & 8.22    & 1.95     & 1.41   & 4.7 & 2.8 \\
O$3$      & 11.5	& 4.39   & 4.62    & 73.66	  & 519.3   & 9.42    & 1.95     & 1.46   & 4.7 & 3.1 \\
O$4$      & 4.04	& 3.78   & 2.14    & 57.00	  & 401.8   & 5.62    & 2.05     & 1.42   & 5.2 & 2.8 \\
O$5$      & 3.23	& 3.54   & 3.31    & 65.98	  & 464.4   & 10.3    & 2.33     & 1.67   & 6.6 & 4.5 \\
O$6$      & 4.68	& 3.80   & 4.09    & 70.75	  & 498.5   & 10.7    & 2.54     & 2.10   & 7.7 & 7.4 \\
\bottomrule
\end{tabular}
\caption{Main properties of our sample of $18$ average,
  intermediate and overdense regions at $z \,=\, 6.2$. Apart from the mass of the
  most massive black hole present in each volume (second column), we
  list the dark matter mass of the underlying FoF group (third
  column), the total virial mass (fourth column), virial radius (fifth
  column), the circular velocity evaluated at $R_{\rm 200}$ (sixth
  column) and the virial temperature (seventh column) of its parent
  halo. We also obtained an estimate of the overdensity with the
  respect to the mean cosmic density on $5 h^{-1} \, \mathrm{comoving
    \, Mpc}$ and $8 h^{-1} \, \mathrm{comoving \,Mpc}$ scales (eighth
  and ninth columns). In the last two columns we provide the
  overdensities at these two scales in terms of standard deviations
  from the cosmic mean. Note that the most massive black holes do not
  necessarily reside in the most overdense regions, but rather in the
  most massive FoF groups.  $\dagger$$R_{\rm
    200}$ is defined as the radius enclosing a density $200$ times
  the mean density of the background Universe.  $\ddagger$A value
  of zero for $T_{\rm 200}$ occurs if a given halo does not have
  a hot shocked gas component.}
\label{table}
\end{table*}

\section{Results}\label{secresults}

\subsection{Black hole growth and cosmic environment}\label{subsec1}

\begin{figure}
\includegraphics[scale = 0.48]{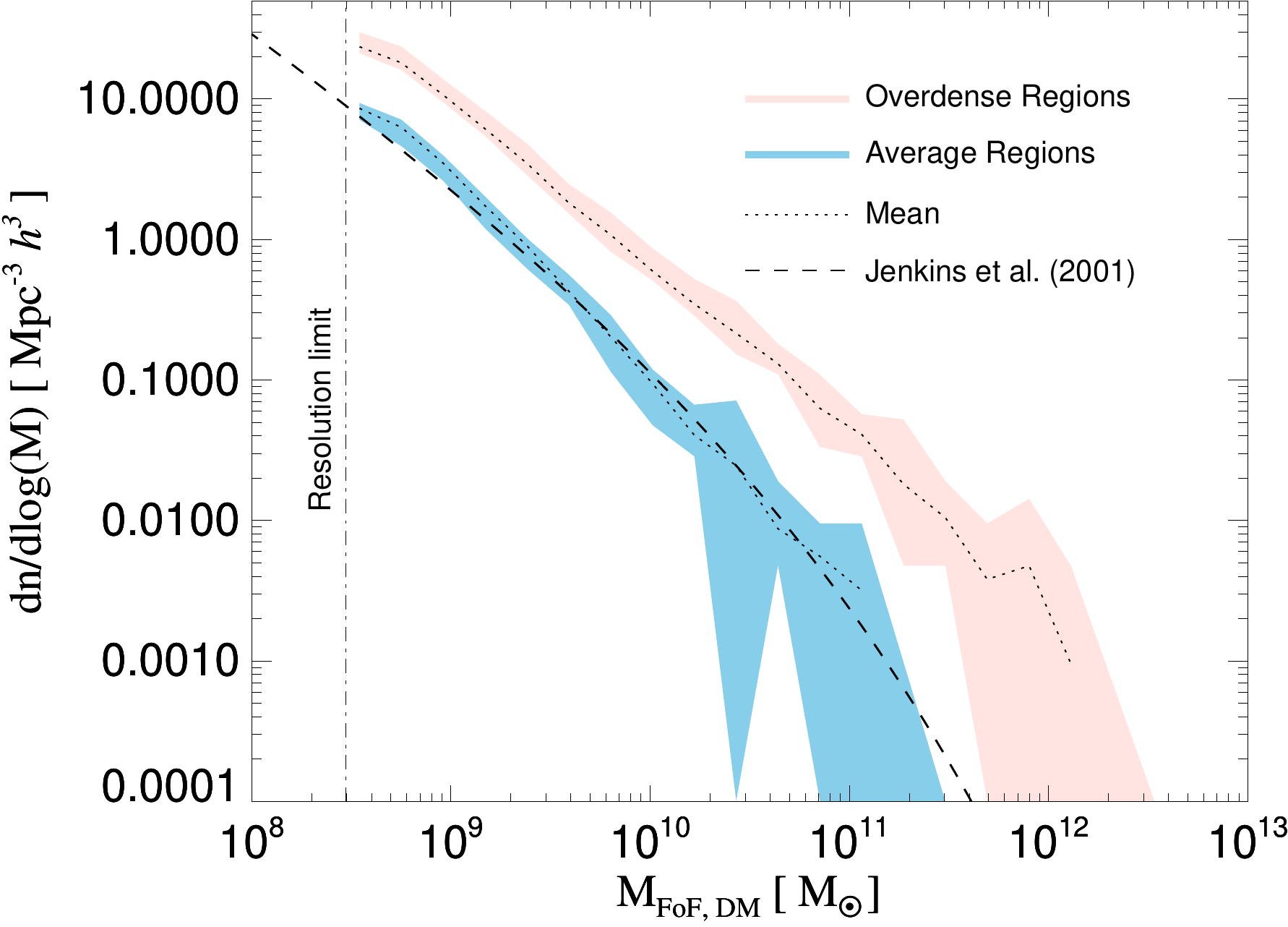}
\caption{Dark matter mass functions of the FoF groups for all
  overdense regions (upper band shaded in pink) and for all average
  regions (lower band shaded in blue) at $z \,=\, 6.2$. These were
  obtained by taking into account all FoF groups within a cube of side
  length $\approx 1.9 \, \mathrm{Mpc} \, (10 h^{-1} \,
  \mathrm{comoving \, Mpc})$ placed at the centre of the
  corresponding resimulation volumes. The dotted lines show the
  average mass function for the two cases. At the high mass end, the
  variance in the mass functions increases due to the small volume of
  the high resolution regions. The low mass end is set by the minimum
  halo mass with at least $32$ dark matter particles which for our
  simulation resolution corresponds to  $\approx 2.96 \times 10^{8} \,
  \mathrm{M_{\odot}}$. The mass function from \citet{Jenkins:01} at
  the same redshift is overplotted as the  dashed curve for
  comparison.}
\label{massfunction}
\end{figure}

Figure \ref{massfunction} shows the dark matter mass functions of all
FoF halos identified in the resimulated regions of our overdense and
average samples. These were constructed by considering the number
counts of FoF groups in a cube\footnote{Even though the high
  resolution region is approximately spherical, we took a sub-region
  whose volume we know precisely.} of side length $\approx 1.9 \,
\mathrm{Mpc}$ ($10 h^{-1} \, {\rm comoving \, Mpc}$) per logarithmic 
mass bin\footnote{Masses of FoF groups
  are often given in units of $h^{-1} \, \mathrm{M_{\odot}}$. Here we
  have explicitly multiplied the FoF group masses by $h^{-1}$.}. In
order to illustrate the variance in the mass functions from region to
region, we shade the area between the highest and lowest mass
functions for both average and overdense samples.  The mass functions
of the average regions are in excellent agreement with the theoretical
mass function by \citet{Jenkins:01}, confirming that our
selected regions indeed provide a good representation of an average
density field at this redshift. In the overdense regions, the mass
functions reveal an excess of halos at all masses when compared to the
average regions and their high mass end extends to higher values. This
is in agreement with what is expected in the $\Lambda$CDM paradigm, where
rare high-$\sigma$ peaks become the first regions to host the most
massive halos at high redshift.

\begin{figure*}
\begin{subfigure}{0.5\textwidth}
\centering \includegraphics[scale = 0.48]{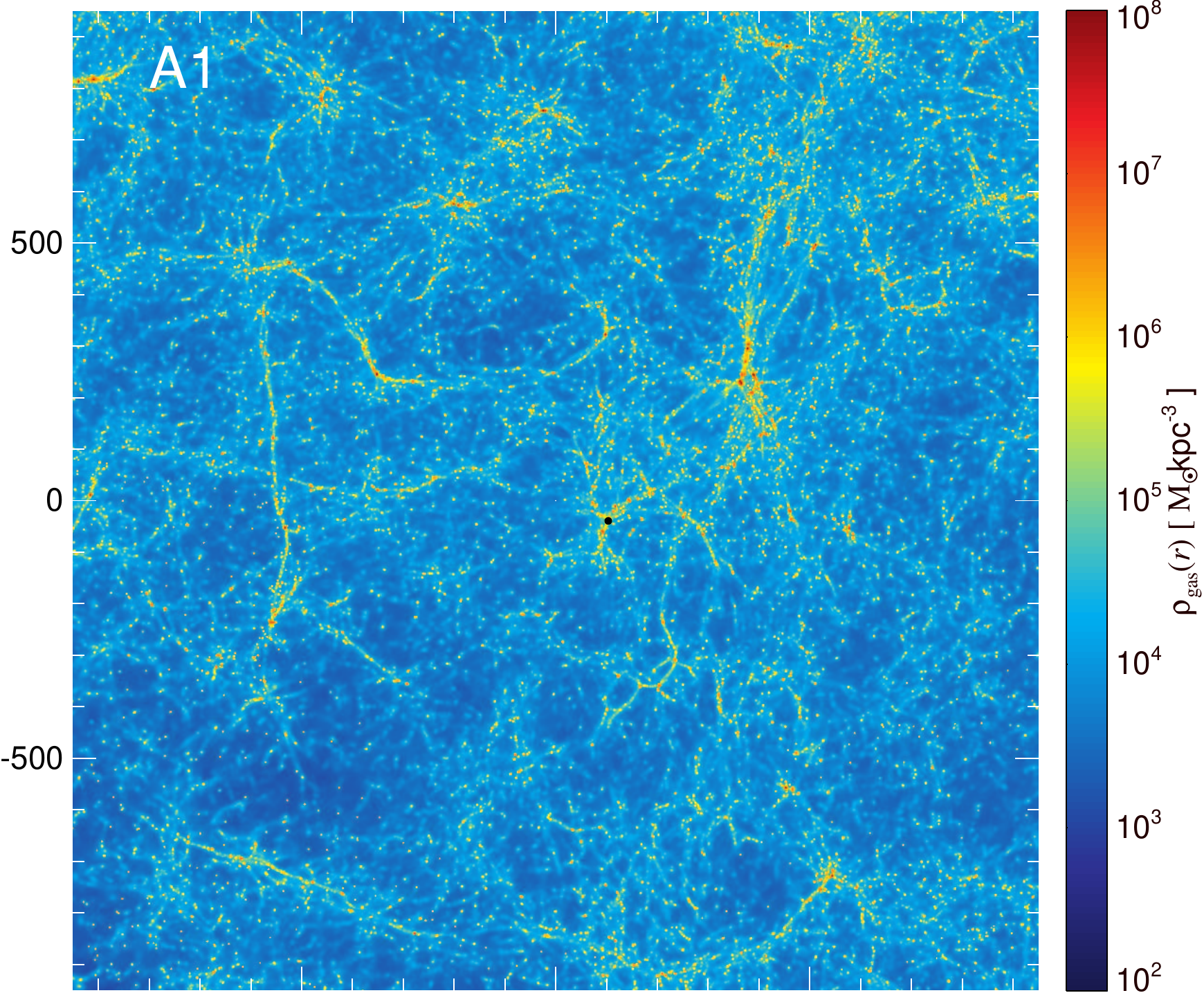}
\end{subfigure}%
\begin{subfigure}{0.5\textwidth}
\centering \includegraphics[scale = 0.48]{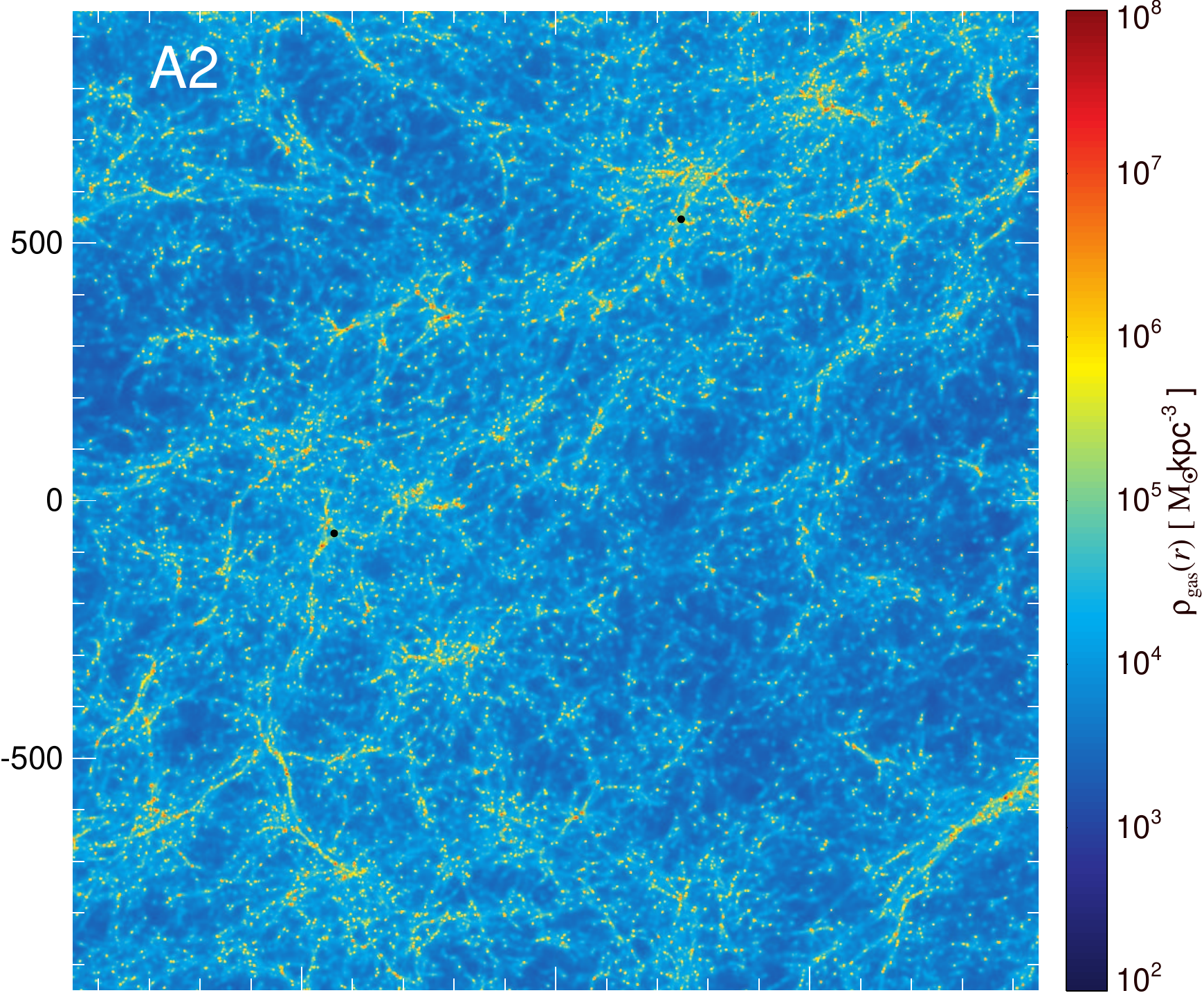}
\end{subfigure}\\

\begin{subfigure}{0.5\textwidth}
\centering \includegraphics[scale = 0.48]{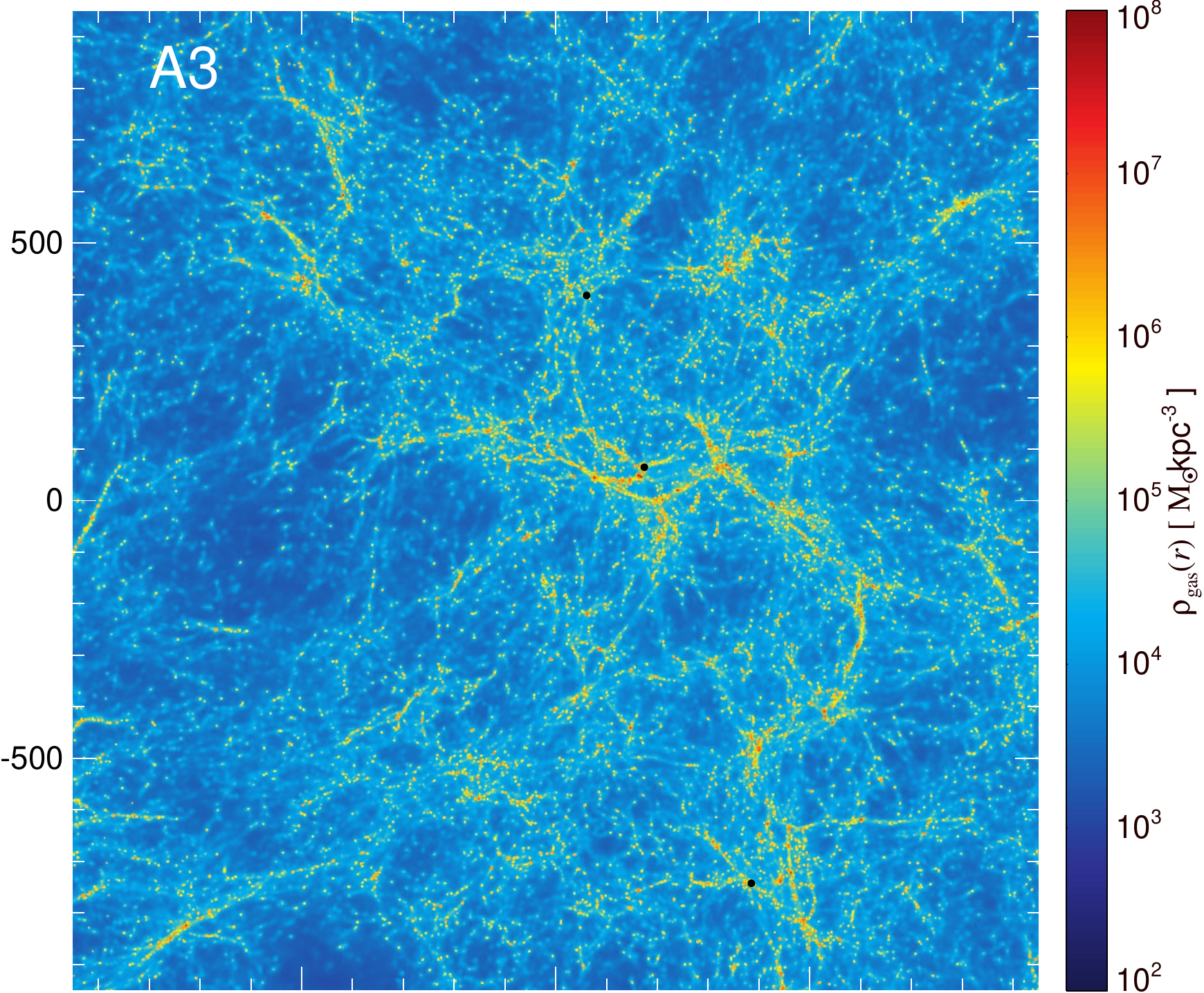}
\end{subfigure}%
\begin{subfigure}{0.5\textwidth}
\centering \includegraphics[scale = 0.48]{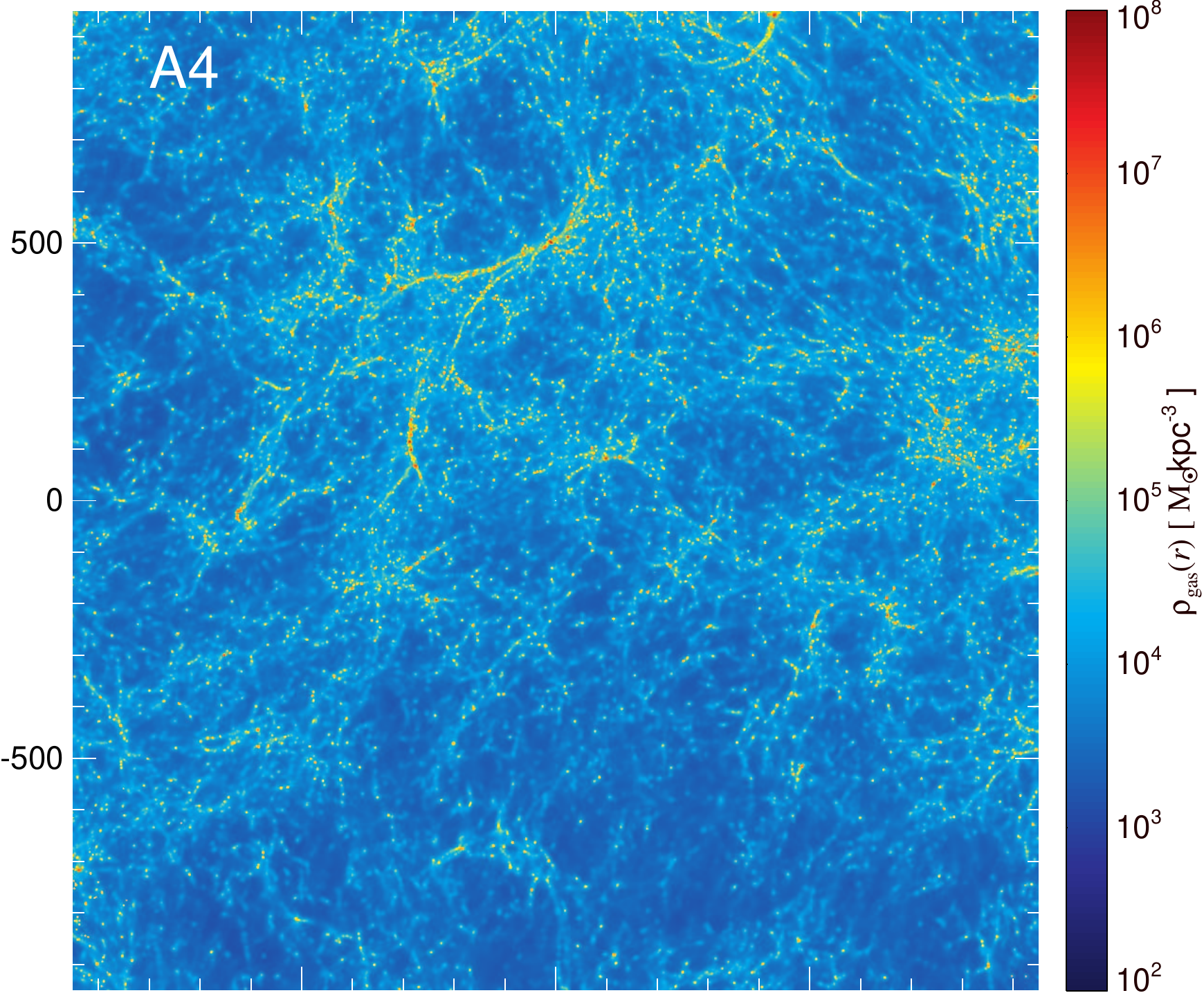}
\end{subfigure}\\

\begin{subfigure}{0.5\textwidth}
\centering \includegraphics[scale = 0.48]{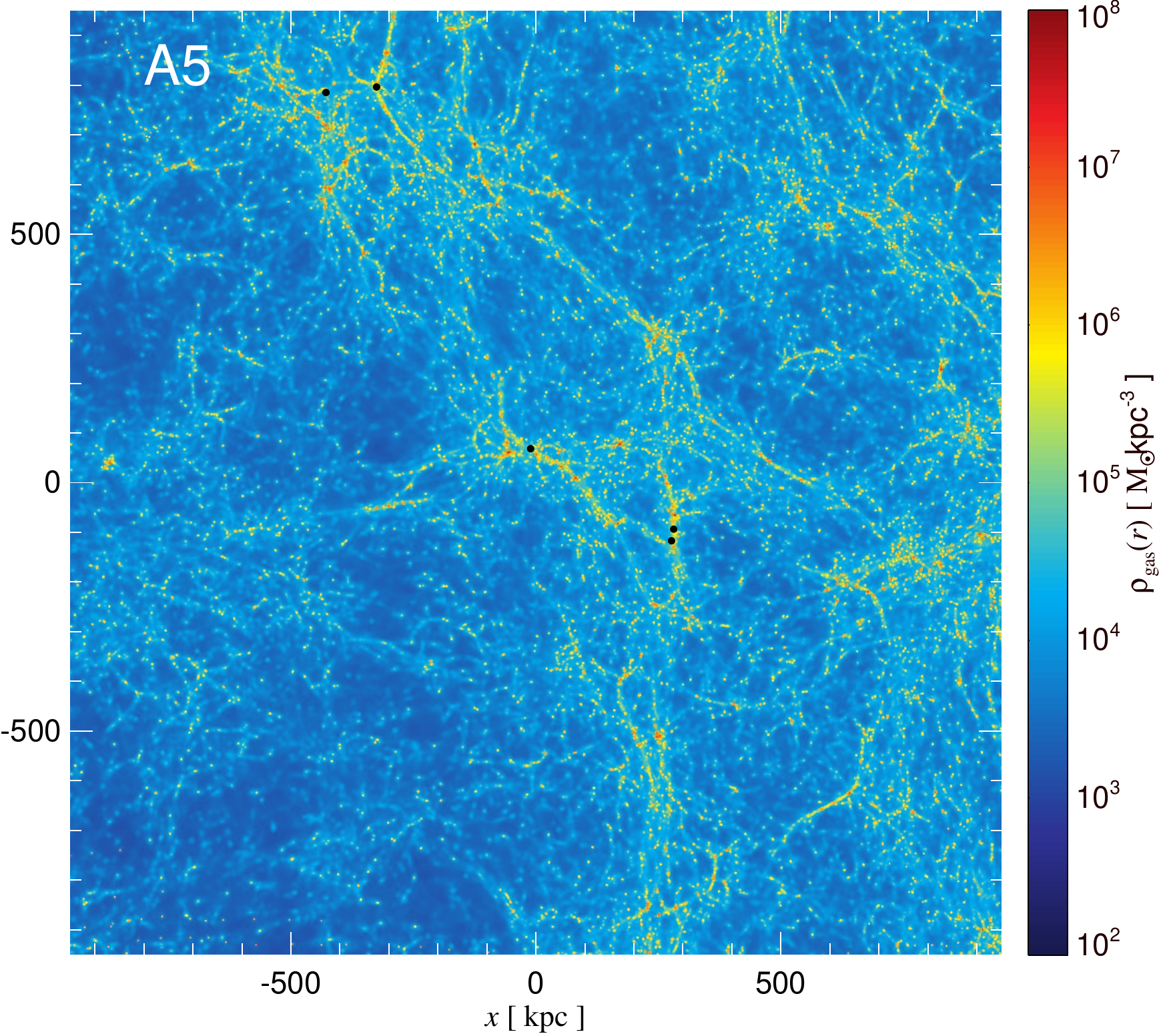}
\end{subfigure}%
\begin{subfigure}{0.5\textwidth}
\centering \includegraphics[scale = 0.48]{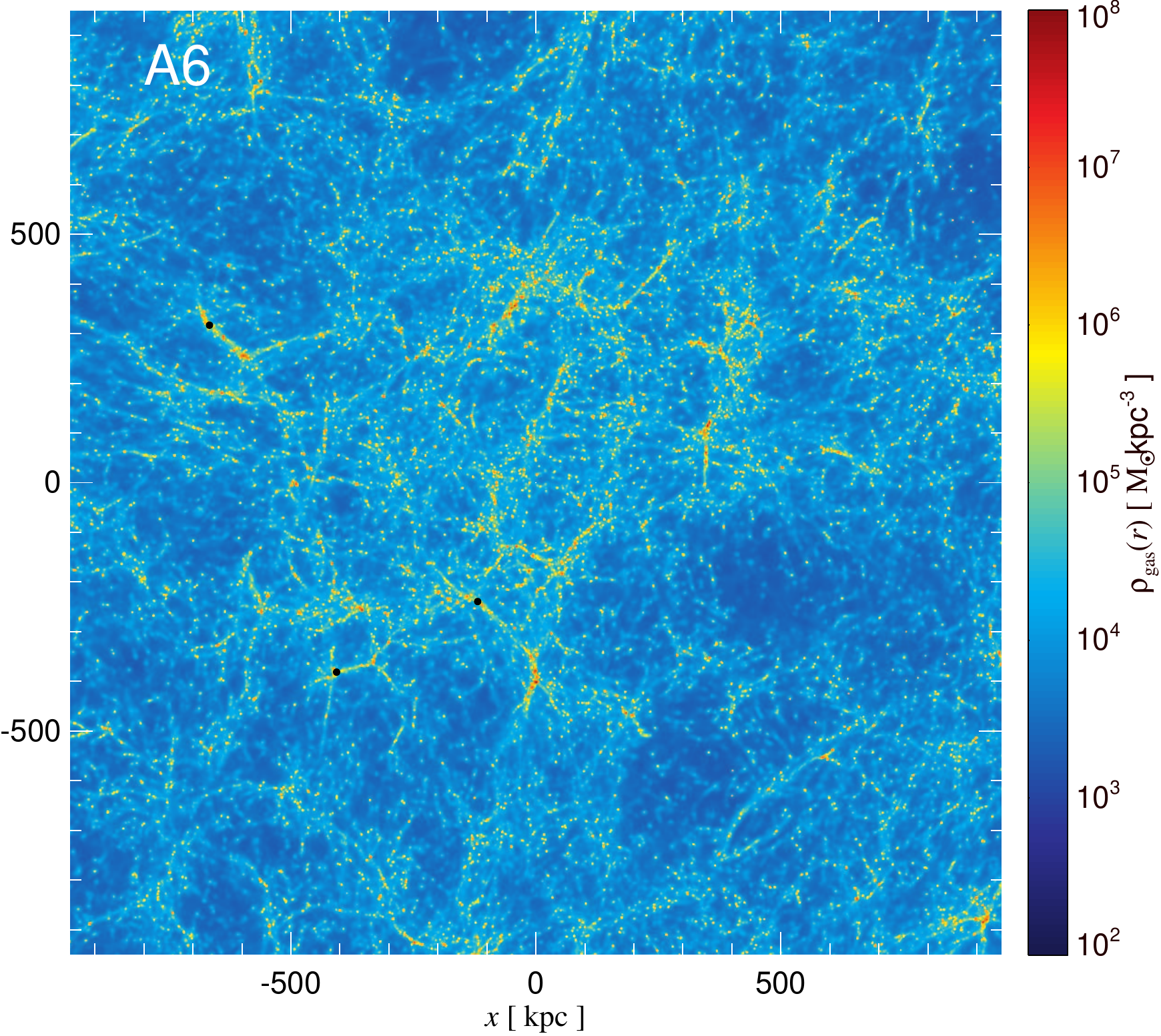}
\end{subfigure}%
\caption{Mass-weighted maps of the gas density projected along a slice
  of thickness $\approx 1.9 \, \mathrm{Mpc} \, (10 h^{-1} \,
  \mathrm{comoving \, Mpc})$ in six regions of average density at $z
  \,=\, 6.2$. Black dots denote the location of black hole particles
  within the projected cube and their size is proportional to black
  hole mass with {\fontsize{4}{4}\selectfont $\bullet[1pt]$}
  representing black holes with mass in the range $10^{6 \-- 7} \,
  \mathrm{M_{\odot}}$ and {\huge$\bullet[-1pt]$}, black holes with
  mass $\geq 10^9 \, \mathrm{M_{\odot}}$. Dark matter halos (and hence
  black holes) form most efficiently along dense filaments that
  characterise the large scale structure. The average density regions
  shown here do not, however, give rise to black holes more massive
  than a few times $10^6 \, \mathrm{M_{\odot}}$. Note 
  the substantial variations of the black hole abundance between the
  different regions due to cosmic variance.}
\label{average_df}
\end{figure*}

\begin{figure*}
\begin{subfigure}{0.5\textwidth}
\centering \includegraphics[scale = 0.48]{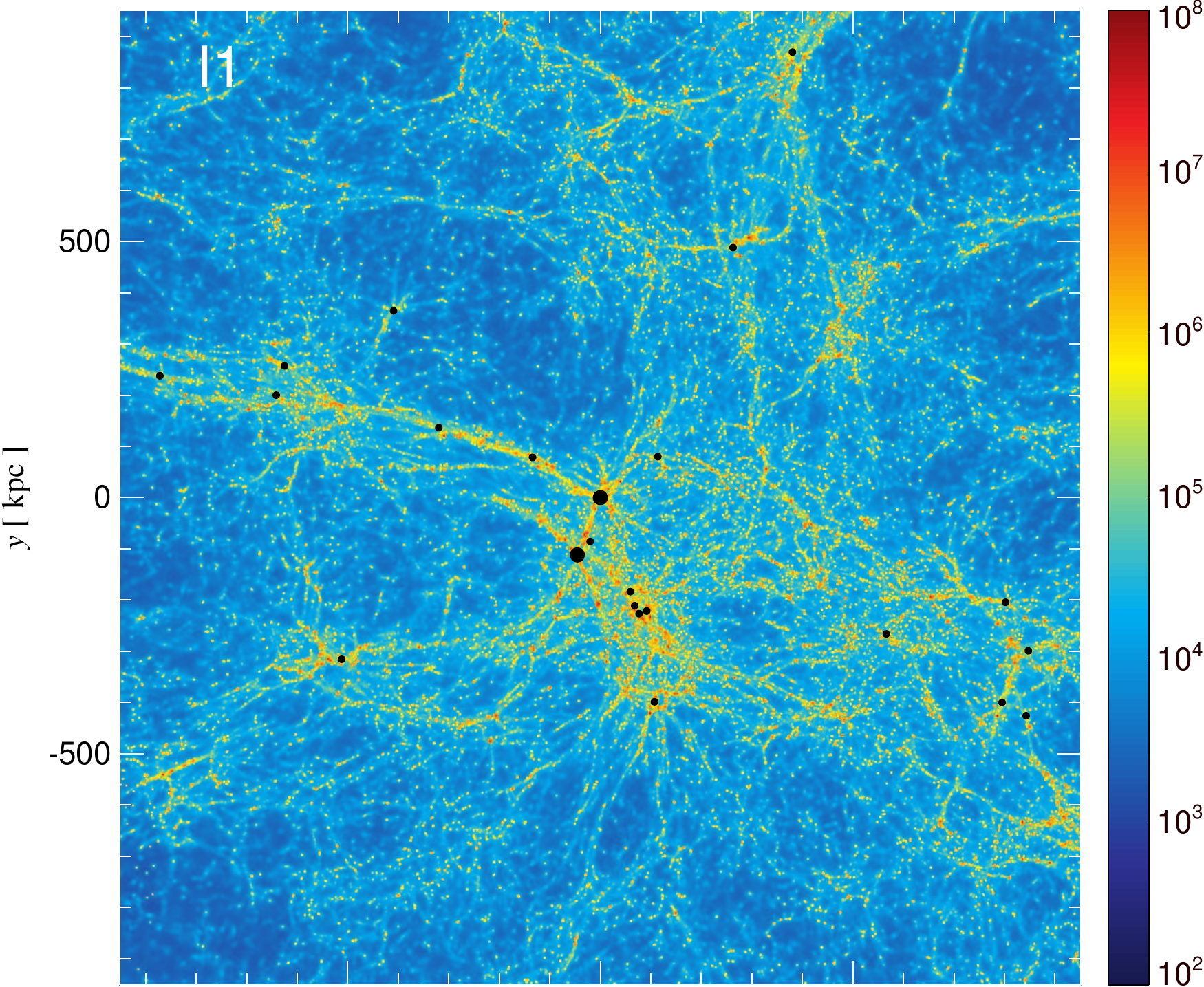}
\end{subfigure}%
\begin{subfigure}{0.5\textwidth}
\centering \includegraphics[scale = 0.48]{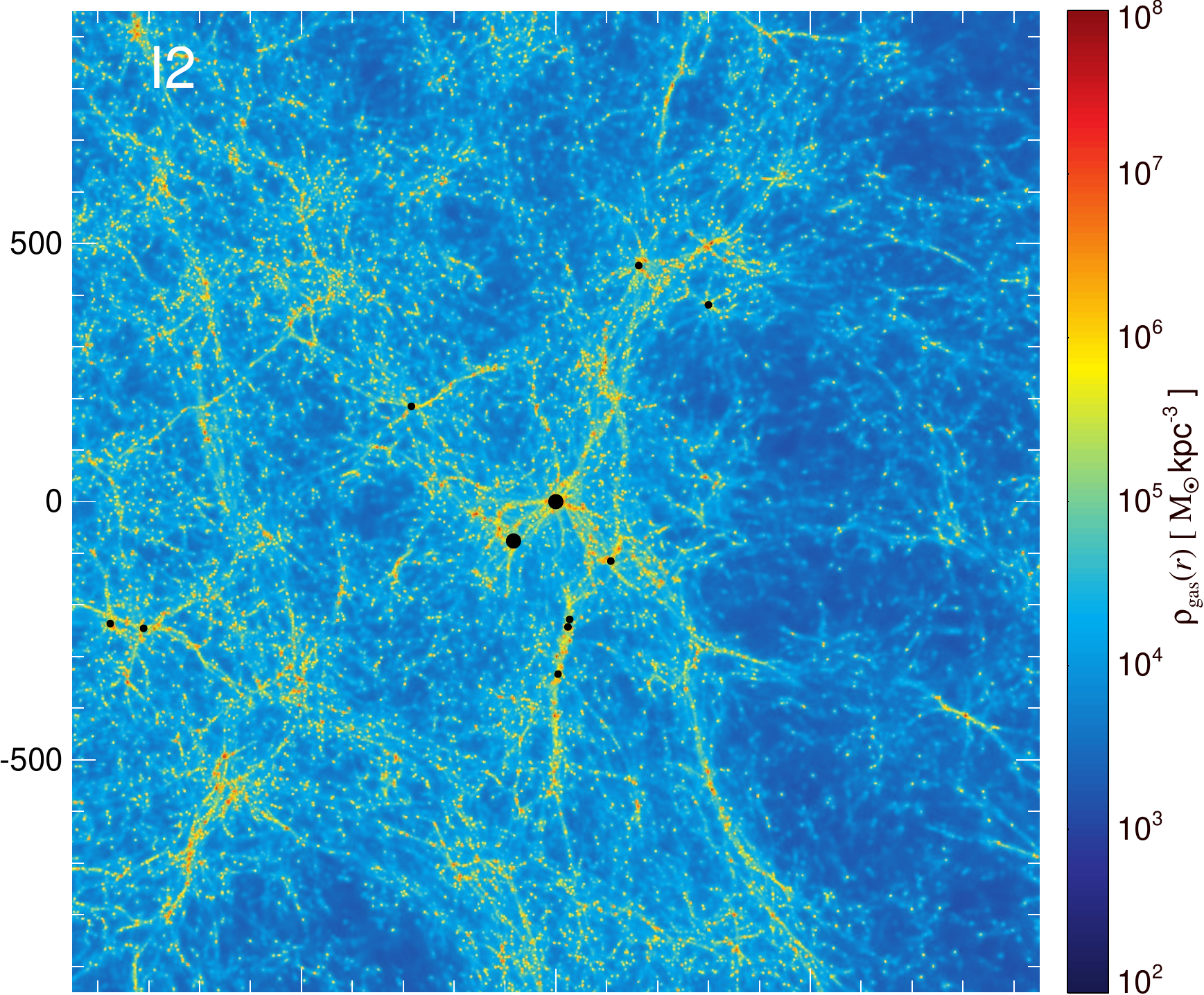}
\end{subfigure}\\

\begin{subfigure}{0.5\textwidth}
\centering \includegraphics[scale = 0.48]{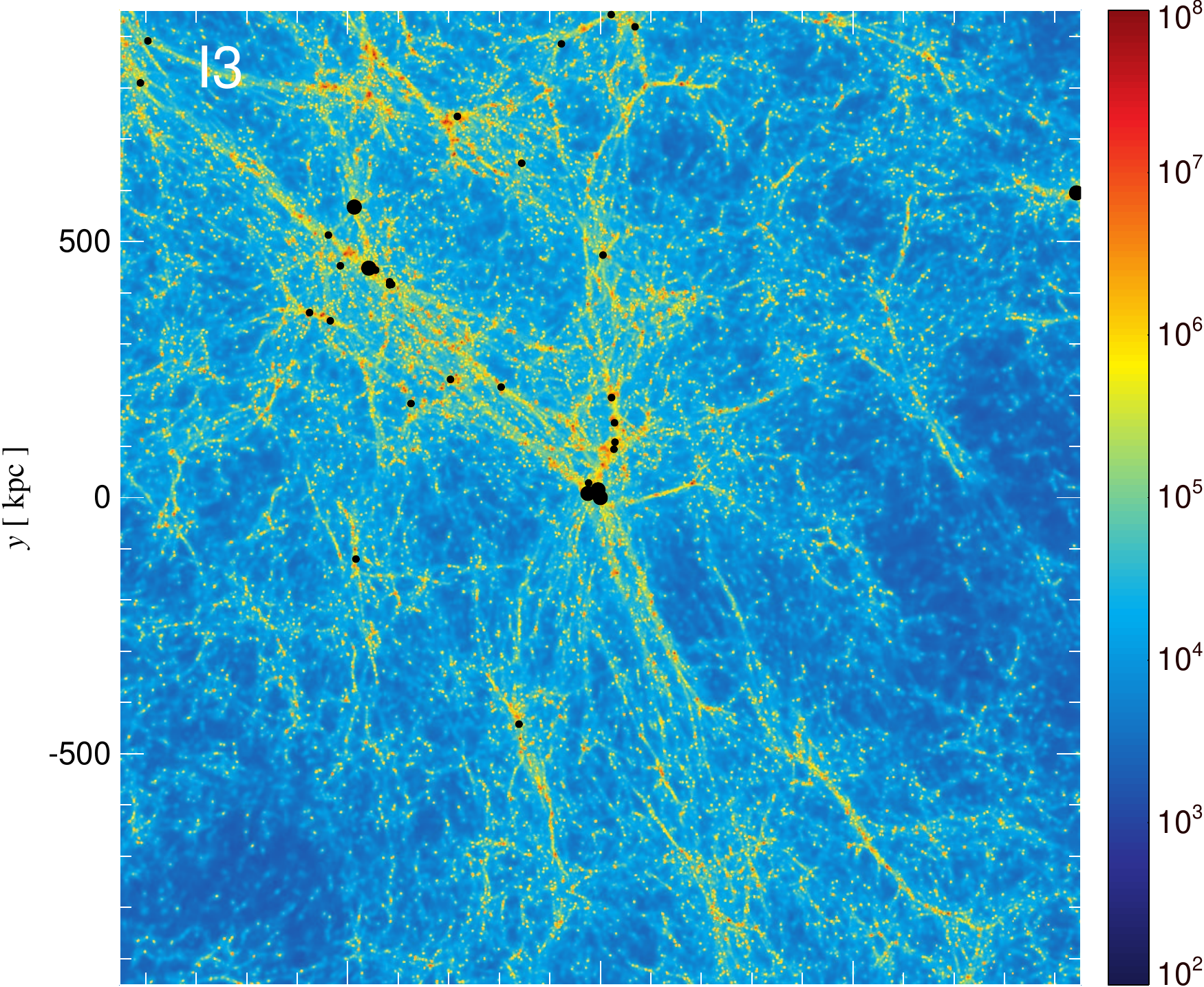}
\end{subfigure}%
\begin{subfigure}{0.5\textwidth}
\centering \includegraphics[scale = 0.48]{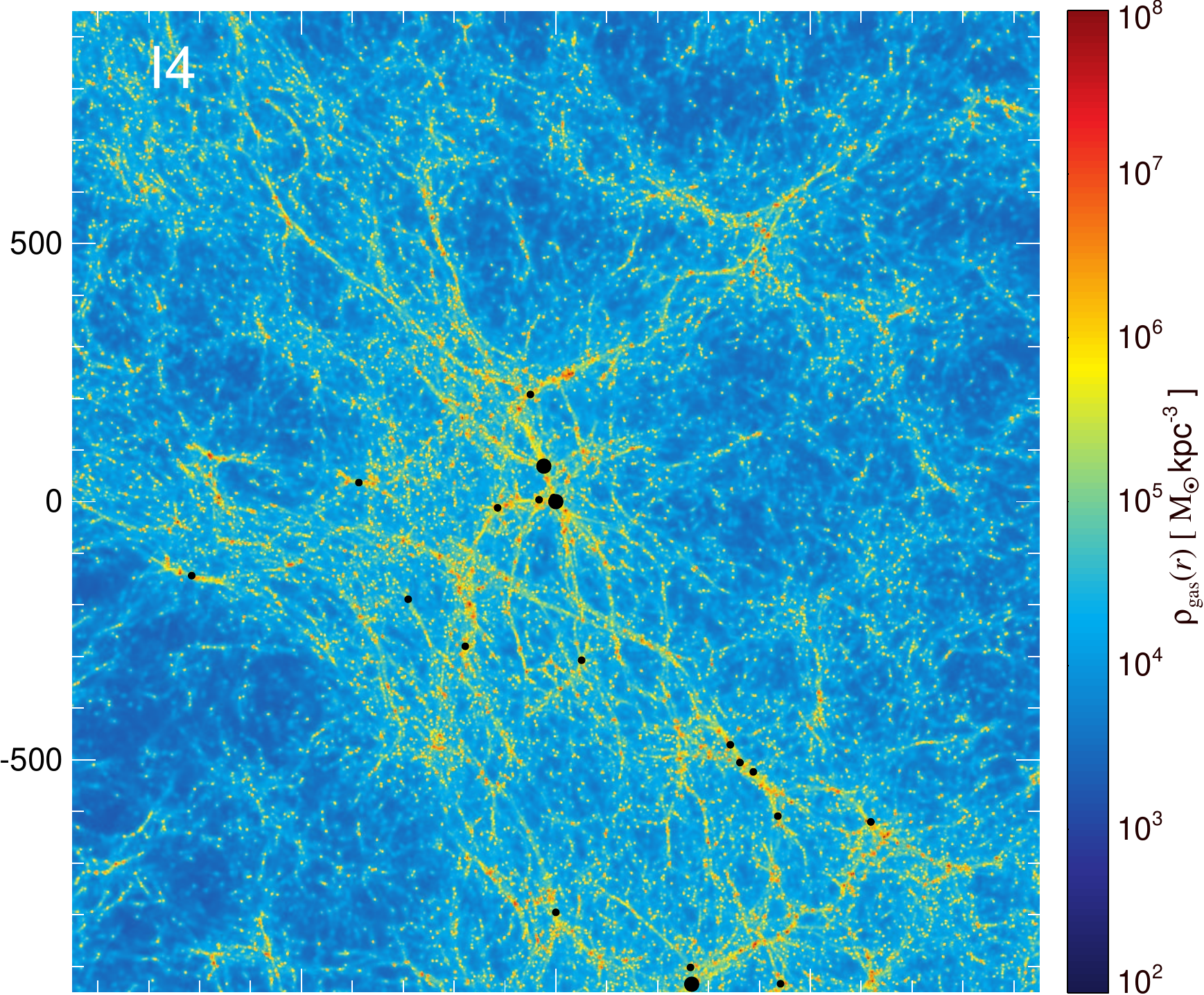}
\end{subfigure}\\

\begin{subfigure}{0.5\textwidth}
\centering \includegraphics[scale = 0.48]{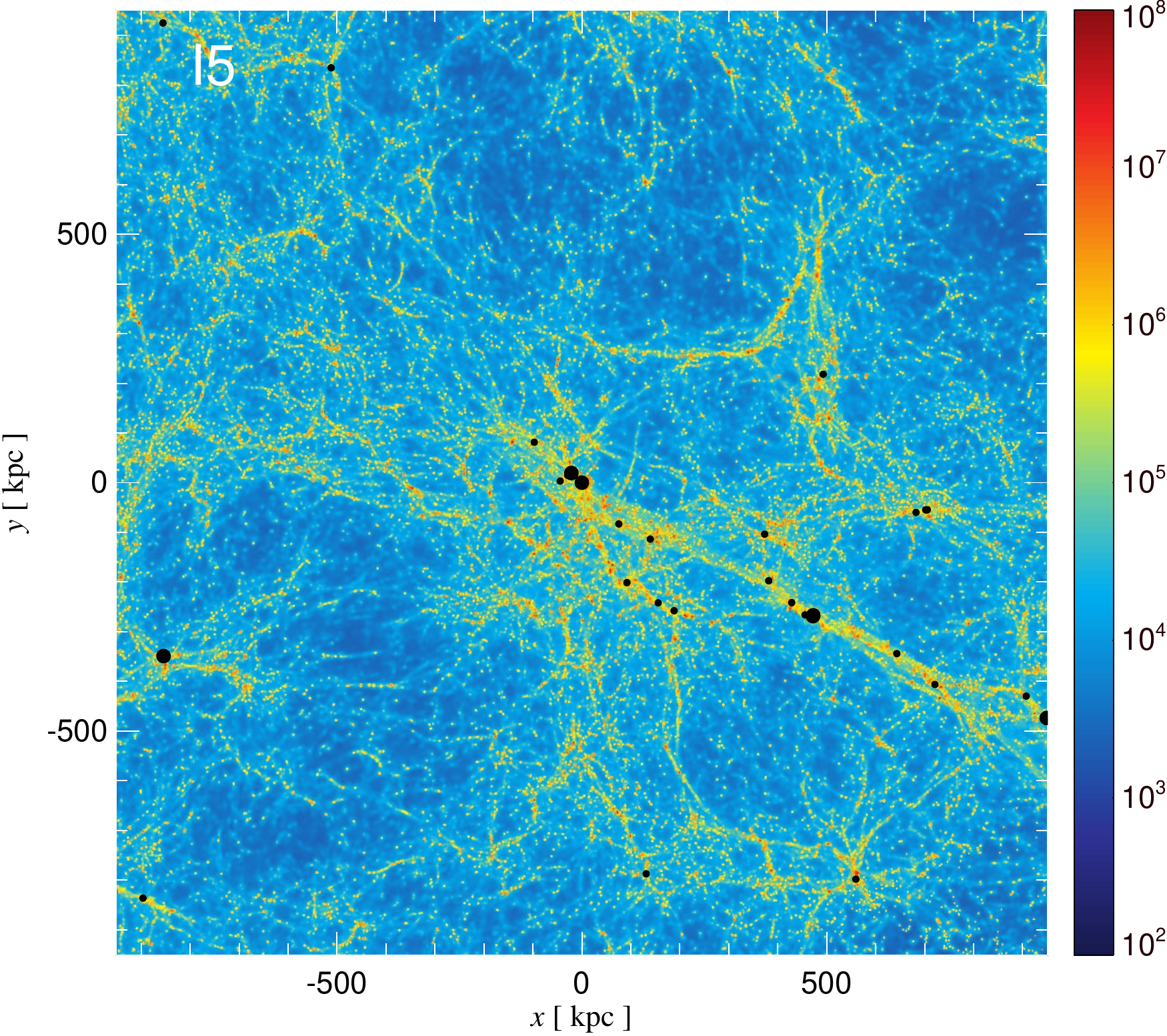}
\end{subfigure}%
\begin{subfigure}{0.5\textwidth}
\centering \includegraphics[scale = 0.48]{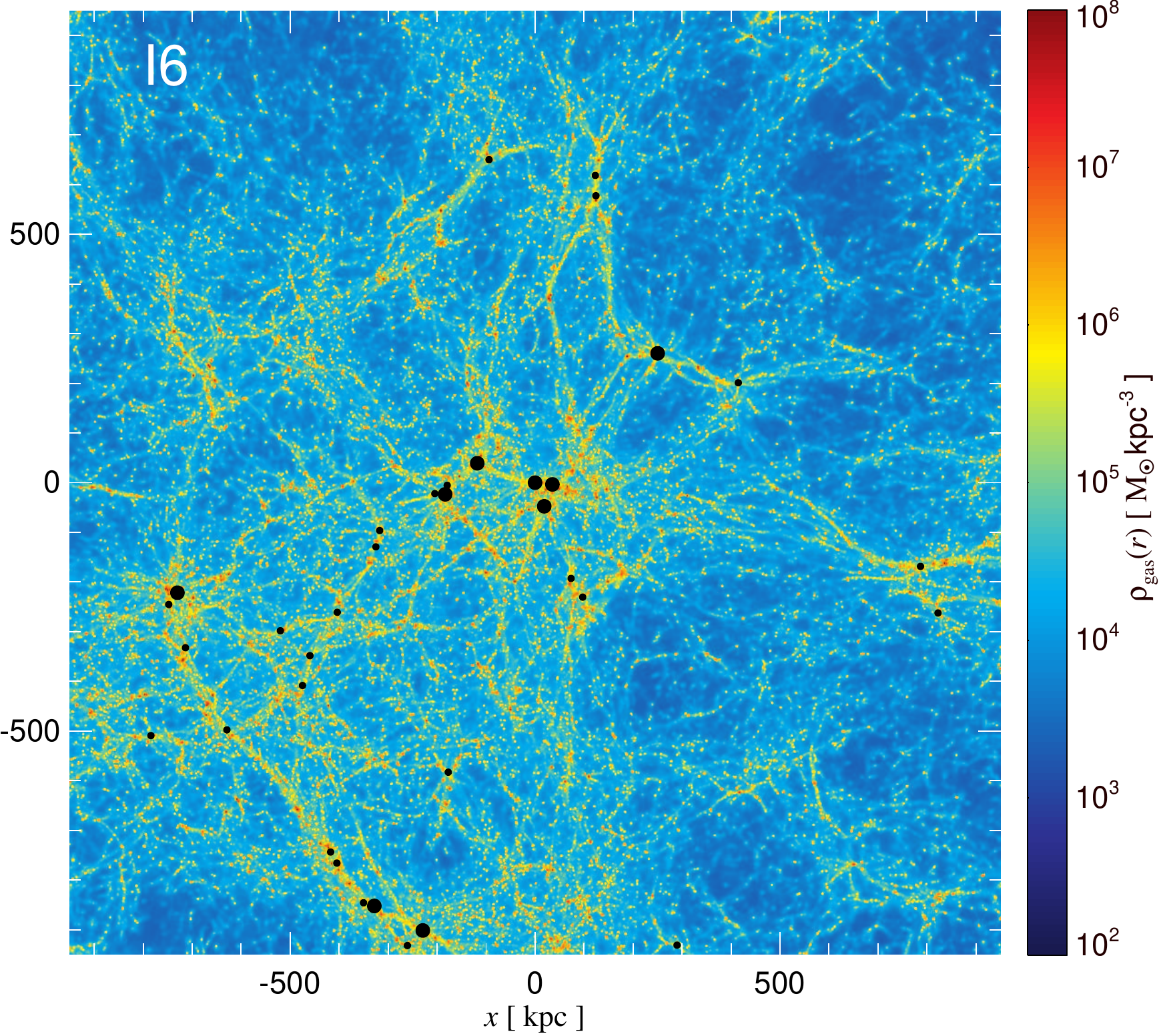}
\end{subfigure}%
\caption{Mass-weighted maps of the gas density projected along a slice
  of thickness $\approx 1.9 \, \mathrm{Mpc} \, (10 h^{-1} \,
  \mathrm{comoving \, Mpc})$ in six regions of intermediate
  overdensity at $z \,=\, 6.2$. Black dots denote the location of
  black hole particles within the projected cube and their size is
  proportional to black hole mass with {\fontsize{4}{4}\selectfont
    $\bullet[1pt]$} representing black holes with mass in the range
  $10^{6 \-- 7} \, \mathrm{M_{\odot}}$ and {\huge$\bullet[-1pt]$},
  black holes with mass $\geq 10^9 \, \mathrm{M_{\odot}}$. The typical
  black hole mass in these regions is a few times $10^7 \,
  \mathrm{M_{\odot}}$, an order of magnitude higher than in the
  regions of average density. Note 
  the substantial variations of the black hole abundance between the
  different regions due to cosmic variance.}
\label{intermediate_df}
\end{figure*}

\begin{figure*}
\begin{subfigure}{0.5\textwidth}
\centering \includegraphics[scale = 0.48]{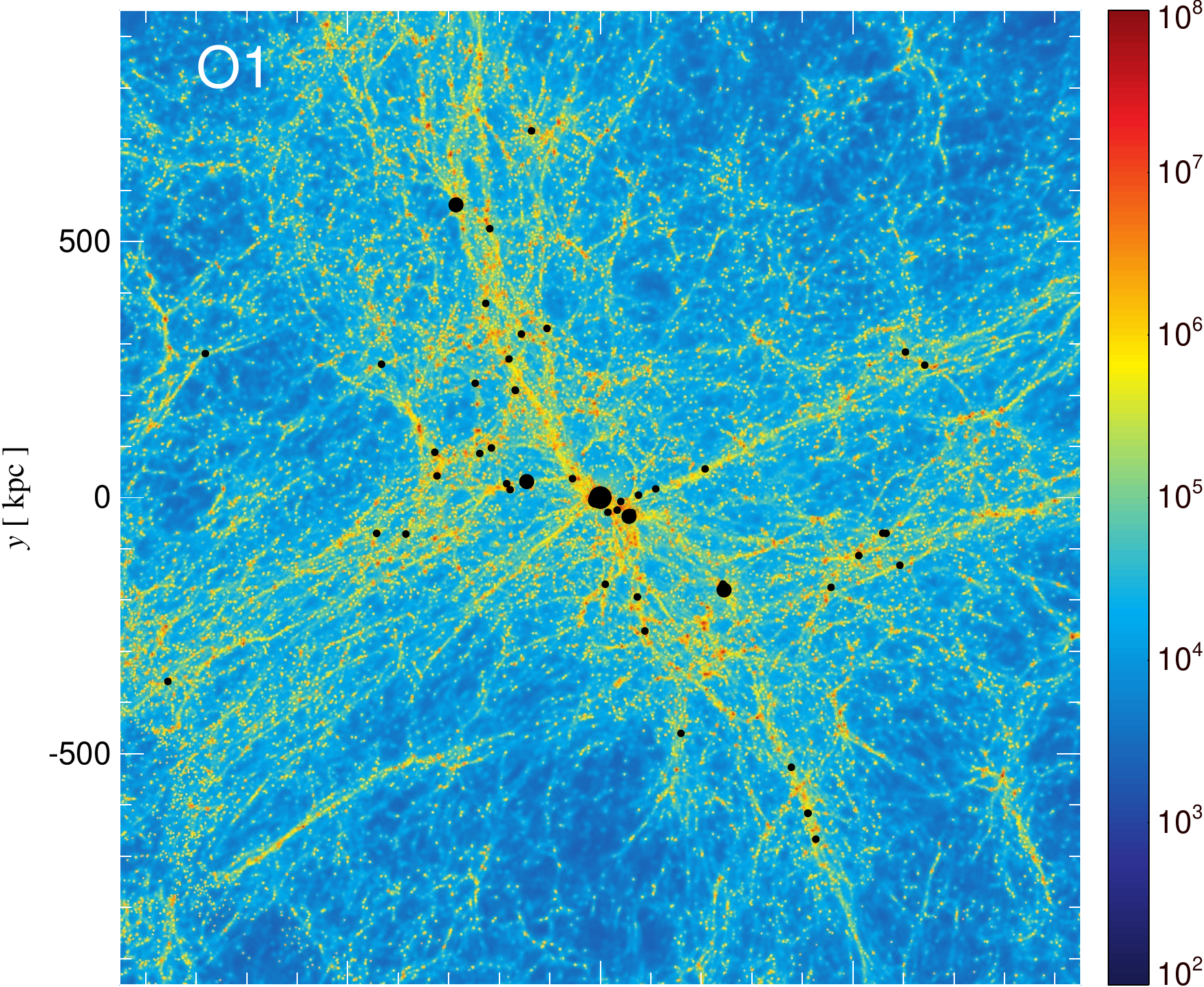}
\end{subfigure}%
\begin{subfigure}{0.5\textwidth}
\centering \includegraphics[scale = 0.48]{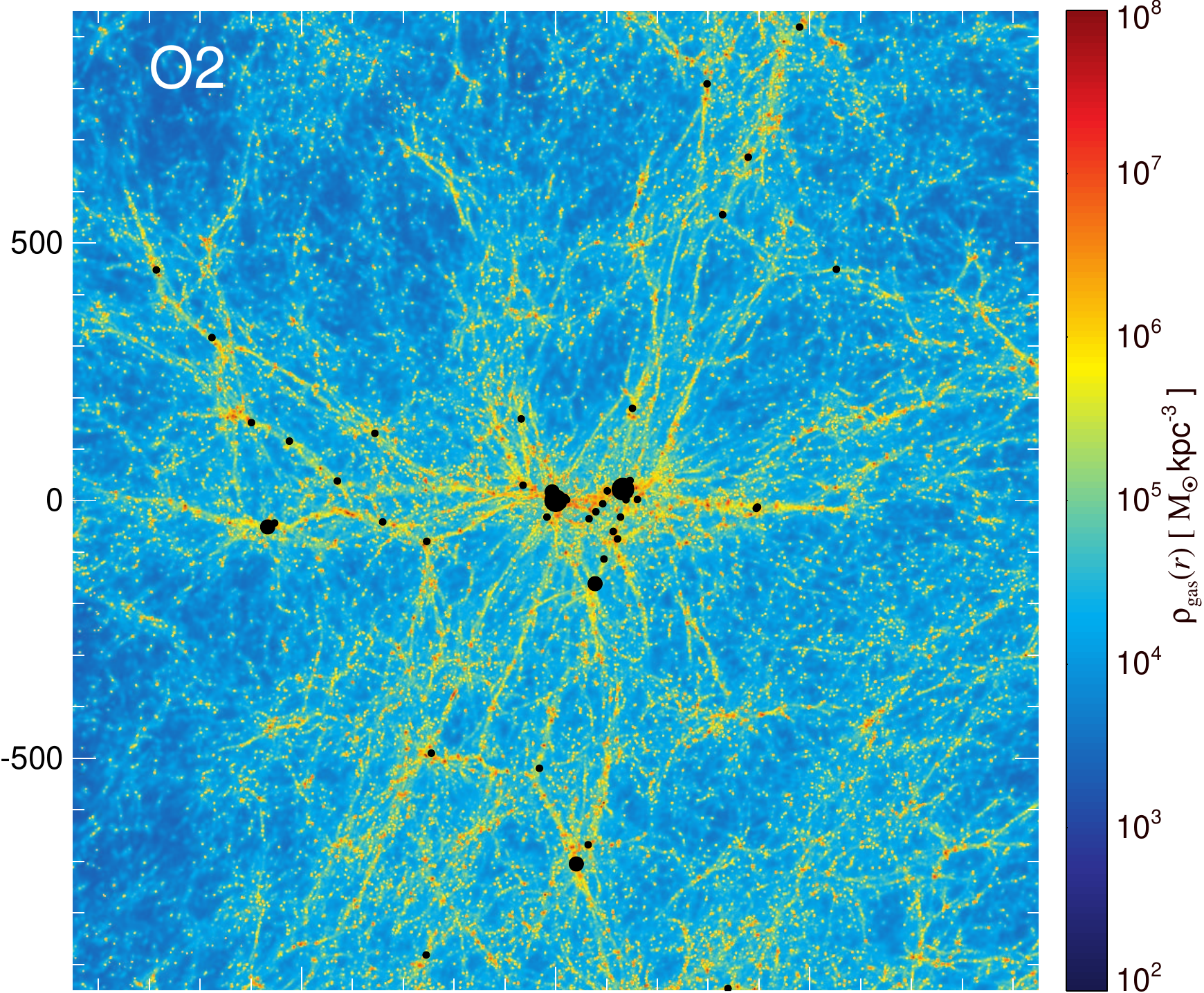}
\end{subfigure}\\

\begin{subfigure}{0.5\textwidth}
\centering \includegraphics[scale = 0.48]{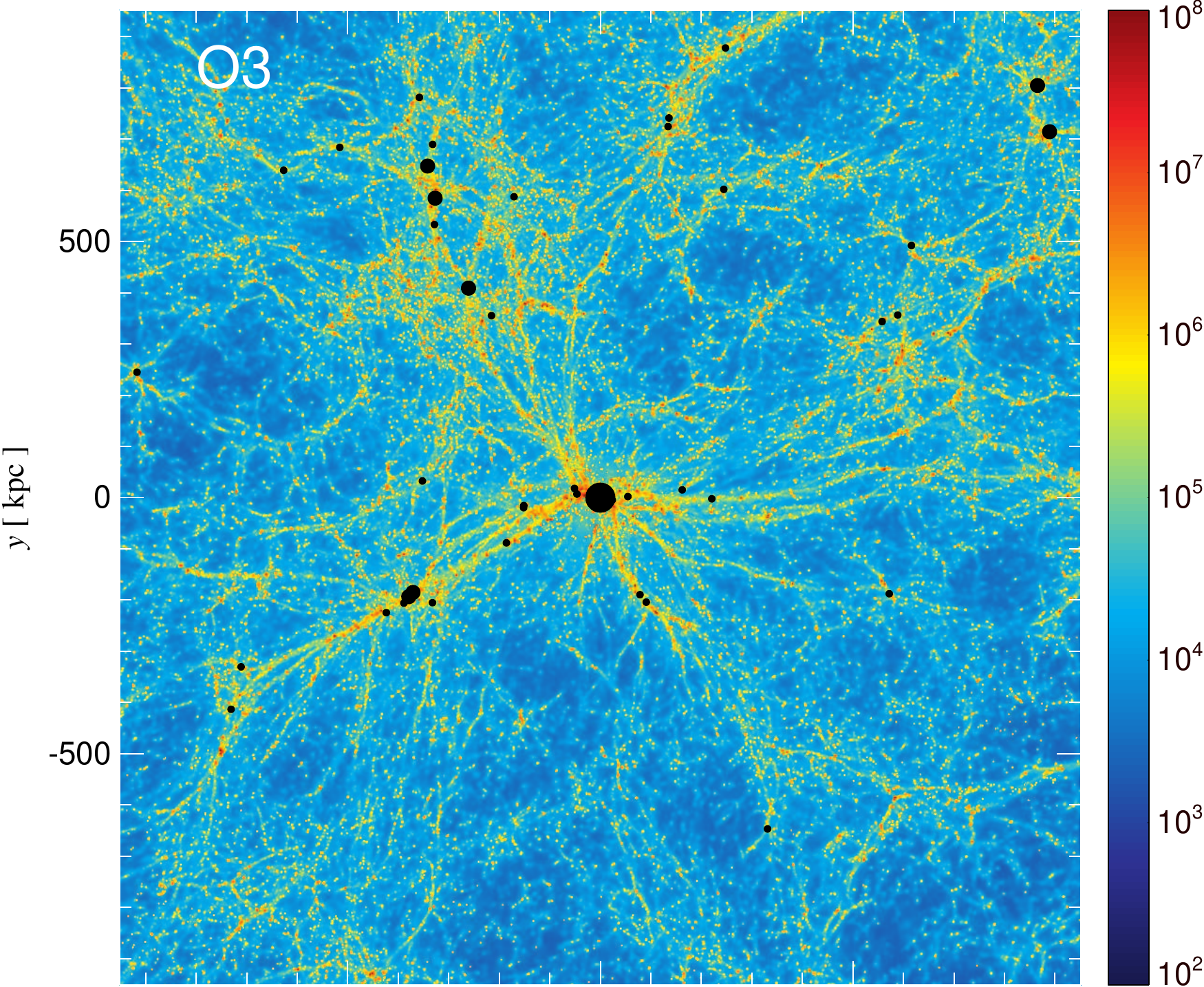}
\end{subfigure}%
\begin{subfigure}{0.5\textwidth}
\centering \includegraphics[scale = 0.48]{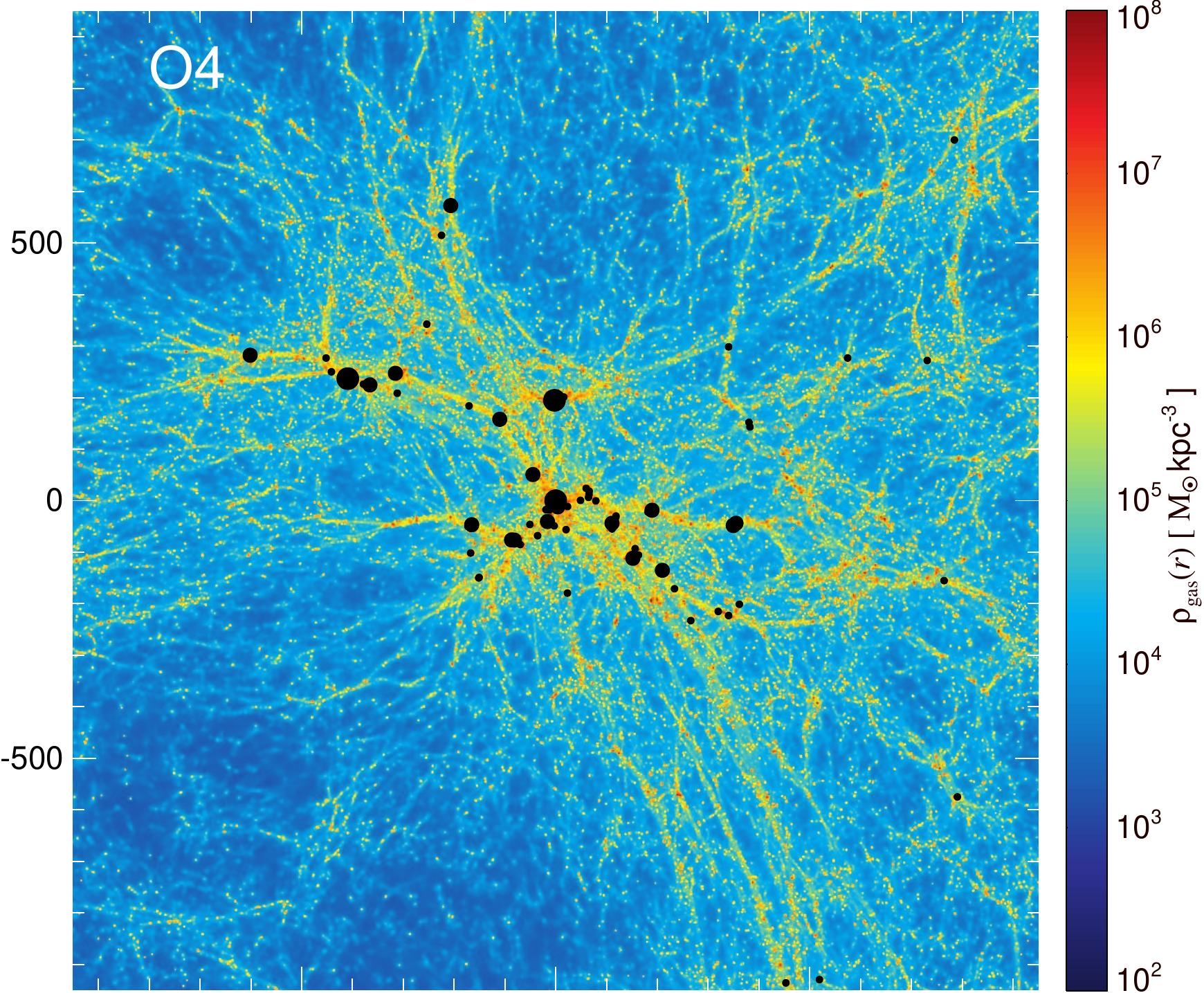}
\end{subfigure}\\

\begin{subfigure}{0.5\textwidth}
\centering \includegraphics[scale = 0.48]{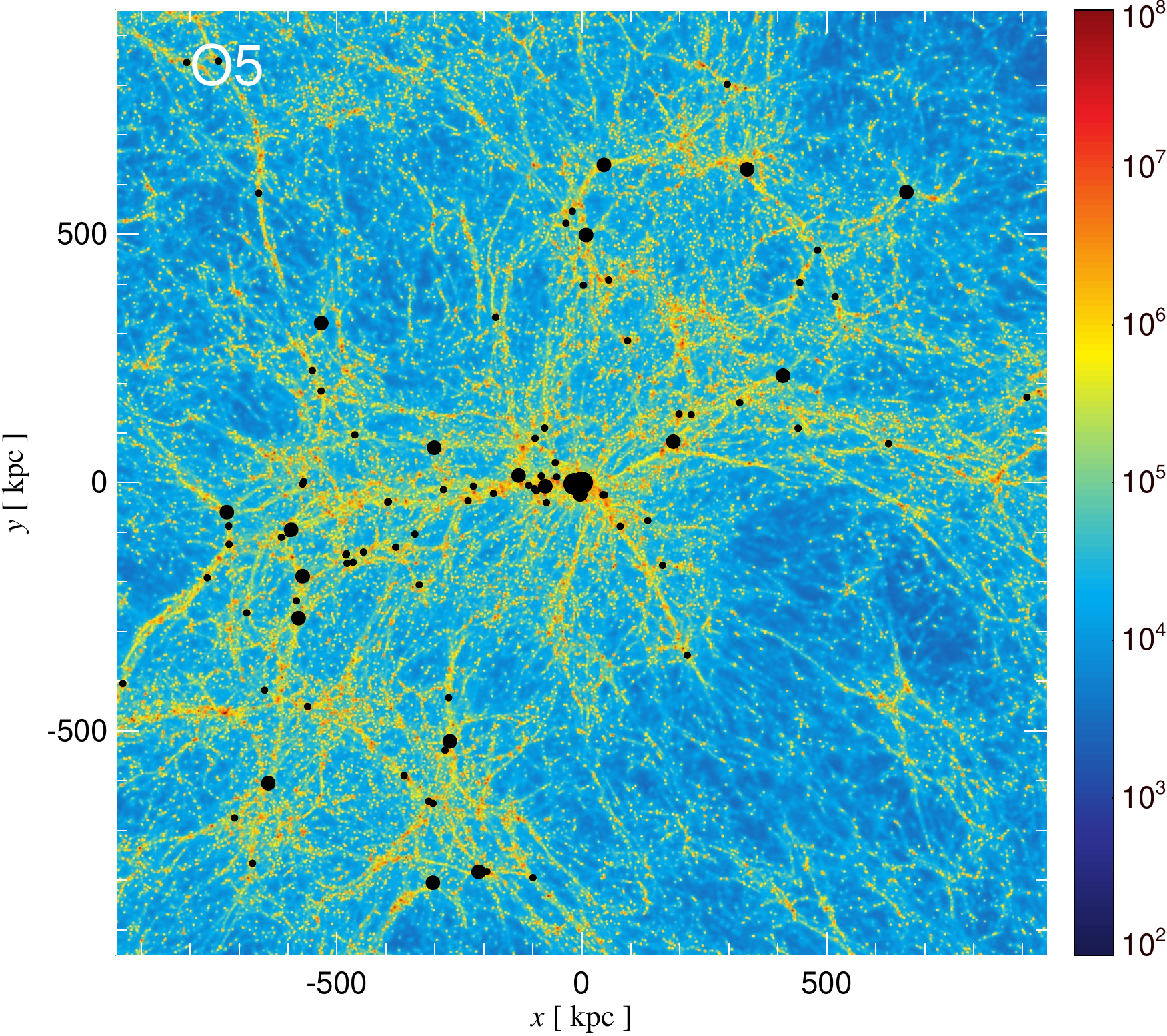}
\end{subfigure}%
\begin{subfigure}{0.5\textwidth}
\centering \includegraphics[scale = 0.48]{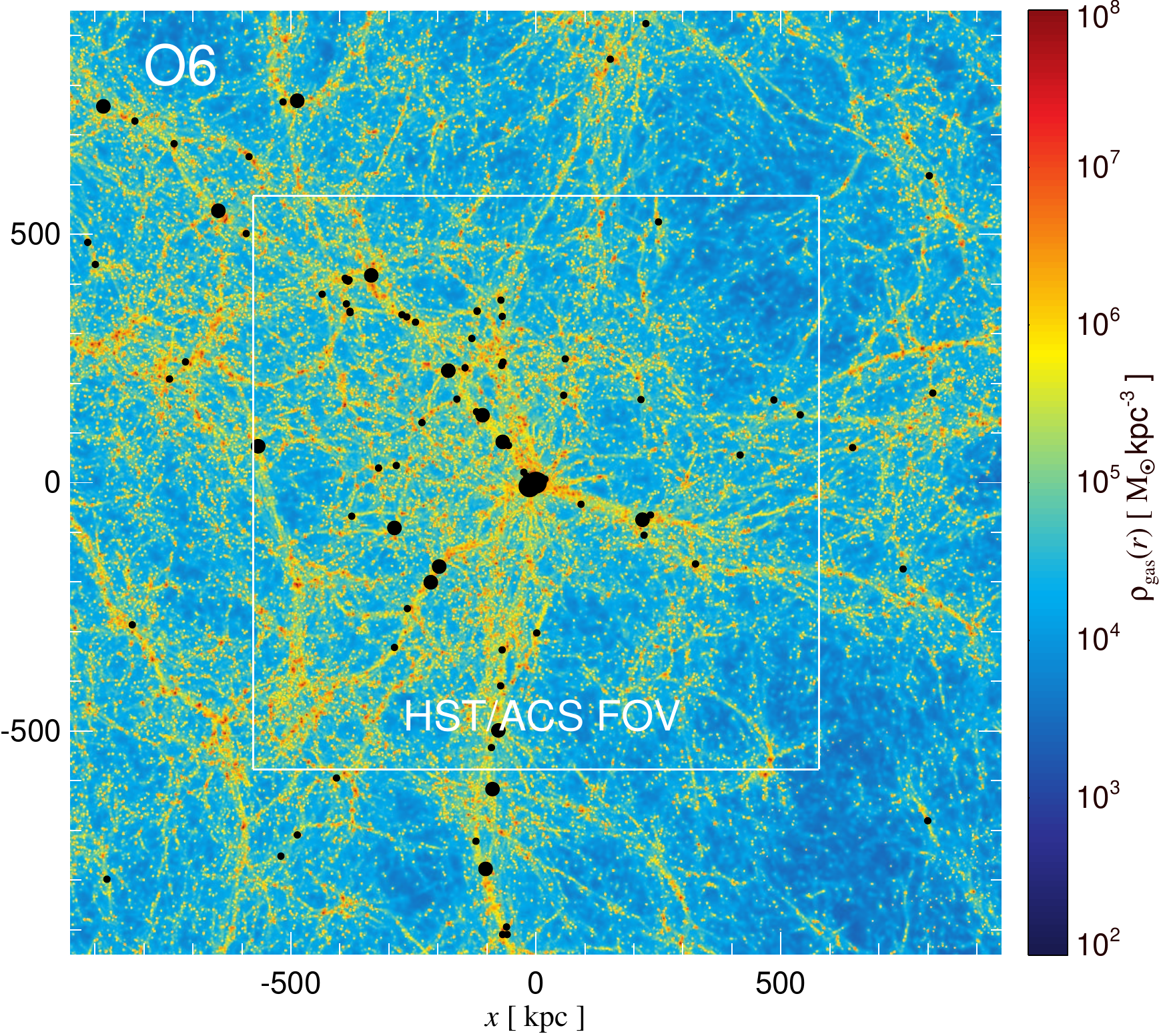}
\end{subfigure}%
\caption{Mass-weighted maps of the gas density projected along a slice
  of thickness $\approx 1.9 \, \mathrm{Mpc} \, (10 h^{-1} \,
  \mathrm{comoving \, Mpc})$ in six overdense regions at $z \,=\,
  6.2$. Black dots denote the location of black hole particles within
  the projected cube and their size is proportional to black hole mass
  with {\fontsize{4}{4}\selectfont $\bullet[1pt]$} representing black
  holes with mass in the range $10^{6 \-- 7} \, \mathrm{M_{\odot}}$
  and {\huge$\bullet[-2pt]$}, black holes with mass $\geq 10^9 \,
  \mathrm{M_{\odot}}$. The most massive black holes form in halos with
  $M_{200} \approx 4 \times 10^{12} \, \mathrm{M_{\odot}}$ at the
  intersection of very prominent filaments. Note 
  the substantial variations of the black hole abundance between the
  different regions due to cosmic variance. The white square in the O$6$ region shows the field
  of view of HST/ACS.}
\label{overdense_df}
\end{figure*}

A visual impression of our average density, intermediate density and
highly overdense regions is shown in Figures \ref{average_df},
\ref{intermediate_df} and \ref{overdense_df}, respectively. The
density field in the average density regions is characterised by thin
filaments hosting low mass halos and large voids with relatively
little structure.  For the intermediate sample, the filaments become
better defined and the voids appear less pronounced as the overdensity
increases.  In the overdense regions, the yet more prominent
filaments intersect in central massive halos with total masses
$M_{200} \approx 2 \-- 5 \times 10^{12} \, \mathrm{M_{\odot}}$. Note
also that the gas density is increased on virtually all scales.  The
more diffuse gas in the highly overdense regions appears also denser
than in the average or intermediate regions. The filled circles
overplotted on the maps denote the positions of black hole particles,
and their size is proportional to the black hole mass. The smallest
circles correspond to a black hole mass of $M_{\mathrm{BH}} \,=\, 10^6 - 10^7
\, \mathrm{M_{\odot}}$ and the largest to $M_{\mathrm{BH}} \ge 10^9
\, \mathrm{M_{\odot}}$. The strong  clustering  of black holes in the
overdense regions reflects the strong clustering of the underlying
dark matter halos which are massive enough to accumulate sufficient
amounts of gas to facilitate efficient black hole growth. The most
massive black holes, typically located in the most massive halos at
the intersection of the most prominent filaments \citep[see also][]{Dubois:12} of our overdense
regions, have masses of $\approx 10^8 \-- 10^9 \,
\mathrm{M_{\odot}}$. Even though our average density regions also
contain black holes, they occur in significantly smaller  numbers (in
the case of the A$4$ region, none in the volume shown) and have much lower
masses. While in the case of the A$1$, A$4$ and A$6$ regions there is one
black hole with a mass of a few times $10^7 \, \mathrm{M_{\odot}}$
(outside of the projected volume shown in Figure~\ref{average_df}),
the most massive black holes forming in the average regions have grown
to a typical mass of only a few million solar masses by $z \sim
6$. Also note that among average, intermediate and overdense regions,
there is considerable scatter in black hole number density due to
cosmic variance. In the case of the O$3$ region, for example, the number
of black holes is smaller than in the other overdense fields, while
the central black hole is the most massive of our resimulated volumes
at $z \sim 6$. The  growth of supermassive black holes appears
therefore not only determined by the larger scale environment in
which they grow, a point to which we will come back later. 

In Table~\ref{table}, we list the masses of the most massive black
holes in each of our resimulated volumes as well as the total mass of
their host dark matter halo. We also provide an estimate of the
overdensity of each region, which we computed by simply taking the
ratio of the average matter density to the average cosmic density
inside a sphere of radius $5 h^{-1} \, \mathrm{comoving \, Mpc}$ and
$8 h^{-1} \, \mathrm{comoving \,Mpc}$ with origin at the centre of the
high-resolution region. Note that in the
  highly overdense regions the overdensities are already (mildly)
  non-linear even on scales of $8h^{-1}$ Mpc and that the ratio of
  these overdensities to the rms fluctuation amplitude  $\sigma$
  ranges from $2.8$ to $7.4$.

A look at Table~\ref{table} further shows a correlation between overdensity and black hole mass,
with more massive black holes generally being found in more overdense environments. However, scatter is significant.
For instance, regions O$2$
and O$3$ contain black holes with masses $M_{\rm BH} \approx 6.5
\times 10^8 \, \mathrm{M_{\odot}}$ and $M_{\rm BH}\approx 1.2 \times
10^9 \, \mathrm{M_{\odot}}$, respectively. These masses are
significantly higher than those found for the most massive black holes
in regions O$5$ and O$6$, even though the former are less overdense. For the latter, the FoF
groups that host these QSOs are also less massive than those of the
most massive black holes in the O$2$ or the O$3$ region. Despite less overdense than other regions, we have verified that the QSO
host halo in the O$3$ region had already collapsed at a higher
redshift, assembling half of its $z \,=\, 6.2$ mass at $z \sim 8.5$ rather 
than at $z \sim 7$ as is the case for the QSO host halos in the remaining 
five overdense regions. In the O$2$ region, instead, the QSO host halo experiences a 
major merger at $z \sim 7$ which boosts both halo and central black hole mass. 
Thus, while a
significant large-scale overdensity is required for the
formation of sufficiently massive host halos, the growth history of individual halos
also plays a role in determining the mass of massive black holes.

In our sample of regions with intermediate overdensity, dark matter
halos with masses in the range $3 \-- 7 \times 10^{11} \,
\mathrm{M_{\odot}}$ form by $z \approx 6$.  Black holes forming in
these halos have masses $4 \times 10^7 \-- 10^8 \,
\mathrm{M_{\odot}}$, up to an order of magnitude higher than the mass
of the most massive black holes found in our sample of regions of
average density. Clearly, despite residing in halos which are
relatively rare with a comoving spatial density of $\lesssim 10^{-3}
h^{-1} \, \mathrm{Mpc^{-3}}$, these black holes are not fed enough
gas to acquire masses comparable to those powering the
observed $z \sim 6$ QSOs, even when growing from massive
initial seeds.  In our model, we thus require QSOs to be located in
very massive halos, that form early enough to provide a sufficiently
deep potential well.  Only under such special circumstances can gas be
transported to the central regions of the halo in sufficient
quantities to grow massive black holes that can power bright QSOs at
$z \sim 6$.

\subsection{Mass growth and emission of the black hole population}\label{secresults2}

\subsubsection{Mass assembly histories}
\begin{figure*}
\begin{subfigure}{0.5\textwidth}
\centering \includegraphics[scale = 0.5]{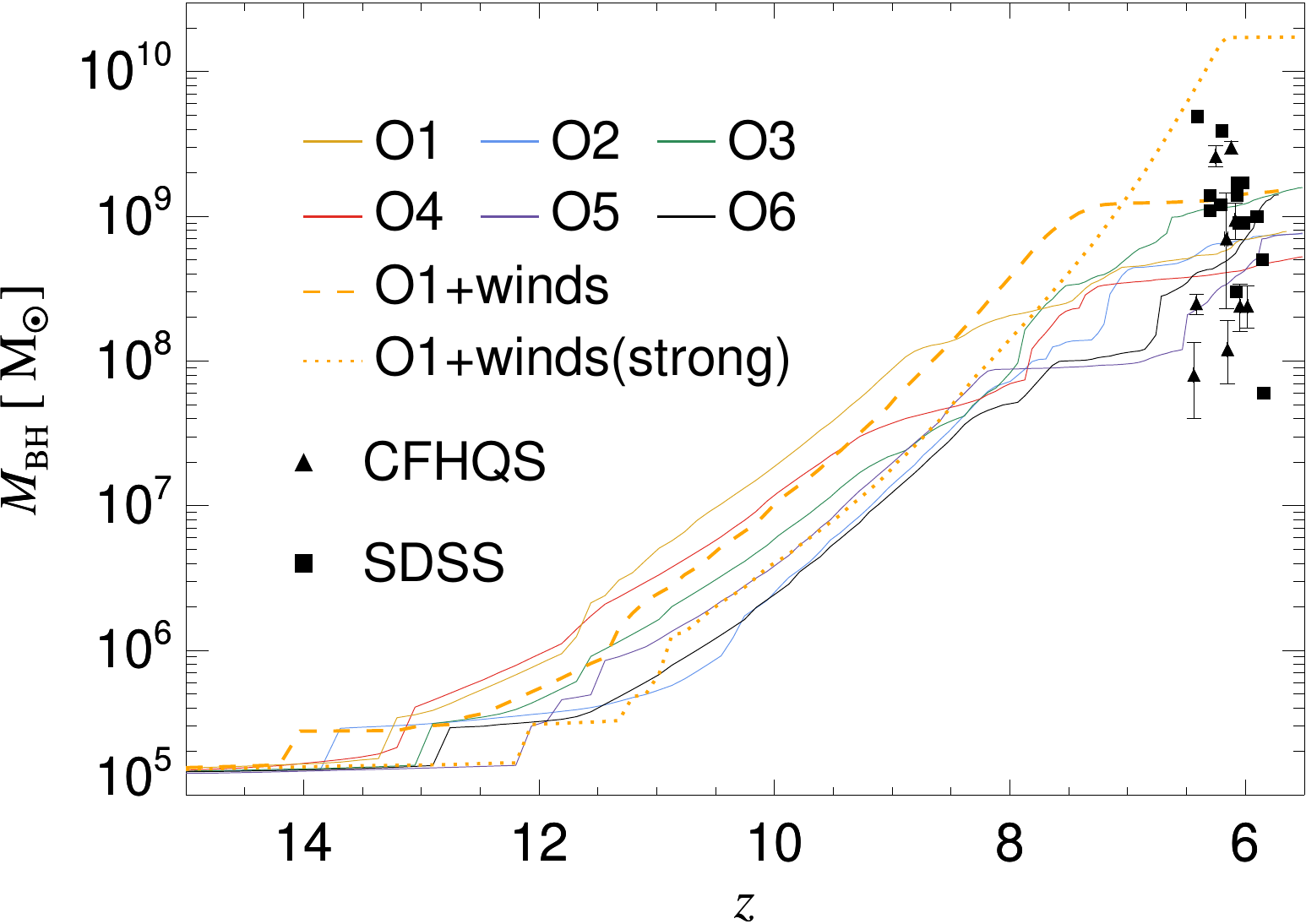}
\end{subfigure}%
\begin{subfigure}{0.5\textwidth}
\centering \includegraphics[scale = 0.5]{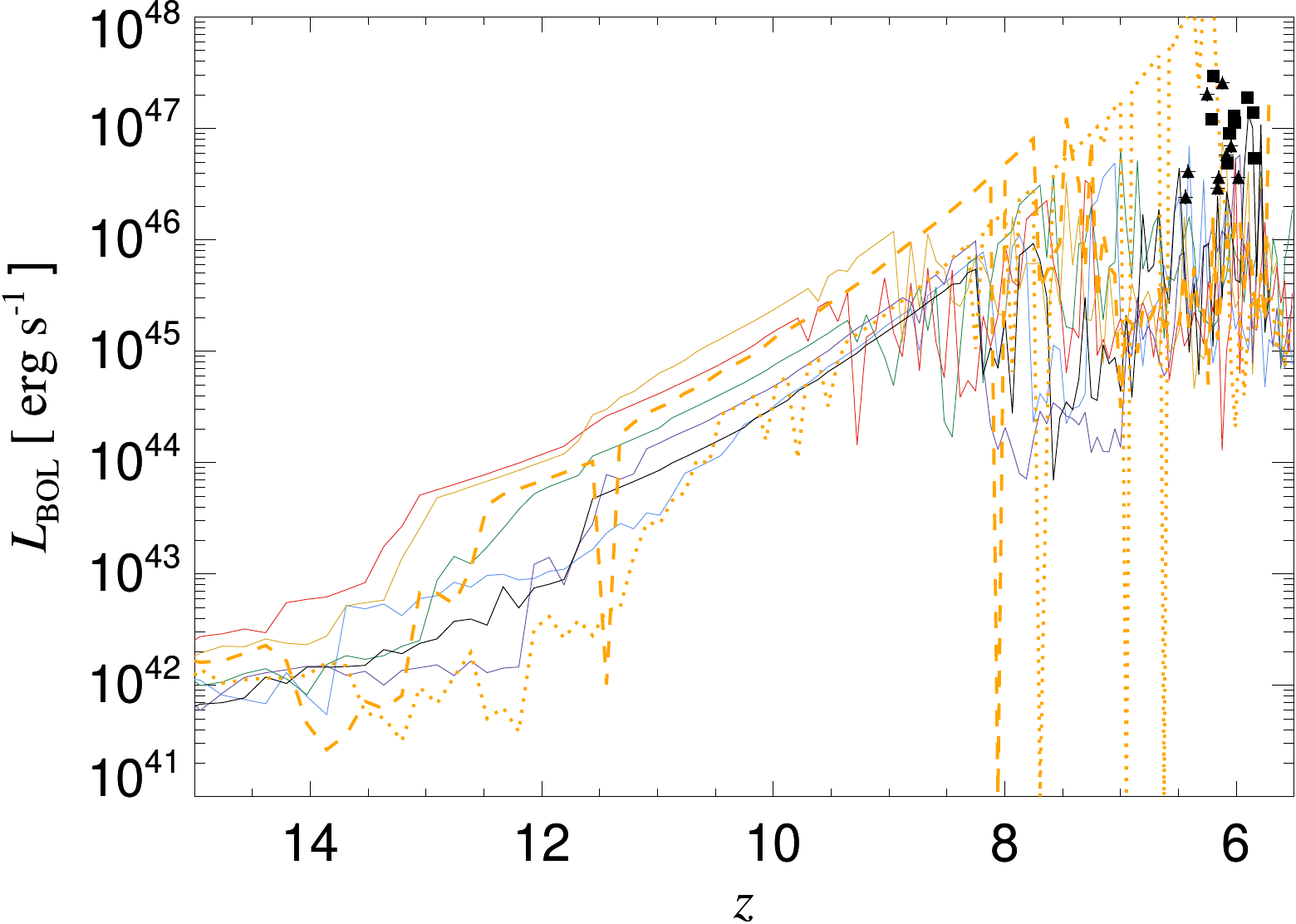}
\end{subfigure}%
\caption{(a) Mass assembly history of the most massive black holes the
  in the six highly-overdense regions.  Results for the O$1$ region
  are also shown for simulations including moderate and extreme
  winds. (b) The evolution of the bolometric luminosity of the QSOs
  powered by these black holes. Since the bolometric luminosity is
  proportional to the black hole accretion rate, the curves trace the
  accretion history of the most massive black holes.  The growth of
  the black holes proceeds in three phases: (i) An initial phase at $z
  \gtrsim 13$ of gas supply limited accretion, in which the host halos
  do not yet accumulate enough gas to reach Eddington limited
  accretion, (ii) an Eddington limited accretion phase in the redshift
  interval $13 \gtrsim z \gtrsim 9$ and (iii) a feedback limited phase
  in which accretion occurs in short Eddington limited bursts followed
  by quiescent periods that result from AGN-driven outflows. We also
  show observed luminosities and inferred black hole masses taken from
  \citet{Willott:10} (CFHQS, triangles) and \citet{deRosa:11} (SDSS,
  squares). Error bars are included when provided in the literature.}
\label{comparewdata}
\end{figure*}

In this section, we investigate the mass assembly history and the
evolution of the bolometric luminosity of the QSOs powered by the most
massive black holes in our simulations. Coloured solid curves  in the
left-hand panel of Figure~\ref{comparewdata} show the mass of the most
massive black holes for our default simulations of overdense regions
as a function of redshift. At $z \approx 15 \-- 14$, dark matter halos
have just become massive enough to acquire a seed black hole at their
centre. Soon thereafter, at $z \approx 13 \-- 12$, the seed black
holes enter a phase of exponential growth which proceeds until $z
\approx 9 \-- 8$, when the black holes reach masses $\gtrsim 10^8 \,
\mathrm{M_{\odot}}$. In this period, most mass is gained through smooth accretion of gas.
 Major mergers would result in collisions of black holes of similar mass and a corresponding
 jump by a factor of about two in mass should be seen in Figure~\ref{comparewdata}. 
While this occurs (note for instance O$2$ at $z  \,\approx\, 10.5$), 
the largest contribution is due to Eddington limited accretion resulting from cold gas brought in to the resolution limit by the large scale filaments. 
Note that due to lack of resolution, we are unable to determine whether clumps 
formed via gravitational disk instabilities also feed the black hole \citep[see also][]{DiMatteo:12, Bournaud:12,Dubois:13}.

At $z \lesssim 8$, black hole growth progresses
more slowly. Short Eddington-limited bursts of accretion are
interdispersed with longer phases of accretion well below the
Eddington rate. By $z \approx 6$, the most massive black holes in our
overdense regions have reached masses in the range $\approx 3 \times
10^8 \-- 2 \times 10^9 \, \mathrm{M_{\odot}}$.

On the same plot, black symbols with error bars denote observational
estimates from CFHQS and SDSS for the mass of black holes powering
the known QSOs at $z \approx 6$ as provided by \citet{Willott:10}
(triangles) and \citet{deRosa:11} (squares),\footnote{For the SDSS
  QSOs we have taken the black hole masses computed with the most
  recent estimator provided in this paper. This is given by \citet{deRosa:11}
as equation $5$ in their paper.} respectively. Overall, the masses of the
most massive black holes in our default simulations (solid lines) are
in reasonable agreement with the observational estimates, even though
perhaps somewhat on the low side. Our default simulations do not
produce black holes with mass substantially higher than $ 10^9 \,
\mathrm{M_{\odot}}$ by $z \approx 6$ as seems to be the case for a few
of the observed QSOs.  If the duty cycle of these QSOs is indeed very
high, this is possibly due to the fact that the Millennium simulation 
has an about a factor 100 smaller volume than probed by current QSO surveys. 
According to the observational data, there is also a population
of QSOs powered by less massive black holes with masses as low as
$M_{\rm BH} \approx 6 \times 10^7 \, \mathrm{M_{\odot}}$ to a few
$10^8 \, \mathrm{M_{\odot}}$. These lower mass observed black holes
may arise as a consequence of cosmic variance, but could also refer to
black holes hosted by less extreme halos found in regions with less
pronounced overdensities. Such a population of black holes
is predicted by our simulations, as can been seen from
Figures~\ref{intermediate_df}, \ref{overdense_df}, \ref{bhsdata} and
also from Table~\ref{table} for our sample of intermediate regions.

We have also investigated whether the inclusion of galactic outflows,
which could be very important already at $z \sim 6$ \citep{Vogelsberger:13}
, impairs the
growth of the black holes by clearing gas away from its sphere of
influence.  For this purpose we used two additional simulations of
region O$1$ which we performed with a prescription for moderate
galactic winds, O$1$+winds, and extreme winds, O$1$+winds(strong), respectively.   
The aim of the
latter was to maximize the possible impact of the effect of galactic
winds (described in detail in Section~\ref{GalacticWinds}).  We found
that galactic outflows can have the counter-intuitive effect of
\emph{aiding} black hole growth.  Early black hole growth, shown as
dashed and dotted lines (moderate and extreme winds respectively) in
Figure~\ref{comparewdata}, is delayed with respect to the default
simulation O$1$ while the host halo potential is relatively shallow. When the
halo becomes massive enough, winds are not able to efficiently eject
gas from its inner regions, which then stays trapped within the halo
and falls back to the central galaxy. Moreover, in smaller galaxies
where star formation efficiency was reduced due to galactic winds a
significant amount of baryons is expelled out of their dark matter
halos. This gas collides with the surrounding intergalactic medium and
is shock heated. With time as the most massive halo in the resimulated
region is growing in mass and the whole region around it is
collapsing, some of this gas will be incorporated into the main halo
boosting its baryon fraction and allowing for more gas to reach the
central supermassive black hole. These effects lead to a factor of about
two higher black hole mass than in the case without galactic outflows
by $z \sim 6$, still in good agreement with observational mass
estimates. Note also, that even if all simulated black holes had a mass higher
by a factor of two, their mass would still fall short of $M_{\rm BH} \approx 5 \-- 6 \times 10^9 \,
\mathrm{M_{\odot}}$ as reported in observations at $z \sim 6$
\citep{deRosa:11}. 
As they boost black hole growth, galactic winds 
also indirectly increase the energy released due to the AGN feedback, leading
to more powerful AGN outflows.
In the simulation with extreme winds,
O$1$+winds(strong), the mass of the most massive black hole is $\sim 2
\times 10^{10} \, \mathrm{M_{\odot}}$  by $z \sim 6$, exceeding the
estimated most massive black hole at this redshift by a factor of two. We
do not regard this particular version of galactic outflows as
the most realistic model and the effect of galactic winds in this simulation can
probably be considered a (generous) upper limit winds of this
type are likely to have on black hole growth.
 
\subsubsection{Bolometric luminosities}

\begin{figure}
\includegraphics[scale = 0.5]{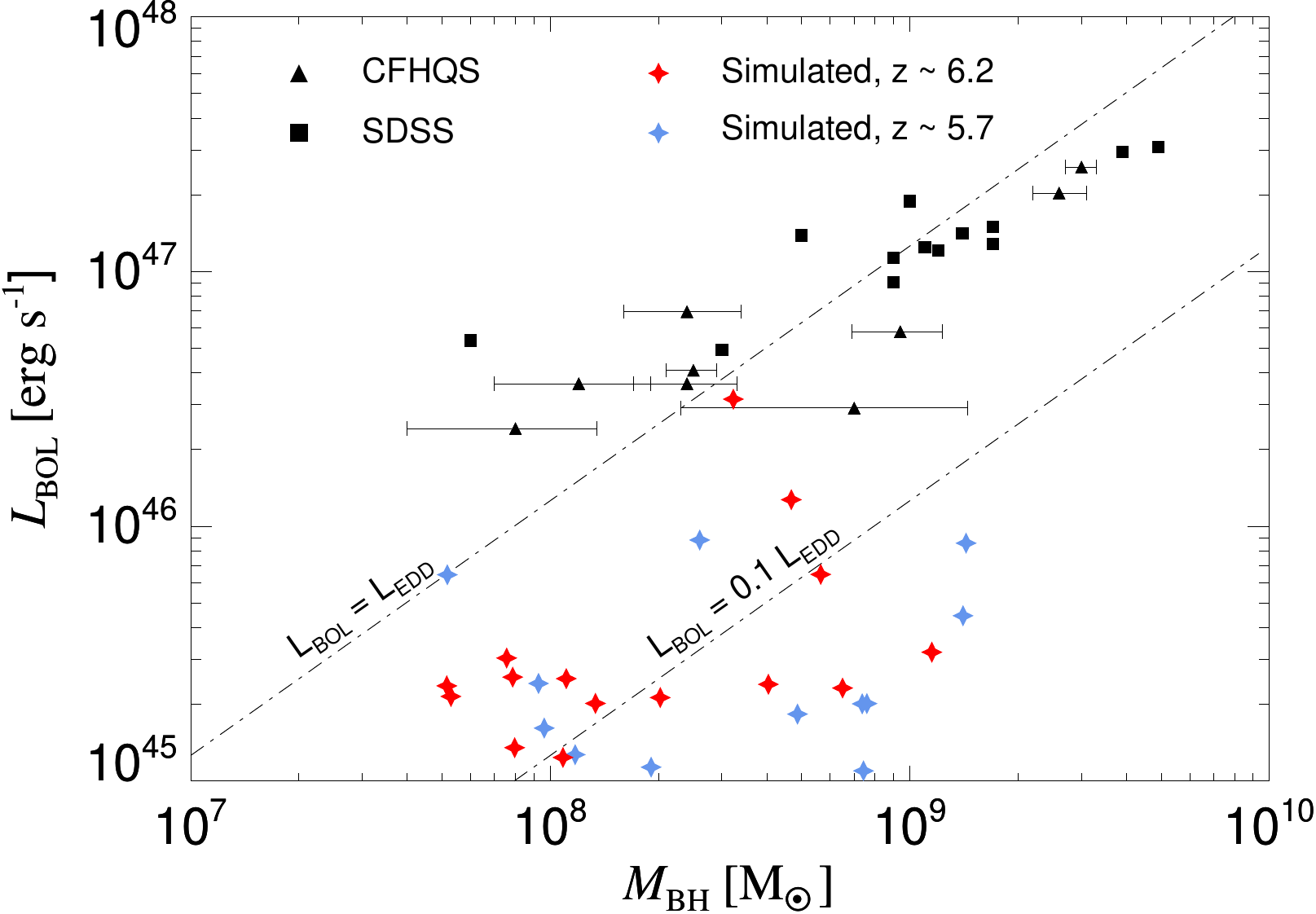}
\caption{The bolometric luminosity of all black holes with masses
  $M_{\rm BH} > 10^7 \, \mathrm{M_{\odot}}$ in our six overdense
  regions at $z \,=\, 6.2$ (red diamonds) and at $z \,=\, 5.7$
  (blue diamonds) as a function of the black hole mass. Observational
  data was taken from \citet{Willott:10} (CFHQS, triangles) and
  \citet{deRosa:11} (SDSS, squares). At $z\, = \,6.2$, accretion onto
  the massive black holes shown in this plot  is limited by thermal
  AGN feedback and therefore proceeds in short Eddington bursts,
  followed by more quiescent accretion periods. Thus, at any given
  time at this redshift, only a small fraction of QSOs will be at
  their luminosity peak and bright enough to have been detected in
  published surveys. Error bars are included when provided in the
  literature.}
\label{bhsdata}
\end{figure}

\begin{figure*}
\begin{subfigure}{0.33\textwidth}
\centering \includegraphics[scale = 0.35]{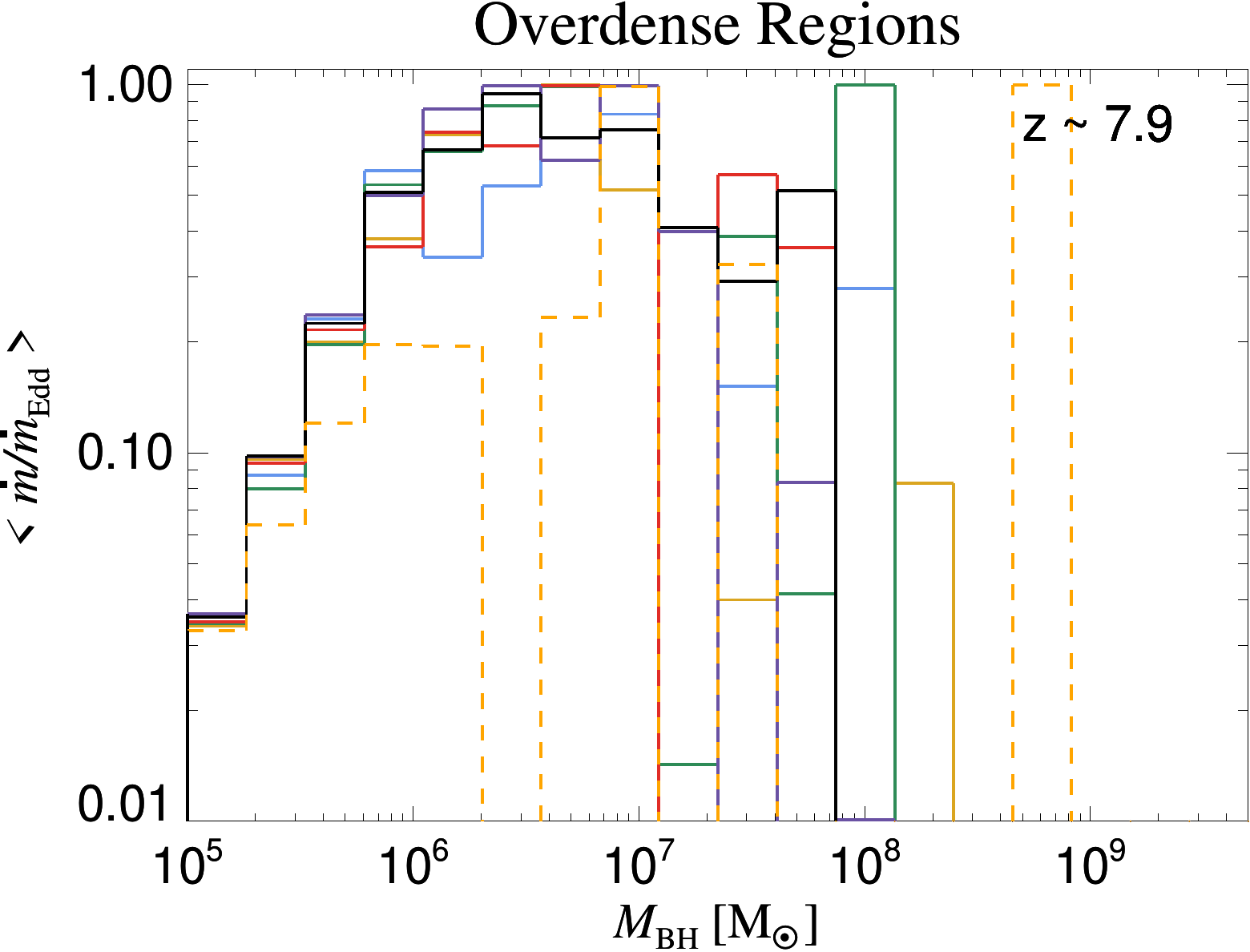}
\end{subfigure}%
\begin{subfigure}{0.33\textwidth}
\centering \includegraphics[scale = 0.35]{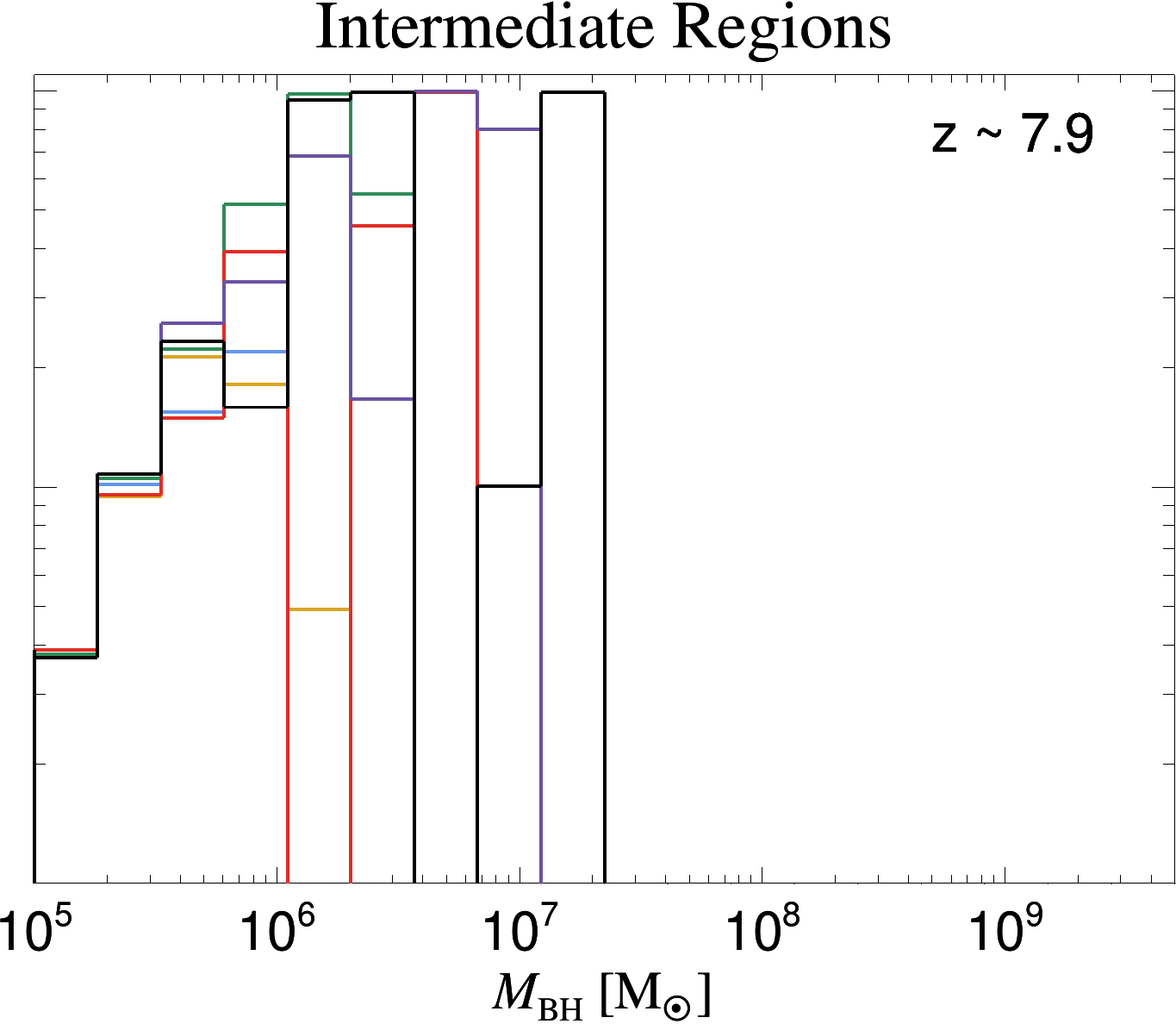}
\end{subfigure}%
\begin{subfigure}{0.33\textwidth}
\centering \includegraphics[scale = 0.35]{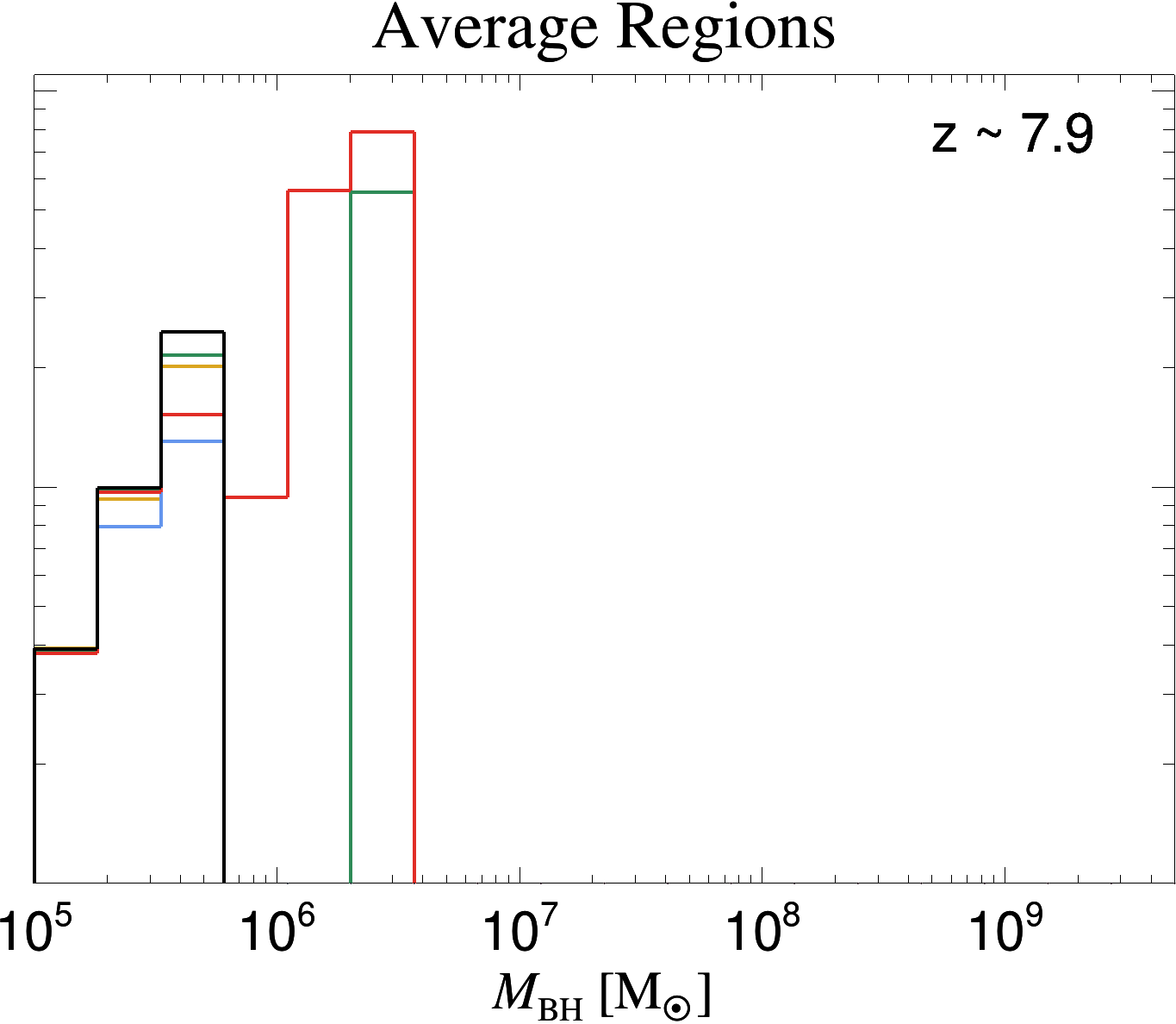}
\end{subfigure}\\%
\vspace{0.5cm}
\begin{subfigure}{0.33\textwidth}
\centering \includegraphics[scale = 0.35]{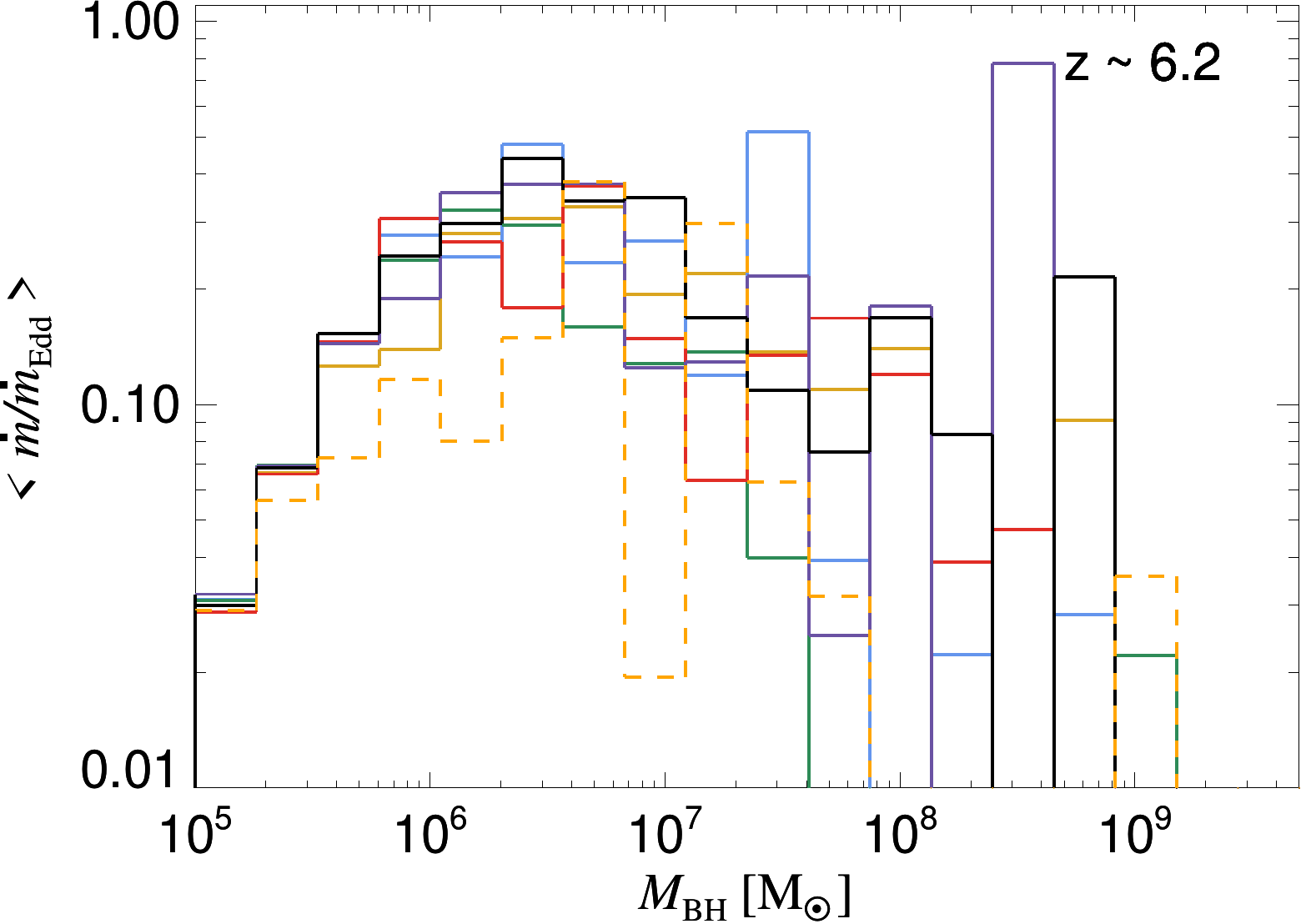}
\end{subfigure}%
\begin{subfigure}{0.33\textwidth}
\centering \includegraphics[scale = 0.35]{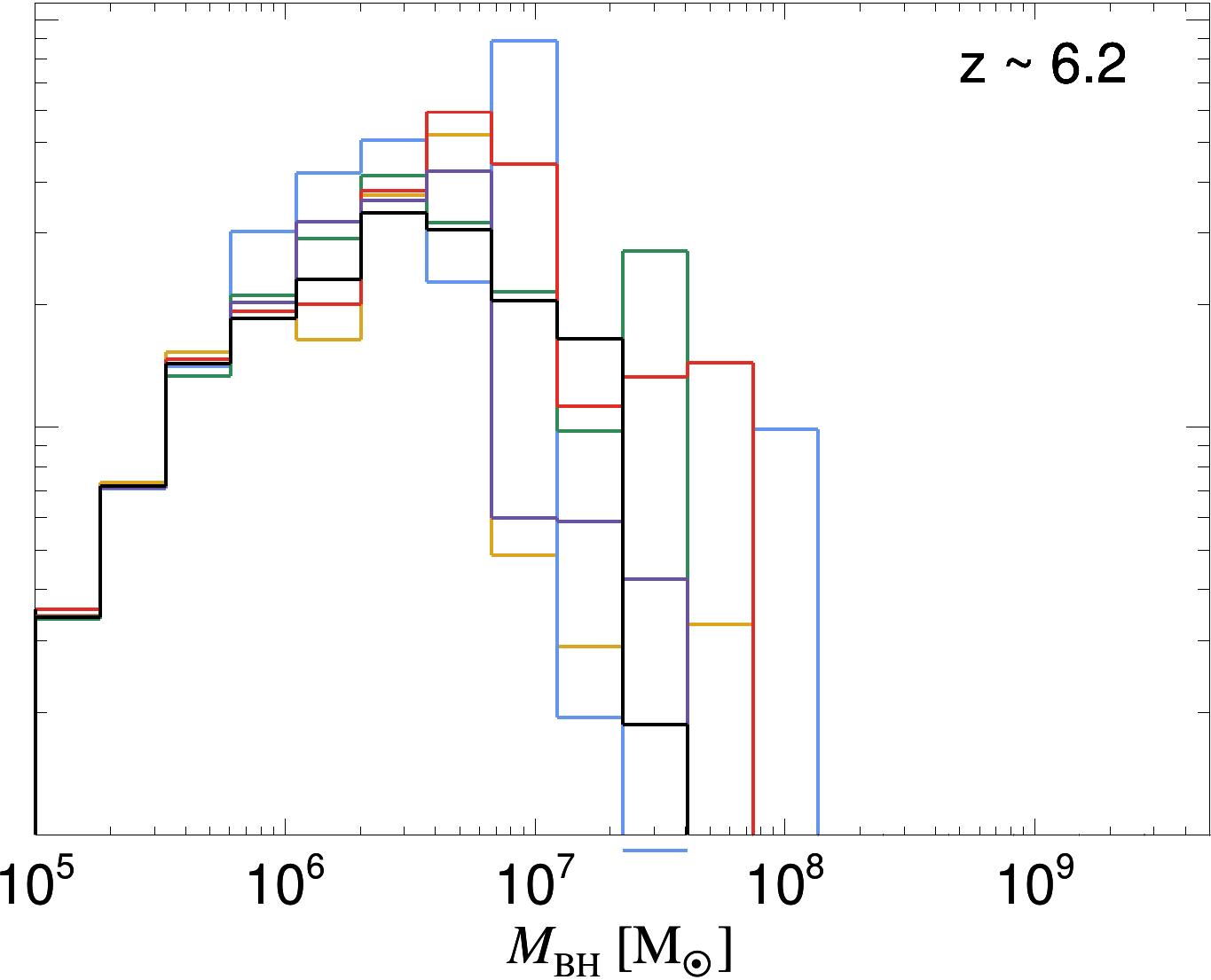}
\end{subfigure}%
\begin{subfigure}{0.33\textwidth}
\centering \includegraphics[scale = 0.35]{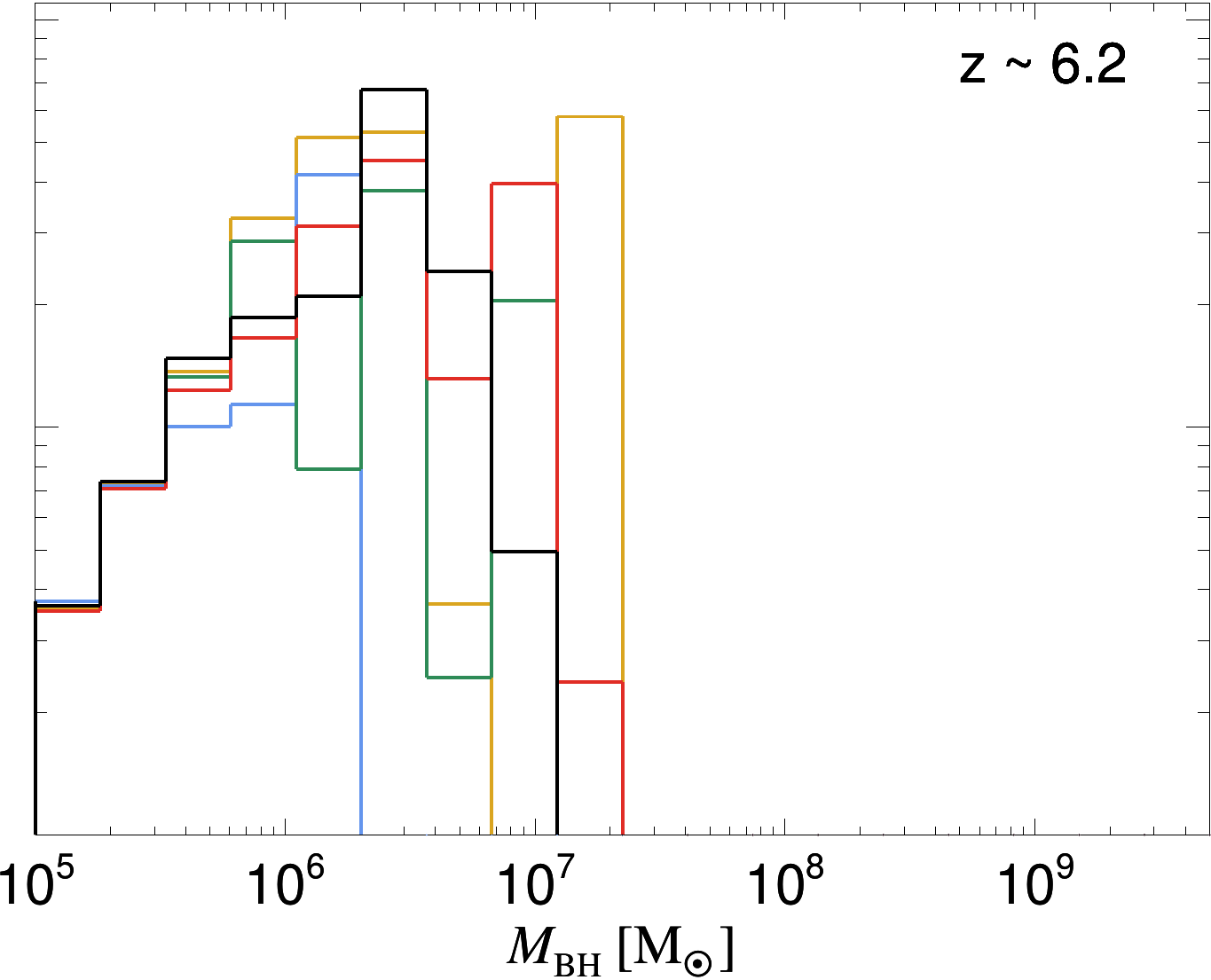}

\end{subfigure}%
\caption{Mean accretion rates relative to the Eddington rate as a
  function of black hole mass for overdense regions (left),
  intermediate regions (middle) and average regions (right) at
  redshifts: $z \,=\, 7.9$ (top) and $z \,=\, 6.2$
  (bottom). The Eddington ratios are large for black holes with
  masses in the range $10^6 \-- 10^7 \, \mathrm{M_{\odot}}$ for all
  regions at both redshifts. At the higher redshift, accretion onto
  black holes in this mass range occurs very close to the Eddington
  limit.  Note further that for the regions of average density, the
  lower gas supply means that on average black holes never quite reach
  Eddington limited accretion.  For the most massive black holes, with
  $M_{\rm BH} \gtrsim 10^7 \, \mathrm{M_{\odot}}$, accretion is
  limited by strong thermal feedback and alternates between quiescent
  and Eddington limited accretion episodes. The accretion rates of
  such massive black holes therefore vary considerably from region to
  region.}
\label{accedd}
\end{figure*}

\begin{figure*}
\begin{subfigure}{0.34\textwidth}
\hspace{-0.8cm}
\centering \includegraphics[width = 0.9\textwidth]{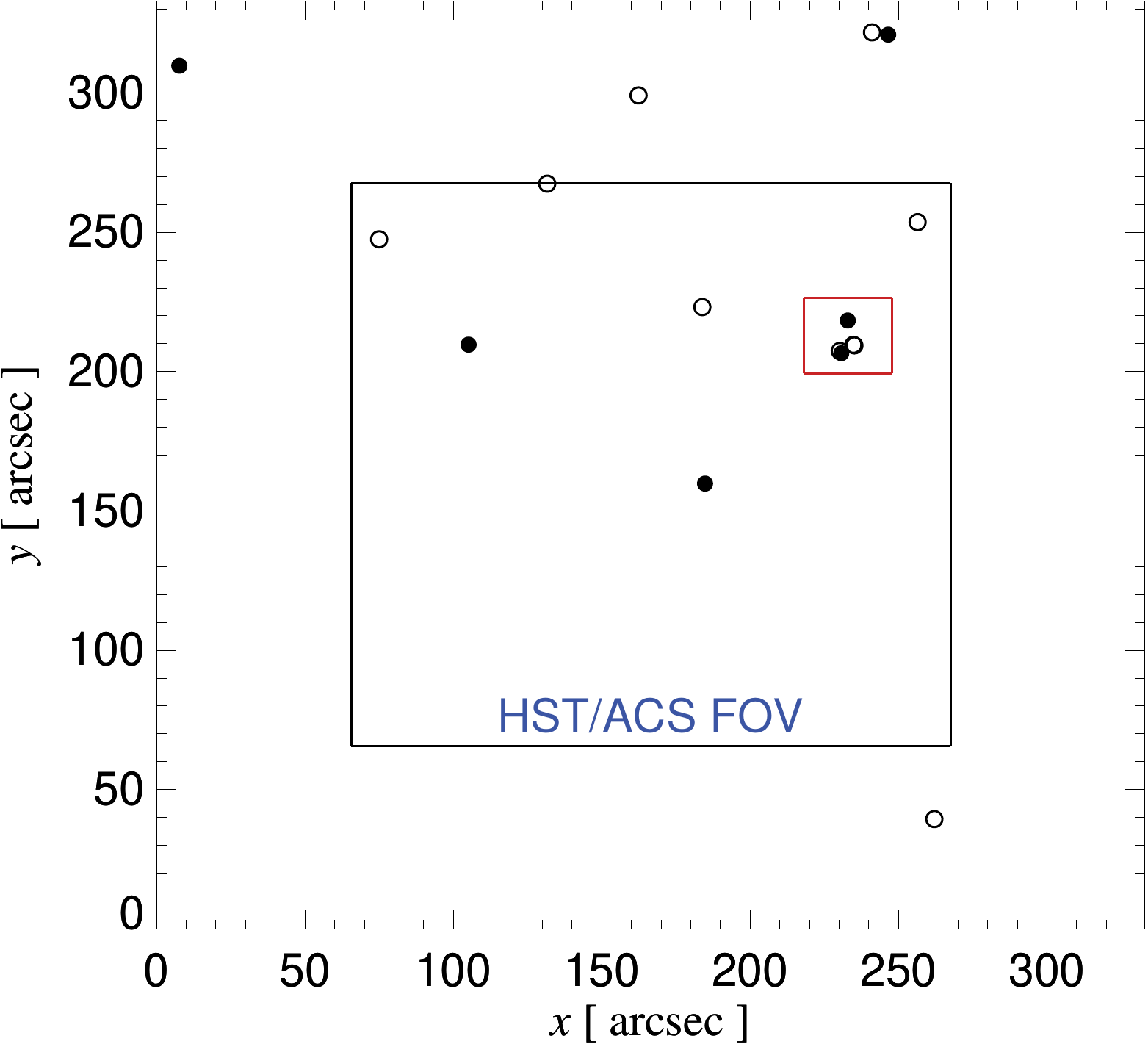}
\end{subfigure}%
\begin{subfigure}{0.34\textwidth}
\hspace{-0.8cm}
\centering \includegraphics[width = 0.9\textwidth]{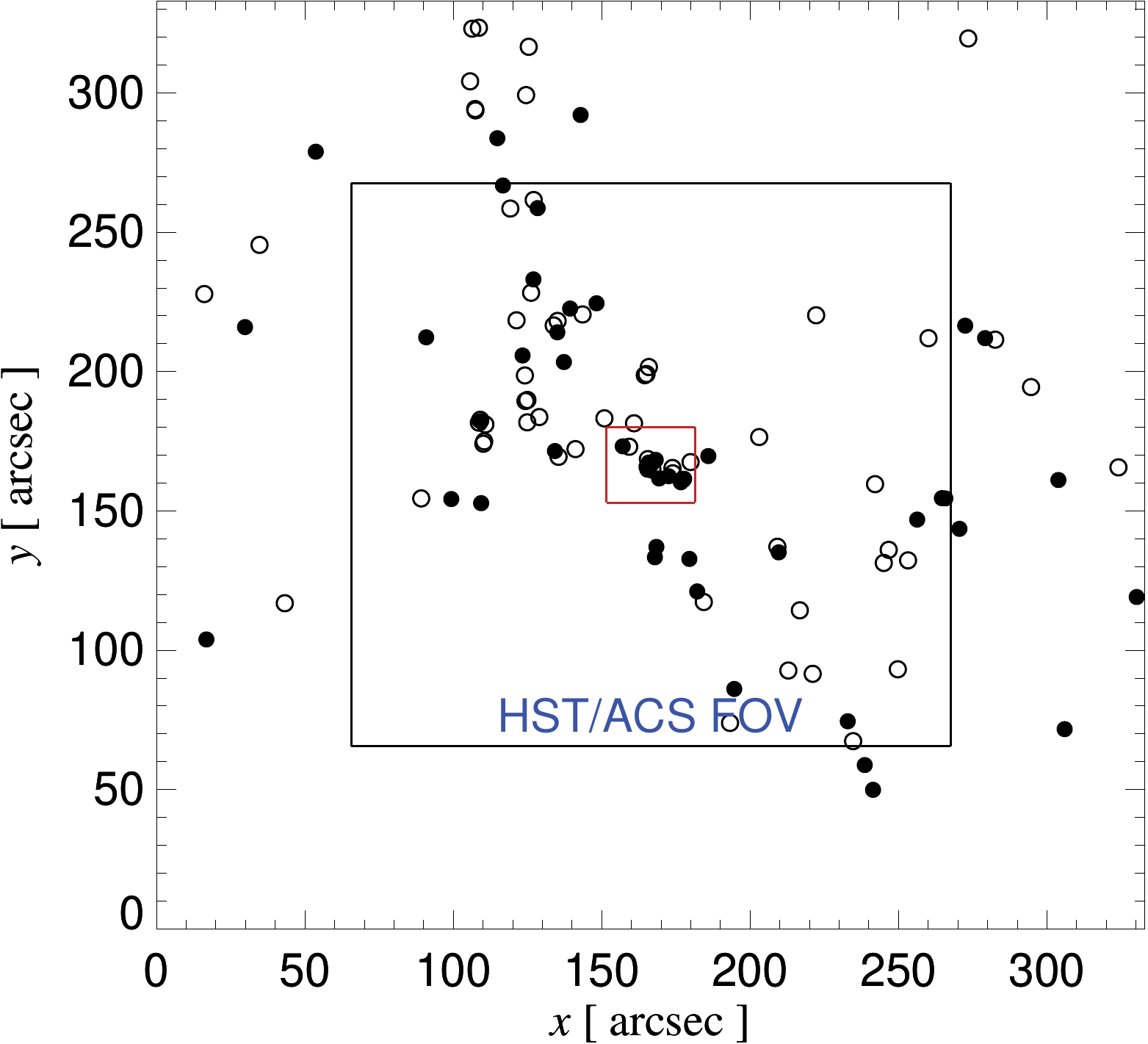}
\end{subfigure}%
\begin{subfigure}{0.34\textwidth}
\hspace{-0.8cm}
\centering \includegraphics[width = 0.9\textwidth]{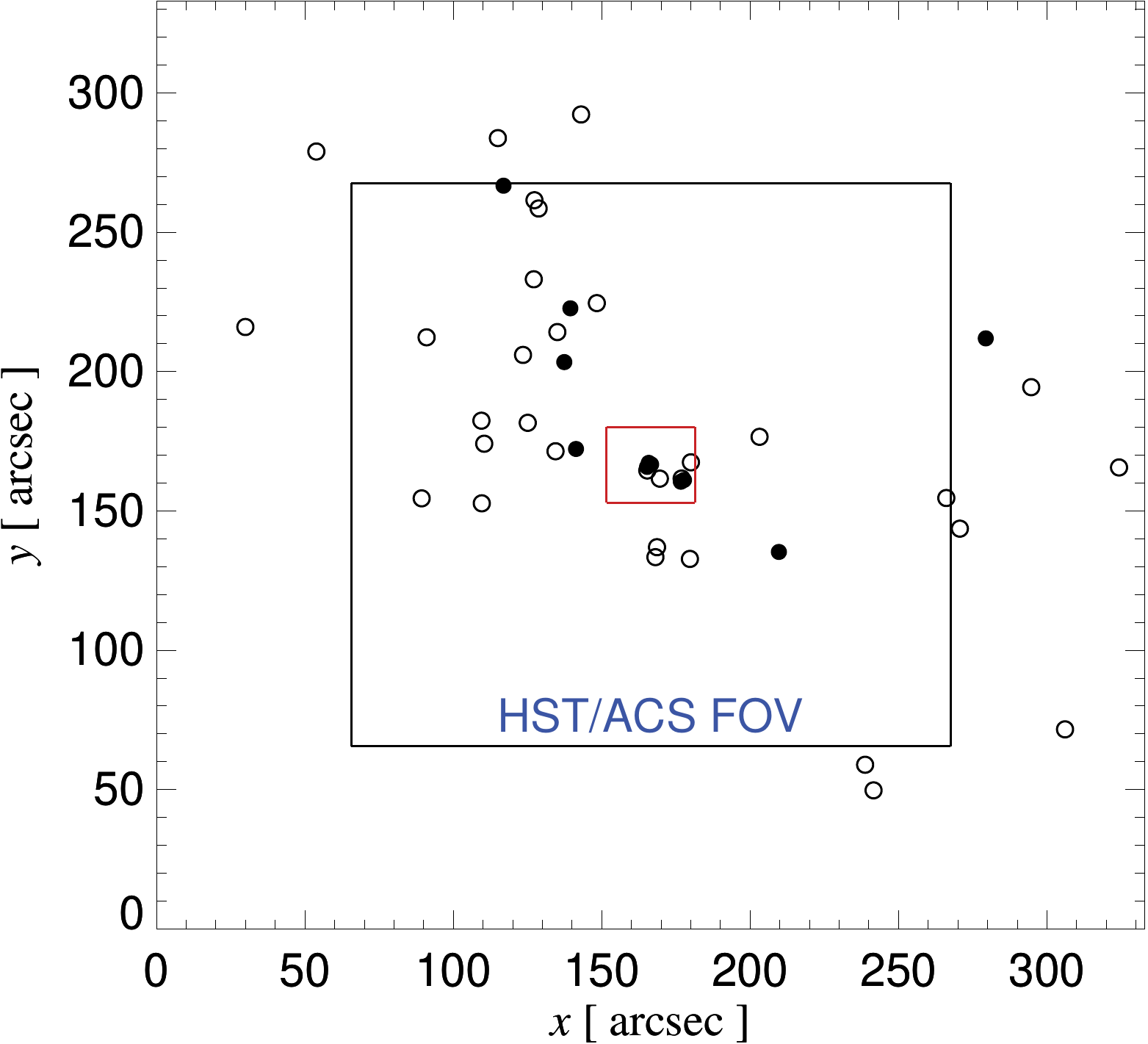}
\end{subfigure}\\%
\vspace{0.5cm}
\begin{subfigure}{0.33\textwidth}
\centering 
\includegraphics[width = 0.9\textwidth]{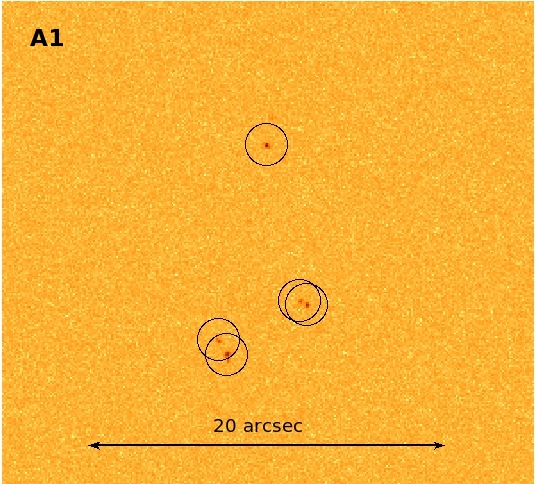}
\end{subfigure}%
\begin{subfigure}{0.33\textwidth}
\centering
 \includegraphics[width = 0.9\textwidth]{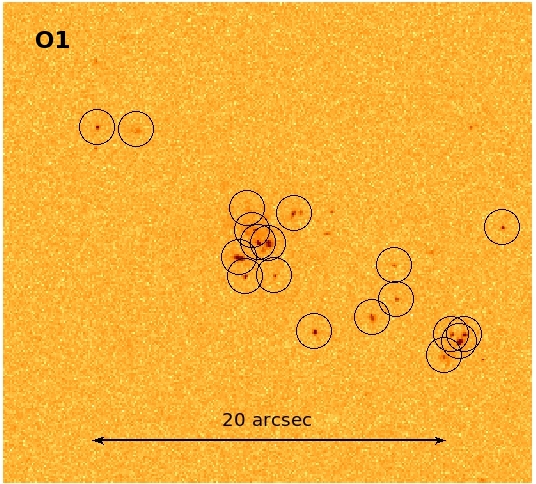}
\end{subfigure}%
\begin{subfigure}{0.33\textwidth}
\centering
 \includegraphics[width = 0.9\textwidth]{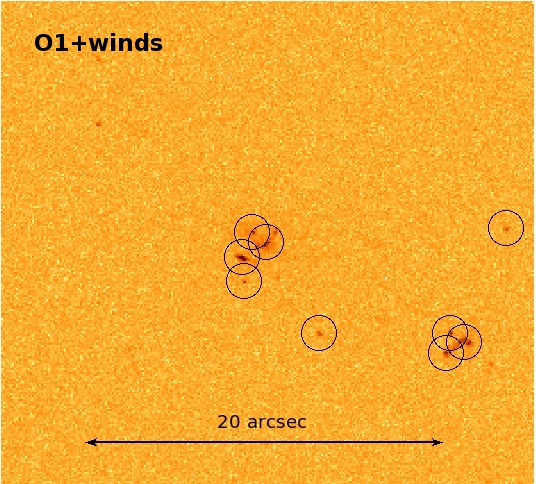}
\end{subfigure}%
\caption{The first row shows the spatial distribution of mock galaxies within a cube with
  side length of $1.9 {\rm Mpc}$ ($10 \, h^{-1} \, \mathrm{comoving \, Mpc}$) from one of our
  simulations of a region of average density, A$1$ (left), compared to one of our overdense regions, O$1$ (middle) and to a simulation of the same overdense region with galactic winds, O$1$+winds (right). We denote the
  galaxies, as identified by SExtractor with $m_{\rm UV} \leq
  26.5$ (solid circles) and $m_{\rm UV} \leq
  27.5$ (open circles). The large black square delimits
  the field of view of HST/ACS. The regions denoted by red rectangles
  -- in the average region focusing on a small group of clustered
  galaxies, and in the overdense region centred on the QSO host -- are
  expanded in the second row and shown as they appear in our mock
  images. As above, open black circles denote the identified galaxies with $m_{\rm UV} \leq
  27.5$. In
  the overdense region, sources surrounding the QSO are brighter and
  far more numerous compared to the average region. Regions with
  diffuse UV emission are also present around the QSO host in the
  overdense region. The mock image of O$1$+winds shows that the number density of galaxies is
  very sensitive to galactic winds and it decreases by a factor of $3.7$ ($2.6$) within the field of view of HST/ACS for $m_{\rm UV} \leq 26.5$ ($m_{\rm UV} \leq 27.5$). Note also the overdensity of galaxies drops quickly beyond the HST/ACS field of view.}
\label{mockfig}
\end{figure*}

In the right-hand panel of Figure~\ref{comparewdata}, we show the
redshift evolution of the bolometric luminosities of QSOs powered by
supermassive black holes in our simulations of the highly overdense
regions shown in the left-hand panel. Since the bolometric luminosity
of a QSO is related to the black hole accretion rate via $L_{\rm BOL}
\,=\, \epsilon_r \dot{M}_{\rm BH} c^2 $, this plot may be equally
interpreted as the accretion rate history of the black holes. During
the initial growth phase, at $z \gtrsim 13 - 12$, accretion onto the
black hole seeds is limited by gas supply. Since the mass of the initial 
seeds is rather high and the potential well of their hosts not yet deep 
enough, gas cannot be accumulated at a sufficiently high rate to power the 
black holes to their Eddington limit. This situation
changes as more massive halos assemble in the interval $13 \gtrsim z
\gtrsim 9$, when accretion becomes Eddington limited for all QSOs
shown. At $z \sim 9$, the accretion rates become high enough for the
first QSOs to reach a bolometric luminosity exceeding
$L_{\mathrm{BOL}} \,=\, 10^{46} \, \mathrm{erg \ s^{-1}}$. The energy
injected via thermal feedback starts to become comparable to the
binding energy of the gas in the QSO host halos, thus pushing gas away
from the QSO and interrupting temporarily the Eddington limited
growth. Further gas accretion gives then rise to a cycle in which
black hole growth alternates between short bursts of Eddington limited
accretion and longer periods of supply limited accretion at
lower rates. During this phase \citep[see also][]{Choi:12}, the bolometric luminosities of the
QSOs can vary by factors 10-100 on times scales much shorter than the
Hubble time.  In the simulations with galactic winds, the phase of Eddington 
limited accretion lasts longer and the black holes
grow to larger masses. The corresponding bolometric luminosities are
also significantly higher. AGN feedback eventually also
starts to shut off the accretion periodically,  introducing similar
variability in the black hole accretion rate and hence QSO  bolometric
luminosity as in the simulations without galactic winds.  The flux
limit of current observational surveys of $z \sim 6$ QSOs
corresponds to $L_{\rm BOL} \approx 2 \times 10^{46} \, \mathrm{erg
  \ s^{-1}}$,  and  only the QSOs that happen to be undergoing a
strong accretion event (shown as peaks in the right-hand panel of Figure~\ref{comparewdata})
can be picked up in the present surveys. Using the estimates for the
Eddington ratio of observed $z \sim 6$ QSOs given in
\citet{Willott:10} and \citet{deRosa:11}, we have computed bolometric
luminosities by simply calculating $\lambda L_{\rm Edd}(M_{\rm BH})$,
where $\lambda$ and $M_{\rm BH}$ are the provided estimates for
Eddington ratio and black hole mass. These are respectively shown as
triangle and square symbols in the right-hand panel of
Figure~\ref{comparewdata} as well. Estimates for the bolometric
luminosities obtained in this way match the peak bolometric
luminosities of the simulated QSOs reasonably well. The very brightest
observed QSOs however require sustained accretion rates onto black
holes which may only be possible in halos which are too rare to
be contained within the Millennium volume.

In Figure~\ref{bhsdata}, we show the instantaneous bolometric
luminosity of the simulated QSOs in our default simulations as a function of
the black hole mass at $z \,=\, 6.2$ (red diamonds) and $z \,=\, 5.7$ (blue diamonds). 
We include all black holes from our default
simulations of overdense regions that have $M_{\mathrm{BH}} > 10^7 \,
\mathrm{M_{\odot}}$. At any given instant, only about one or two black
holes have bolometric luminosities comparable to $10^{46} \,
\mathrm{erg s^{-1}}$ due to variability in the accretion rate in the
feedback limited regime. The remaining black holes, even those with
masses exceeding $10^9 \, \mathrm{M_{\odot}}$ have luminosities at $z
\,=\, 6.2$ in the yet observationally inaccessible range $10^{45} -
10^{46} \, \mathrm{erg s^{-1}}$. A key prediction of our simulations
is therefore a steep rise in the number of observed QSOs with
bolometric luminosities in the range $10^{45} \lesssim L_{\rm BOL}
\lesssim 10^{46} \, \mathrm{erg s^{-1}}$ and masses spanning the wide
range  $5 \times 10^7 \-- 2\times 10^9 \, \mathrm{M_{\odot}}$. 
Note that the existence of such a population of QSOs at $z \sim 6$ is not trivial,
 as it depends on assumptions made for the black hole seeds at such early times.
We also
predict the galactic hosts of these QSOs to show signs of strong
AGN-driven outflows (see Section \ref{secoutflow}).

\subsubsection{Eddington ratios}

Figure~\ref{accedd} shows the mean  black hole accretion rates
relative to the Eddington rate as a function of black hole mass. The
top panels are for the six overdense regions (left), the six
intermediate regions (middle) and the six average regions (right) at
$z \,=\, 7.9$, while the bottom panels are for the same regions at $z
\,=\, 6.2$. Note that results for O$1$+winds are also shown as the
dashed orange histogram in the left column. At $z \,=\,7.9$, the
highest Eddington ratios occur mainly for black holes with masses $10^6
\lesssim M_{\odot} \lesssim 10^7 \, \mathrm{M_{\odot}}$.  For
overdense and intermediate regions, the accretion rate onto black
holes in this mass range is Eddington limited.  In the average
regions, the accretion rates onto black holes in this mass range do
not quite reach the Eddington limit due to a lower supply of gas.  At
$z \,=\,6.2$, the spatial density of halos hosting black holes with
masses $10^6 \lesssim M_{\odot} \lesssim 10^7 \, \mathrm{M_{\odot}}$
is higher and these halos therefore no longer provide the rare
potential wells which efficiently collect gas from a large region. As
a consequence, typical accretion rates become lower. At both redshifts
shown, efficient AGN feedback leads to a decrease in the typical
Eddington ratios for black holes with mass $\gtrsim 10^8 \,
\mathrm{M_{\odot}}$.  For the overdense regions, where several very
massive black holes can be found, the accretion rates are variable due
to feedback limited episodic accretion and the most massive black
holes accrete at only a few percent of their Eddington rate.  Note how
in our run with moderate galactic winds the accretion rates are
suppressed for all the black holes but the most massive.  The latter
are the only black holes that reside in halos which are massive enough
to trap sufficient quantities of   gas for efficient growth in their
potential wells despite the influence of the galactic winds.

\begin{table*}
\centering
\begin{tabular}{lcccc}
\toprule Simulation   & $N (m_{\rm UV} \leq 26.5)$  & $N (m_{\rm UV} \leq 27.5)$ & $N (m_{\rm UV} \leq 26.5)$  & $N (m_{\rm UV} \leq 27.5)$\\ 
    & $6.86 {\rm Mpc}^3$  & $6.86 {\rm Mpc}^3$ & $ 2.48 {\rm Mpc}^3$  & $2.48 {\rm Mpc}^3$\\ 
\midrule
A$1$ & $6$ & $16$ & $4$ & $11$ \\   
A$2$ & $2$ & $4$  & $2$ & $3$ \\  
A$3$ & $4$ & $13$ & $2$ & $6$ \\  
A$4$ & $1$ & $6$  & $1$ & $4$ \\  
A$5$ & $4$ & $12$ & $2$ & $5$ \\  
A$6$ & $4$ & $8$  & $3$ & $5$ \\  
\midrule 
A$1$+winds & $1$ & $5$ & $0$ & $3$\\  
A$1$+winds(strong)& $0$	& $0$ & $0$ & $0$ \\  
\midrule 
I$1$ & $21$    & $64$ & $16$ & $49$ \\   
I$2$ & $12$    & $23$ & $9$ & $14$ \\    
I$3$ & $24$    & $69$ & $18$ & $51$ \\   
I$4$ & $21$    & $56$ & $14$ & $32$\\  
I$5$ & $32$    & $81$ & $22$ & $52$ \\   
I$6$ & $37$    & $104$ & $18$ & $54$\\  
\midrule 
 O$1$ & $50$ & $110$ & $37$ & $83$ \\     
O$2$  & $40$ & $116$ & $27$ & $77$\\      
O$3$  & $46$ & $130$ & $26$ & $77$\\  
O$4$  & $90$ & $138$ & $73$ & $113$\\  
O$5$  & $96$ & $199$ & $62$ & $127$\\  
O$6$  & $91$ & $254$ & $59$ & $159$\\  
\midrule 
O$1$+UV & $53$	& $112$ & $39$ & $86$ \\  
O$1$+UV(strong) & $59$	& $127$ & $43$ & $95$\\  
O$1$+winds & $11$ & $43$ & $10$ & $32$ \\  
O$1$+winds(strong) & $3$ & $4$ & $3$ & $4$ \\  
O$1$nobh & $56$	& $111$ & $43$ & $83$ \\  
O$1$lowres & $4$ & $5$ & $4$ & $5$ \\  
\bottomrule
\end{tabular}
\caption{Predicted number of galaxies at $z \,=\, 6.2$ inside a cube
  with side length $1.9 \, \mathrm{Mpc} \, (10 h^{-1} \,
  \mathrm{comoving \, Mpc})$ (second and third columns) and inside the 
  fraction of the volume probed by the HST/ACS field of view (fourth and fifth columns). 
  We list the total number of selected mock galaxies in both volumes for the two magnitude limits: $m_{\rm UV} \leq 26.5$ as
  in \citet{Kim:09} and $m_{\rm UV} \leq 27.5$. Without feedback, excesses in galaxy numbers of about one
  order of magnitude are predicted in the simulations of the overdense
  regions compared to simulations of  regions of average
  density. Changing the reionisation history and including a stronger
  UV background and/or AGN feedback in the simulations does not
  attenuate this difference. Galactic outflows, however, which we
  included in our simulations of regions O$1$ and A$1$ led to a
  significant decrease of the number of surrounding galaxies  as well
  as to a much smaller excess of predicted galaxies in overdense
  regions compared to regions of average density.}
\label{tablecounts}
\end{table*}

\subsection{Star-forming galaxies}\label{secresults3}

\subsubsection{Number counts from mock images}

\begin{figure*}

\begin{subfigure}{0.5\textwidth}
\centering \includegraphics[scale = 0.5]{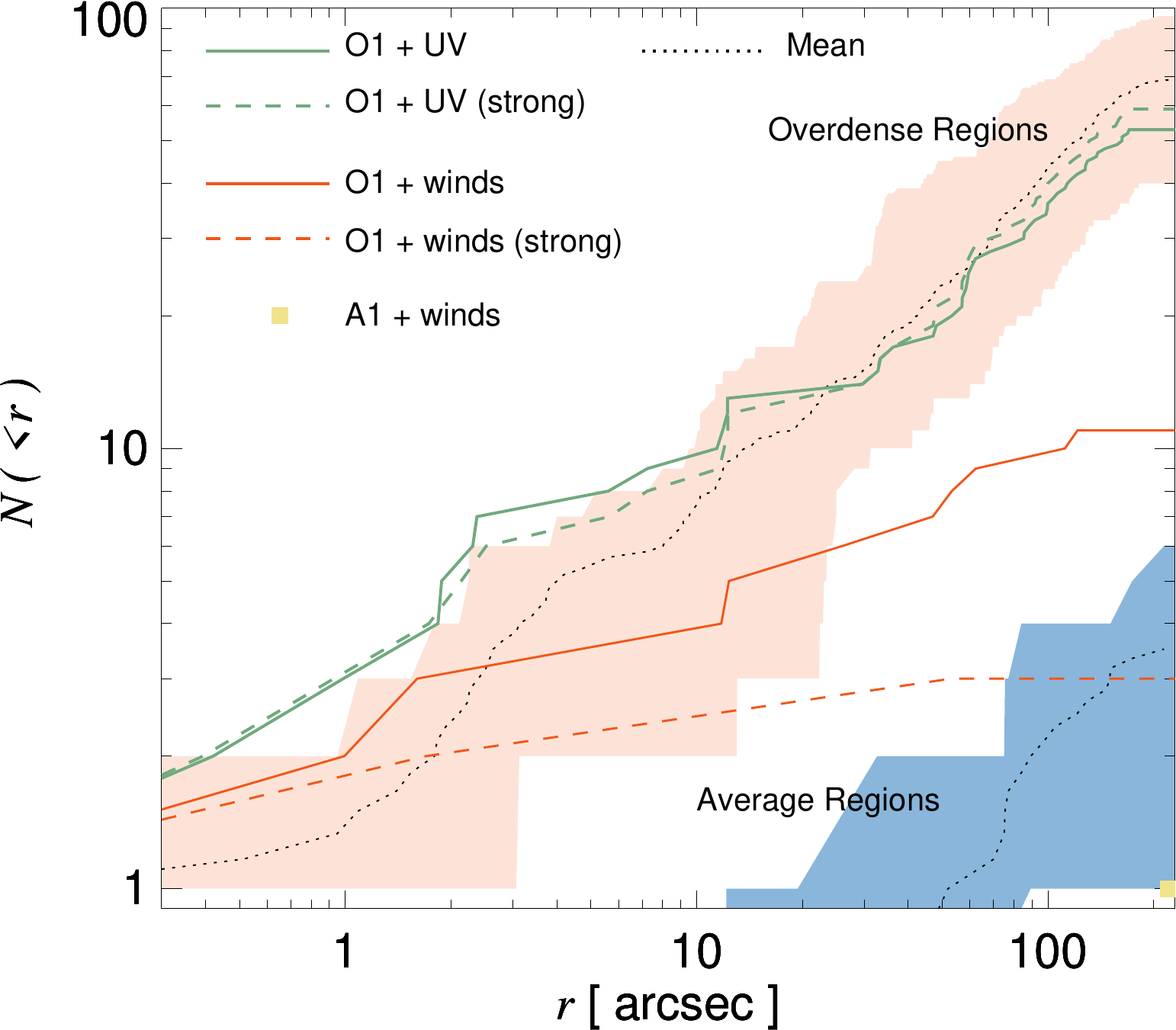}
\end{subfigure}%
\begin{subfigure}{0.5\textwidth}
\centering \includegraphics[scale = 0.5]{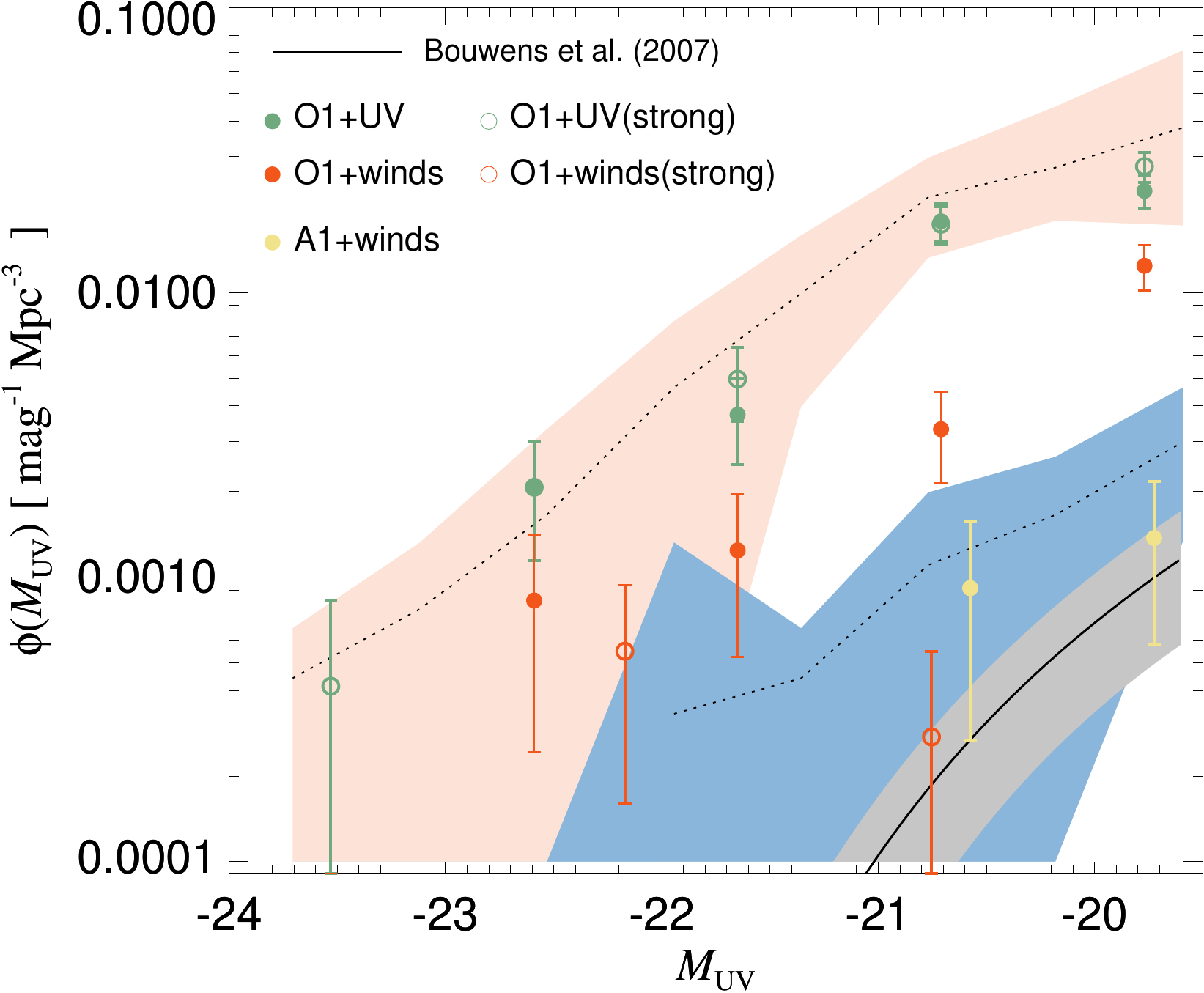}
\end{subfigure}%
\caption{(a) Cumulative number of galaxies within a projected radius
  $r$ for galaxies of apparent brightness $m_{\rm UV} \leq 26.5$. The
  pink shaded band shows the cumulative number counts obtained for our
  overdense regions, and the dotted line passing through it is the
  mean of the six overdense regions. We also show the results for two
  simulations with modified UV backgrounds (green continuous and
  dashed lines; see Section~\ref{Reionisation} for further
  description) and for the simulations with supernova-driven winds of
  different strength (red continuous and dashed lines; see
  Section~\ref{GalacticWinds} for further details). The blue shaded
  band refers to the average regions, and the dotted line passing
  through it is the mean of the six average regions. Since for
  A$1$+winds, there is only one source in the region of interest, we
  mark the radius at which this source is located with a yellow
  square. Overdense regions typically contain an order of magnitude
  more galaxies than average regions. Galactic winds decrease the
  number counts in both overdense and average regions. (b) Luminosity
  function of detected galaxies down to $m_{\rm UV} \,=\, 27.5 \,
  (M_{\rm UV} \,=\, -19.3)$ in the overdense regions (shaded pink) and
  in the average regions (shaded blue). Dotted lines passing through
  the pink and blue shaded regions of the plot denote the mean
  luminosity function of overdense and average regions, respectively.
  The luminosity functions for the simulations with a modified UV
  background is given by the green circles. Red circles denote the
  luminosity functions for the simulations including supernova-driven
  galactic outflows. Error bars illustrate the effect of Poissonian
  fluctuations in number counts in each bin shown. The inclusion of
  galactic outflows brings the luminosity function of one of our
  average regions in reasonable agreement with the \citet{Bouwens:07}
  luminosity function (shown as black line, together with $1 \sigma$
  error level as a light grey shade) and strongly decrease the
  expected overdensity of galaxies around the luminous QSOs in our
  simulations.}
\label{counts_lum}
\end{figure*}

Having discussed the evolution of the (most massive) black holes in our
simulations, we now focus on the number counts of bright star-forming galaxies 
at $z \, = \, 6.2$. In order to identify
galaxies in the different fields in a way as close to the observations
as possible, we adopt the procedure outlined in Section
\ref{secmethod}. 

An example of the spatial distribution of simulated mock galaxies is shown
in Figure~\ref{mockfig}, where the large-scale field is shown for the
A$1$ average region (left), for the O$1$ overdense region
(middle) and for the O$1$ overdense region with moderate galactic winds,
  O$1$+winds (right). We consider two magnitude limits  for galaxies to be
included in our catalogues: {\it i)} $m_{\rm UV} \leq 26.5$ \citep[as
  in][]{Kim:09} and {\it ii)} a selection going one magnitude deeper,
$m_{\rm UV} \leq 27.5$.
The galaxies identified by SExtractor as included in our
$m_{\rm UV} \leq 26.5$ ($m_{\rm UV} \leq 27.5$) sample are marked by
closed (open) 
black circles in Figure~\ref{mockfig}, and in the second row they are visible as
concentrated clumps against the noisy background. In the average
region, galaxies appear evenly distributed within the field of view of
ACS, with the exception of a small group of clustered galaxies. In the
overdense region, galaxies clearly trace the dense filaments, where
high gas density favours galaxy formation (see O$1$ in
Figure~\ref{overdense_df} for comparison). In the second row of
Figure~\ref{mockfig}, we zoom into the regions marked by the red
rectangles, to show how they appear in our UV mock observational
images. Apart from containing both brighter and more numerous sources,
O$1$ (and O$1$+winds in the right panel) also contain diffuse regions with significant UV emission,
particularly surrounding the QSO host in the centre. Fainter galaxies
whose apparent magnitude satisfies $m_{\rm UV} > 27.5$ are also
visible as lighter coloured clumps. The absence of such faint clumps
in A$1$ indicates that the galaxy overdensity persists at lower
magnitudes than considered with our magnitude selection, in agreement
with the enhancement of the mass function at all resolved scales shown
in Figure~\ref{massfunction}.

In Table~\ref{tablecounts}, we list the number of bright galaxies
identified in the mock images of each of our simulations and for both
of our magnitude selections. The excess of bright galaxies with $m_{\rm UV} \leq 26.5$ around our
bright QSOs is quantified in the left-hand panel of
Figure~\ref{counts_lum}. Here, we show the cumulative
number of galaxies within a projected radius $r$ for our different
simulations.  We shade the area enclosed between the minimum and
maximum curves for both average and overdense regions in order to
illustrate cosmic variance.  Within a projected radius of $200 \,
\mathrm{arcsec}$, the number of bright galaxies in our six average
regions, as indicated by the blue band, lies within the range $1 \--
5$. For the six overdense regions, this number lies in the
range $40 \-- 95$ (pink band).

The right-hand panel of Figure~\ref{counts_lum}  compares the  luminosity functions 
obtained from our mock images down to a magnitude $m_{\rm
  UV} \,=\, 27.5$ to the observed (field) luminosity function
by \citet{Bouwens:07}. Note, that despite matching the dark matter mass
functions of average regions to \citet{Jenkins:01}, our average
regions contain significantly more  bright galaxies than observed. 
This mismatch is not surprising given that we have not yet
considered any strong stellar feedback processes e.g. due to reionisation
and/or supernova-driven winds. Especially the latter will be very
efficient given the high star formation rates of the galaxies we are
investigating.

\subsubsection{Galactic outflows}\label{GalacticWinds}

\begin{figure*}
\begin{subfigure}{0.33\textwidth}
\centering \includegraphics[scale = 0.35]{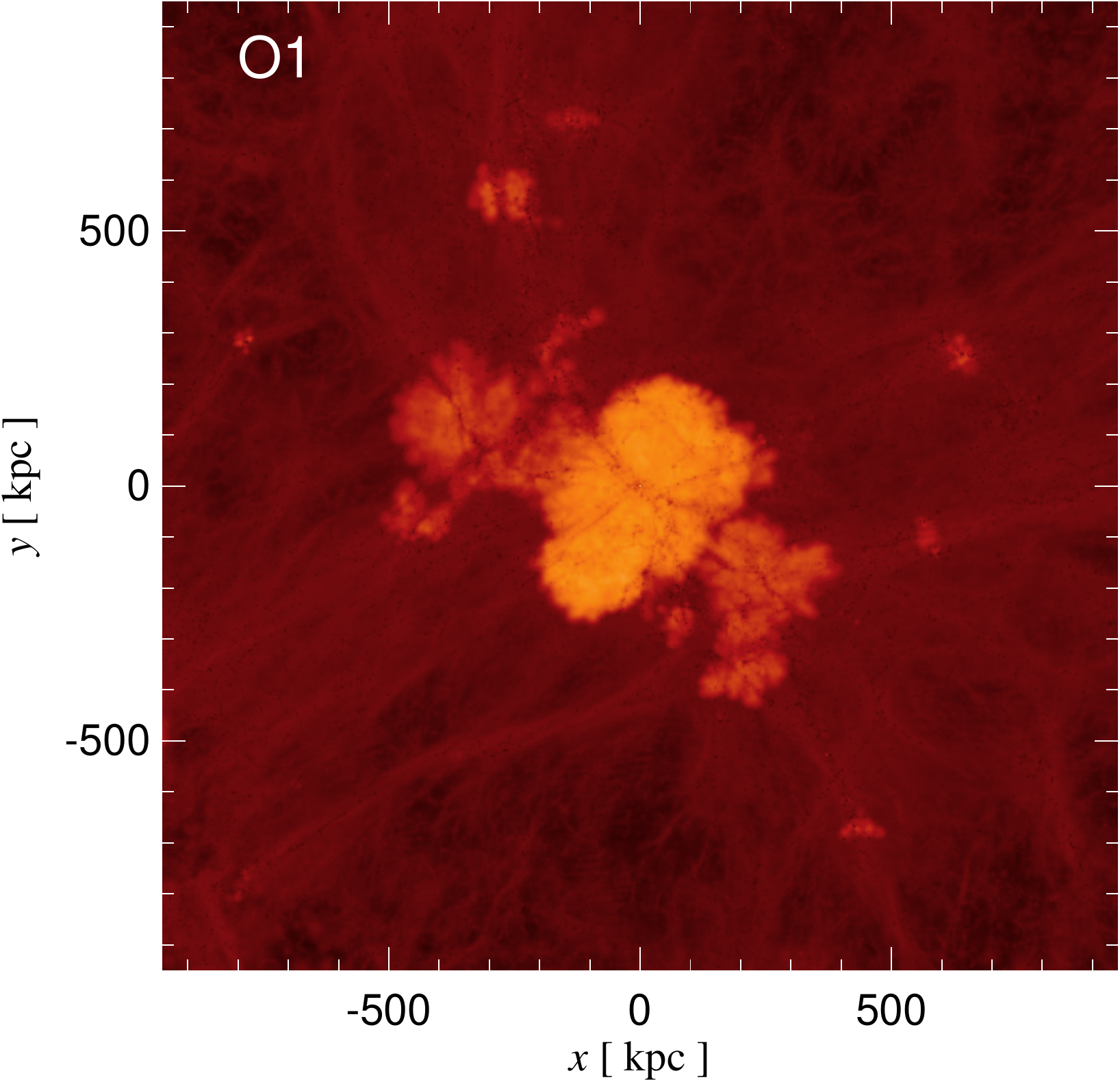}
\end{subfigure}%
\begin{subfigure}{0.33\textwidth}
\centering \includegraphics[scale =
  0.35]{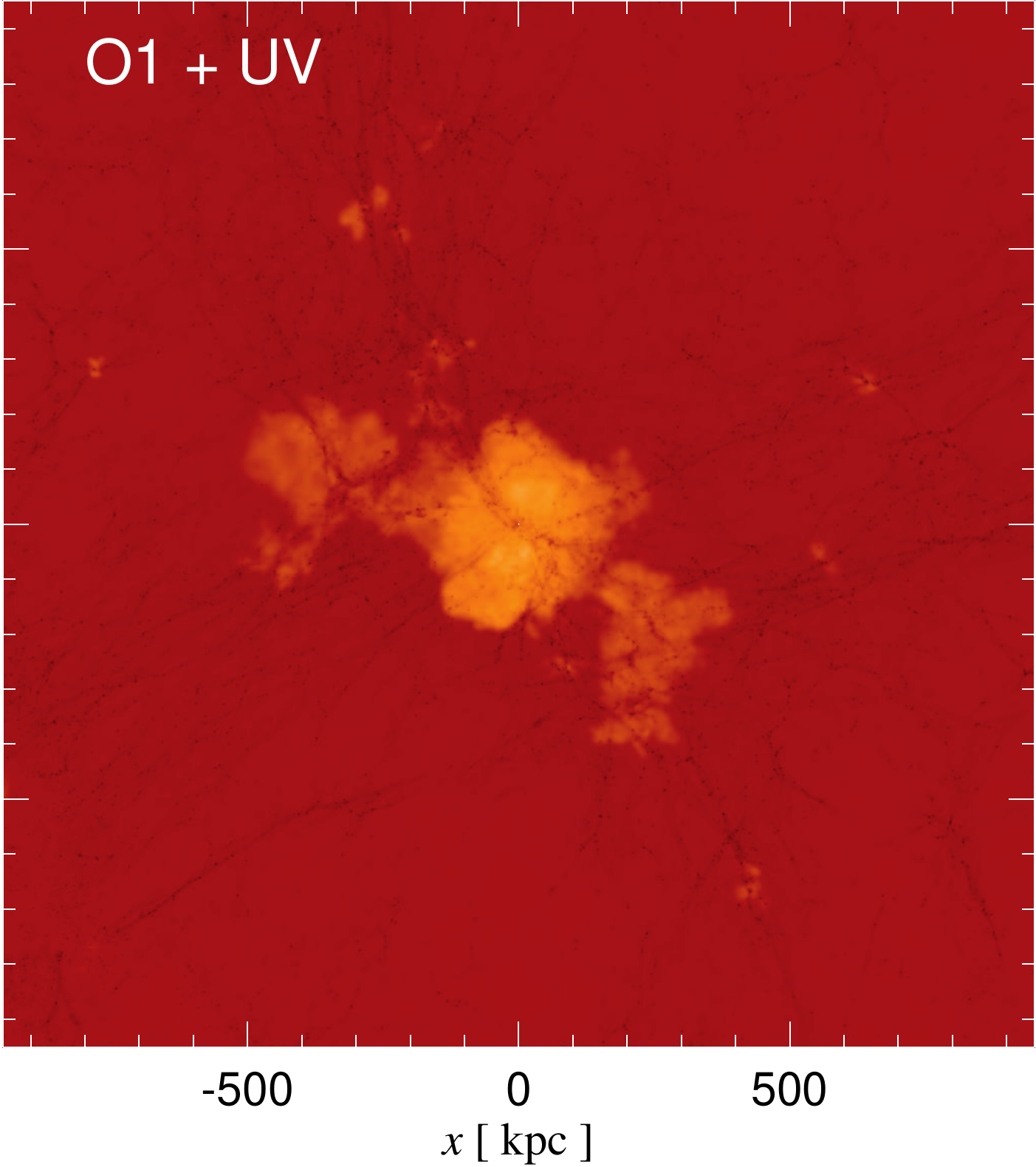}
\end{subfigure}%
\begin{subfigure}{0.33\textwidth}
\centering \includegraphics[scale =
  0.35]{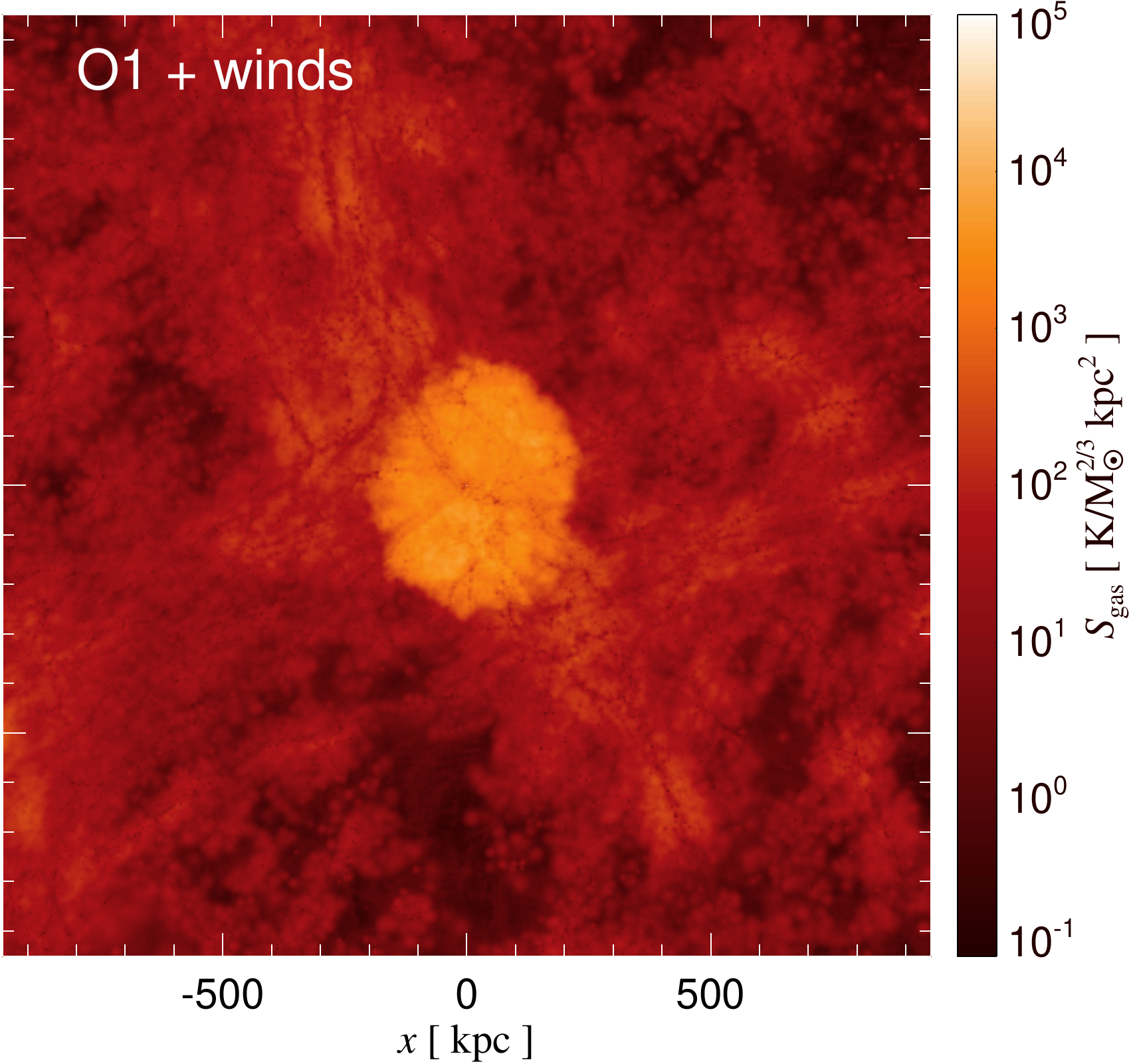}
\end{subfigure}%

\caption{Mass-weighted maps of the entropy projected along a slice of
  thickness $\approx 1.9 \mathrm{Mpc} \, (10 h^{-1} \,
  \mathrm{comoving \, Mpc})$ at $z \, = \, 6.2$ in O$1$ (left),
  O$1$+UV (middle) and O$1$+winds (right) simulations. The entropy was
  estimated as $S \,=\, T/\rho^{2/3}$, where $T$ and $\rho$ are the
  gas temperature and density, respectively. The entropy distribution
  is strongly dependent on the input physics and is mainly generated
  in the virial shocks that form in very massive halos (seen in all
  figures), but also by photoionisation of diffuse IGM gas (middle
  panel) and by dissipation of kinetic energy in galactic outflows
  (right panel).}
\label{entropymaps}
\end{figure*}

The high star formation rates of bright high-redshift galaxies
suggests that energy output from numerous supernova events is
an important source of feedback. Supernova explosions drive
galactic outflows that can drastically reduce the gas content of
galaxies, especially from host halos with shallow potential wells. To
test this in our simulations, we performed two runs for which we
switch on the \citet{Springel:03} prescription for supernova-driven
galactic winds.  In this model, the gas outflow rate $\dot{M}_{\rm
  out}$ is assumed to be proportional to the star formation rate
$\dot{M}_{\rm SFR}$ such that $\dot{M}_{\rm out} = \eta \dot{M}_{\rm
  SFR}$, where the mass loading factor $\eta$ is assumed constant for
every halo. In these two runs, we left the wind speed constant at
$\approx 480 \, \mathrm{km/s}$ and we only varied $\eta$, considering
in turn the typical case where $\eta \,=\, 2$ (O$1$+winds) and a more
extreme case where $\eta \,=\, 10$ (O$1$+winds(strong)), which
should bracket the possible suppression of star formation due to
supernova-driven winds.

In the right-hand panel of Figure~\ref{entropymaps}, we show how the
spatial entropy distribution changes when  supernova-driven outflows
are included in our simulations. Additional entropy is generated due
to dissipation of kinetic energy of the outflow leading to overall
higher levels of entropy around star forming regions.
The mock image of O$1$+winds in the right panel of  Figure~\ref{mockfig}
shows that the number density of galaxies is very sensitive to
galactic winds and it decreases by a factor of $3.7$ ($2.6$) within the
field of view of HST/ACS for $m_{\rm UV} \leq 26.5$ ($m_{\rm UV} \leq
27.5$). Note also that the overdensity of galaxies drops 
quickly beyond the HST/ACS field of view.
Figure~\ref{counts_lum} shows the corresponding number counts of
bright galaxies. Galactic winds affect the number counts in both overdense
and average regions by a similar factor. Within $6.86 {\rm Mpc}^3$ this factor is $\sim 5 \-- 6$
for $\eta \,=\, 2$ and $\sim 17$ for $\eta \,=\, 10$. In the O$1$+winds simulation, the
number of galaxies falls from $50 \, (110)$ to $11 \, (43)$ for
$m_{\rm UV} \leq 26.5 \, (\leq 27.5)$. In O$1$+winds(strong), the
winds are so efficient at ejecting gas out of dark matter halos that
only four bright galaxies remain within the projected volume. For the
A$1$+winds simulation, the observed number counts drop from $6 \,
(16)$ to one (four) and to zero for A$1$+winds(strong). We however
caution that in our simulations with $\eta \,=\, 10$, we are
intentionally considering an extreme scenario that represents
very strong supernova-driven winds which are probably too strong to be
realistic. In Figure~\ref{sfrgal} we show that the inclusion of galactic
winds suppresses star formation across the entire range of dark matter
halo masses. As expected, this effect is more pronounced for
O$1$+winds(strong). Note thereby that the star formation rate is
suppressed comparatively mildly for the FoF group mass range $10^{12}
\-- 10^{13} \, \mathrm{M_{\odot}}$. This is a consequence of extremely
efficient suppression of star formation in lower mass halos. The
surplus gas is instead incorporated in the most massive halos at a later time and
accumulates there. With moderate galactic winds with $\eta \,=\, 2$,
the luminosity function of the A$1$ region is only slightly
higher than the observed luminosity function of \citet{Bouwens:07}
(see the right-hand panel of Figure~\ref{counts_lum}). As A$1$ is the average region with the
highest number of sources of our sample of average regions, it is
likely that other average regions would have luminosity functions
similar or slightly lower than that of \citet{Bouwens:07}, leading to
overall good agreement with observations \citep[see also numerical studies by][]{Finlator:11, Jaacks:12}.
Note however that whereas good agreement with observations is achieved for global properties of simulated galaxies, 
higher resolution is necessary in order to investigate whether currently
 adopted sub-grid models lead to internal properties of simulated galaxies
in line with observations.

We have also verified the impact of AGN feedback on the predicted
number counts of galaxies by comparing to the simulation O$1$nobh,
identical in setup to O$1$ but without black holes.
Including thermal AGN
feedback leads to a drop from $56$ to $50$ galaxies for the $m_{UV}
\leq 26.5$ magnitude limit. The number counts are almost
unchanged when considering the number of galaxies down to $m_{UV}
\,=\, 27.5$, dropping from $111$ to $110$. 
The small impact of AGN feedback on number counts can be attributed to the fact 
that only the most massive halos (with masses above $\sim 10^{11} \, \rm M_\odot$) hosting the brightest galaxies experience efficient enough black hole growth.
Feedback-driven outflows in these galaxies reduce the star formation rates and consequently lower their UV luminosity. However, galaxies with mass $\lesssim 10^{11} \, \rm M_\odot$ contain black holes with masses of at most $\sim 4 \times 10^6 \, \rm M_\odot$ and have therefore not released enough energy to expel gas from their interior. Since detectable galaxies reside in halos with masses down to about $10^{10} \, \rm M_\odot$ and AGN feedback affects only the most massive objects, the number of galaxies in our sample is not significantly affected by this process.

\subsubsection{Reionisation history}\label{Reionisation}

\begin{figure*}
\begin{subfigure}{0.5\textwidth}
\includegraphics[width = \textwidth, height =
  7cm]{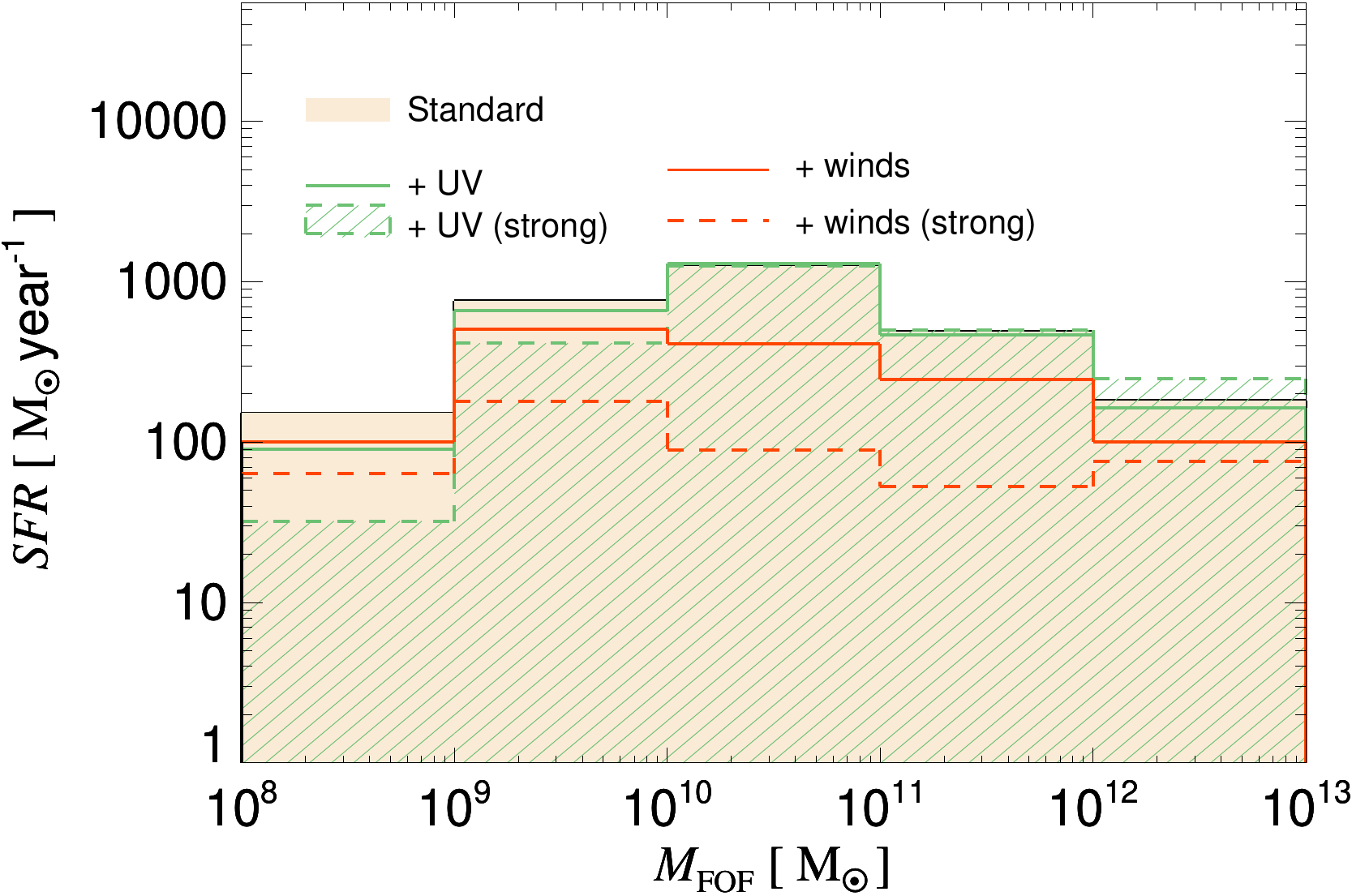}
\end{subfigure}%
\begin{subfigure}{0.5\textwidth}
 \includegraphics[width = \textwidth, height = 7cm]{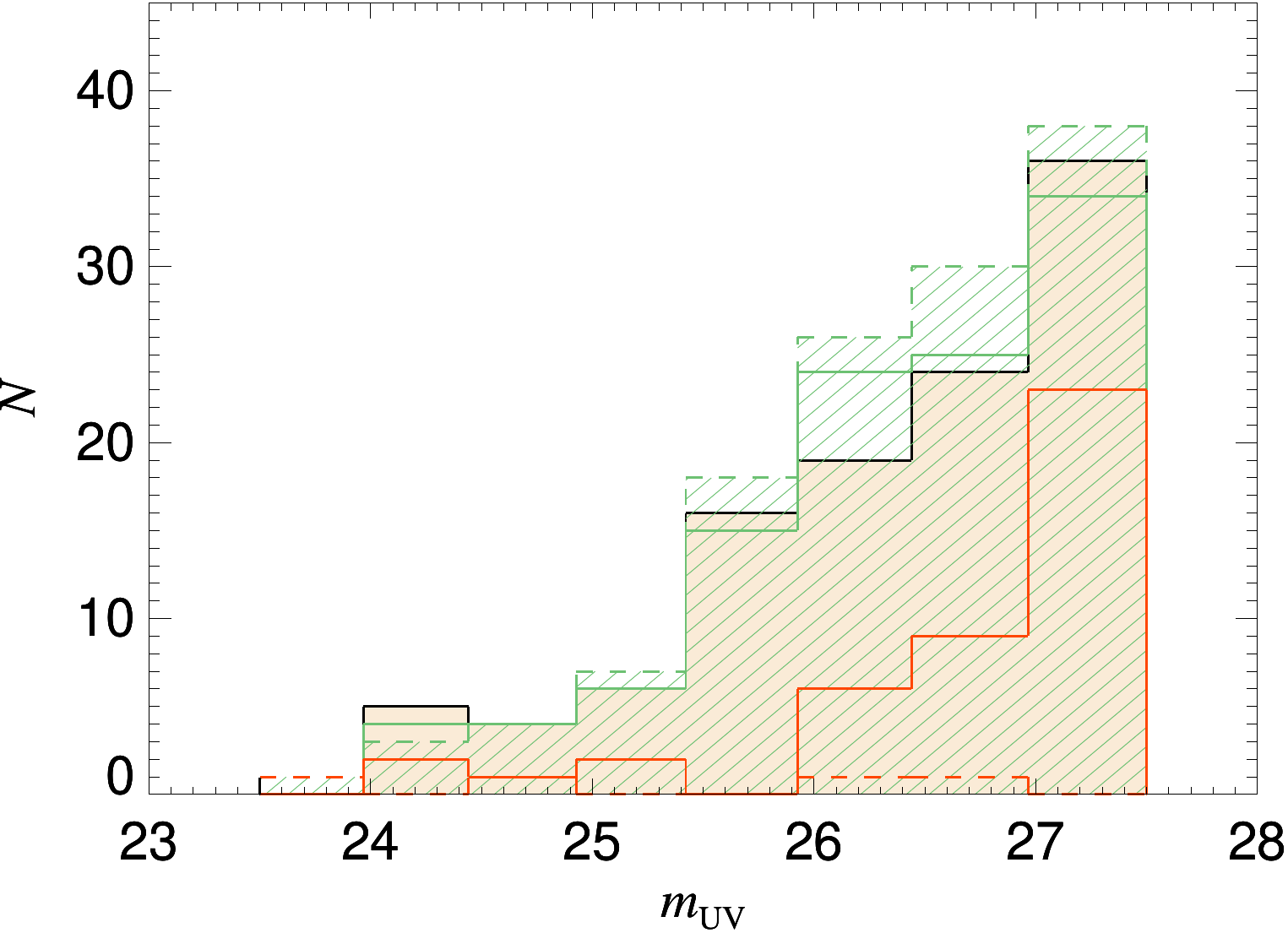}
\end{subfigure}%
\caption{(a) The total star formation rate as a function of FoF group
  dark matter mass for the simulations of region O$1$ including
  different input physics at $z \,=\, 6.2$. Histograms shaded in pink
  give the star formation rates for our default simulation. Results
  from our simulations using different UV backgrounds are given by the
  green histograms; the solid line shows results for the default
  \citet{Faucher-Giguere:09} tables and the dashed-shaded histogram
  shows results for the same tables, but with the ionising flux
  increased  by a factor  $100$. The solid and dashed orange
  histograms are for our simulations of O$1$ including moderate and
  strong galactic outflows, respectively. The biggest contribution to
  star formation in the default O$1$ simulation is provided by halos
  with masses in the range $10^{10 \-- 11} \, \mathrm{M_{\odot}}$. The onset 
of earlier reionisation in the simulation with the
\citet{Faucher-Giguere:09}  UV background suppresses star formation in
halos with $M_{\rm FoF} < 10^{10} \, \mathrm{M_{\odot}}$. This effect
is more pronounced for O$1$+UV(strong), where the additional gas that
does not collapse onto lower mass halos falls at lower redshifts into
more massive halos instead leading to a moderate increase  in star
formation for these halos. Inclusion of  galactic outflows leads to
suppression of star formation in halos of any mass and also shifts the
mass range of the biggest contribution  to star formation to halos
with masses in the range $10^{9 \-- 10} \, \mathrm{M_{\odot}}$.  (b)
Number of identified sources in different magnitude bins for the same
simulations. Earlier reionisation slightly alters the predicted
magnitude of detected galaxies without changing their number
counts. The inclusion of galactic outflows on the other hand clearly
reduces the predicted number of galaxy detections.}
\label{sfrgal}
\end{figure*}

We now investigate the effect of changing the reionisation history.
We have run two additional simulations which are identical to our default
simulation setup apart from the UV background, which we have adopted
from \citet{Faucher-Giguere:09} instead of \citet{Haardt:96}. The
purpose of the two additional simulations ran with the
\citet{Faucher-Giguere:09} UV tables was twofold: to test if
a different reionisation history, with the UV background becoming important at a higher
redshift, or if a stronger ionising background due to the higher number star forming galaxies and QSOs expected in overdensities,
 can lead to lower number counts of bright galaxies.
To obtain a lower limit for the number of
galaxies in our overdense regions, we selected the region O$1$ since
this is the field with the lowest count of galaxies to begin with. In
the first test run, which we shall refer to as O$1$+UV, we replaced
the  UV background of \citet{Haardt:96} with that of
\citet{Faucher-Giguere:09}. In the second test run, O$1$+UV(strong),
we increase the ionising flux by a (large) factor $100$,
in order to reproduce the enhanced UV background due to the presence
not only of a bright QSO but also brighter and more numerous galaxies.
These very
different choices for the reionisation history should  bracket the
possible effects of local, UV bright sources on the star formation
rates of simulated galaxies. Figure~\ref{counts_lum} shows that the
number counts of observed galaxies is not significantly affected by
the choice of the reionisation history. In order to understand this
we have examined the time evolution of the neutral hydrogen
fraction. As expected, the neutral hydrogen fraction drops at higher
redshift in the O$1$+UV and O$1$+UV(strong) simulations, indicating
that reionisation occurs at higher redshift with the
\citet{Faucher-Giguere:09} UV background, especially in the simulation
where  we boost the UV flux by a large factor. The projected
mass-weighted entropy maps shown in Figure~\ref{entropymaps} for the
O$1$ (left) and O$1$+UV (middle) simulations at $z \,=\, 6.2$ show that
the earlier onset of photo-heating alters the spatial entropy
distribution considerably. In the O$1$ simulation, reionisation occurs
at the slightly lower redshift of $z \approx 6$ and entropy is mostly
produced within the virial shock of the central QSO halo. In the
O$1$+UV simulation, however, the spatially uniform heat injection due
to reionisation leads to higher entropy in the entire volume.

The left-hand panel of Figure~\ref{sfrgal} shows that an earlier
reionisation redshift leads to efficient suppression of star formation
only for halos residing in FoF groups with  masses  in the range $10^8
\-- 10^9 \, \mathrm{M_{\odot}}$. For more massive halos, the effect of
the heat injection during reionisation is marginal. Galaxies hosted by
these small mass halos are not bright enough to meet our magnitude
selection criterion and our predictions for the number counts are
therefore not very sensitive to the heat injection during
reionisation. The right-hand panel of Figure~\ref{sfrgal} shows that
the suppression of star formation in the lowest mass halos only
somewhat lowers the prediction of the observed number of galaxies with
magnitude in the range $27 \-- 27.5$ for the O$1$+UV simulation. In
the O$1$+UV(strong) run, star formation is suppressed by about an
order of magnitude in halos in the mass range $10^8 \-- 10^9 \,
\mathrm{M_{\odot}}$. The excess gas which fails to collapse onto these
low mass halos falls later on into the deeper potential wells of more
massive halos. The resulting boost in star formation rates leads to
more UV emission and hence overall brighter galaxies, thus leading to
the opposite effect than needed to suppress the number of galaxies.
 Thus, for the magnitude
limits investigated here even a very strong ionizing UV background has
very little effect on the predicted number counts of galaxies in QSO
fields.

\subsubsection{Predicted  (over-)densities of star-forming galaxies in  ACS/HST-fields}

\begin{figure*}
\begin{subfigure}{0.5\textwidth}
\centering \includegraphics[scale = 0.5]{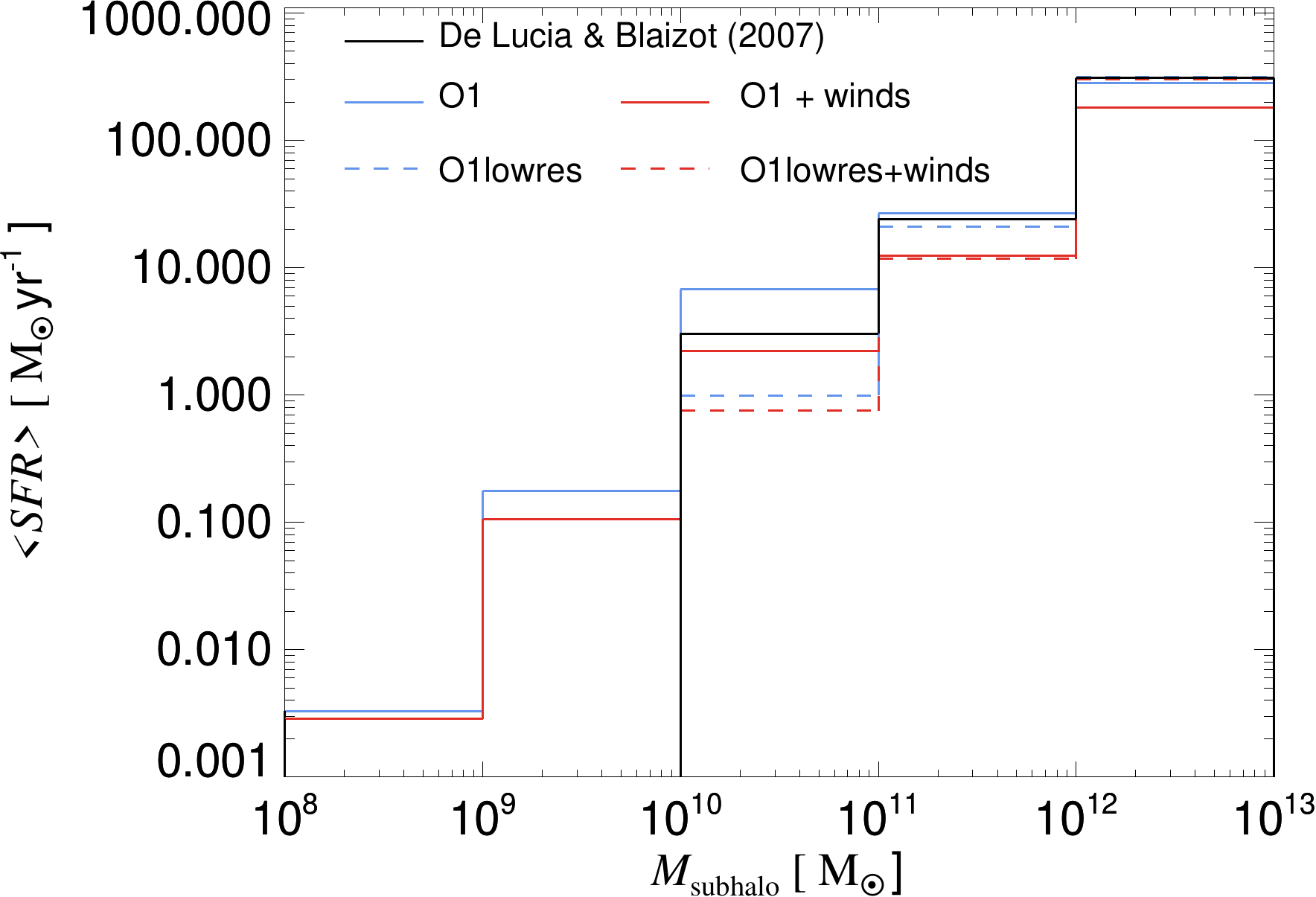}
\end{subfigure}%
\begin{subfigure}{0.5\textwidth}
\centering \includegraphics[scale = 0.5]{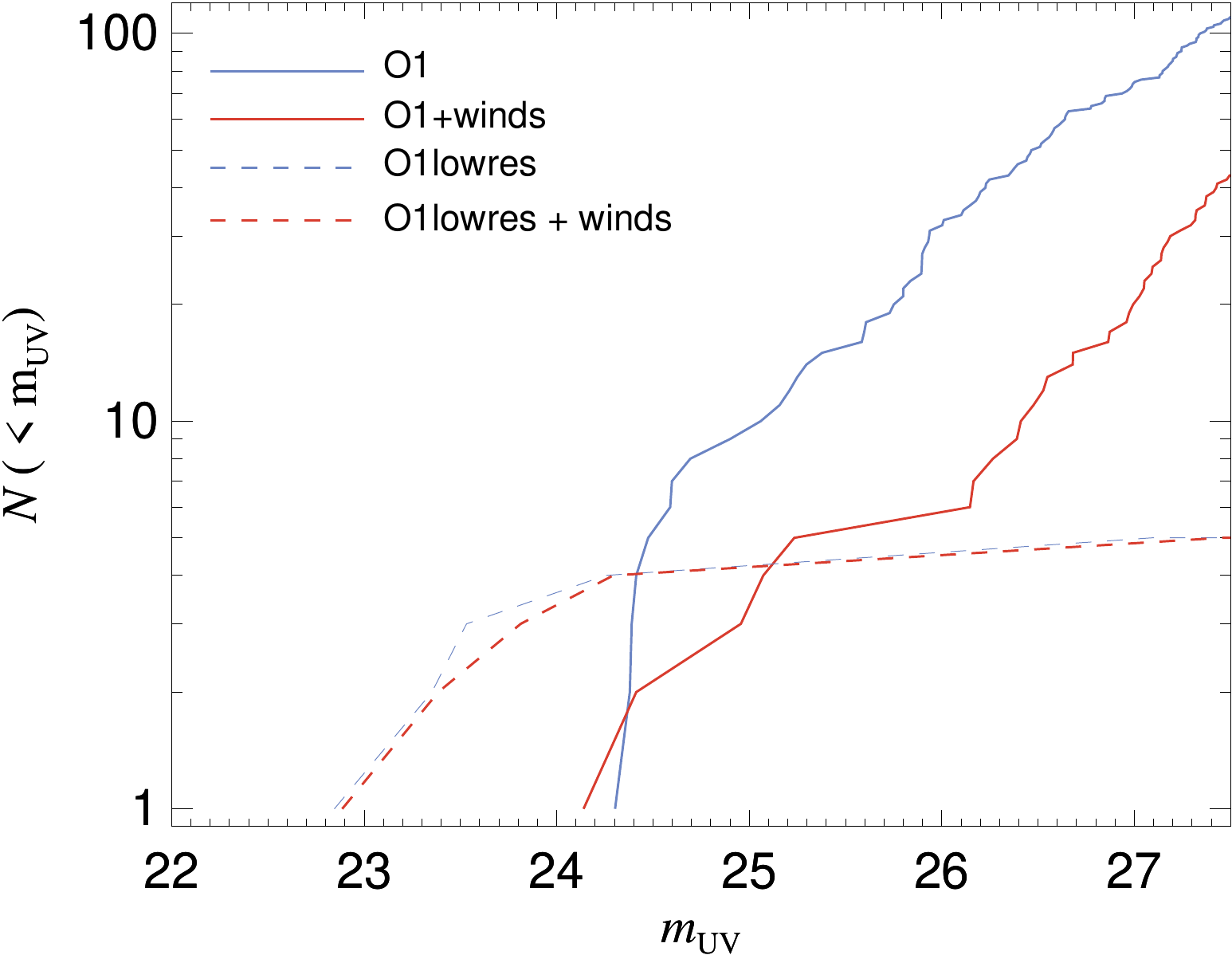}
\end{subfigure}%
\caption{(a) The average star formation rate of subhalos as a function
  of subhalo mass (dark matter only) for our O$1$, O$1$+winds, O$1$lowres
  and O$1$lowres+winds simulations as well as according to the 
  semi-analytic catalogues of 
  \citet{DeLucia:07}. Agreement between these three models is very
  good for subhalos with mass $> 10^{11} \, \mathrm{M_{\odot}}$, but
  becomes poor for lower mass subhalos in O$1$lowres, O$1$lowres+winds simulations and 
  in the semi-analytic catalogue. For low-mass halos  star formation
  is underestimated when compared to our high resolution simulation,
  O$1$. (b) Cumulative plot of number counts as a function of apparent
  magnitude in  O$1$, O$1$+winds, O$1$lowres and O$1$lowres+winds simulations. 
  In O$1$lowres,
  halos hosting observable galaxies are not well resolved leading to a
  drastic under-prediction of sources with respect to O$1$. Incorporating galactic 
  winds does not affect this conclusion.}
\label{milcomp}
\end{figure*}

We are now in a position to estimate the number of star forming
galaxies in the environment of bright $z \sim 6$ QSOs as well as in average density regions.
In order to estimate how many sources an
HST/ACS observation employing the usual drop-out technique 
would be expected to detect, we consider a pencil beam with an area given by the
field of view of ACS, i.e. $202 \times 202 \,
\mathrm{arcsec^2}$. Following \citet{Romano-Diaz:11}, i.e. assuming a
redshift range  of $\Delta z \,=\, 1$, we get $V \approx 87.1 \,
\mathrm{Mpc^3} \, (11628 h^{-3} \, \mathrm{comoving \, Mpc^3})$ for
the volume probed by the \citet{Kim:09} observations. Due to incompleteness 
as well as contamination by foregrounds, the effective volume probed by these 
observations is likely to be equal to half of $V$ \citep{Romano-Diaz:11}. From Table~\ref{tablecounts}, we first compute the mean number of sources
for our simulations without strong stellar feedback. Within $6.86 {\rm Mpc}^3$ and for $m_{\rm UV} \,\le\, 26.5$ we obtain $N_{\mathrm{avg}} \,=\, 3.5 \pm 0.7$ 
and $N_ {\mathrm{QSO}} \,=\, 68.8 \pm 10.6$ for our samples of average density
and  highly-overdense  regions, respectively.  
The space density of galaxies is strongly enhanced 
by a factor of about twenty around the  most massive halos in the
Millennium simulation. The expected number of sources in an average field as seen by HST/ACS, 
can be estimated   as  $N_{\mathrm{ACS}} \,=\, 0.5 \times \frac{N_{\mathrm{avg}}}{V_{\mathrm{cube}}} \times
V \, = \, 20.3 \pm 4.1$, where $V_{\mathrm{cube}}$ is the
volume of the cube used for  producing the mock images.  As expected, this is 
significantly  larger than the number of observed drop-out galaxies to
the same magnitude limit in a field of this size (according to \citet{Kim:09} 4-8
per ACS field depending on the exact colour selection criterion), as we have used our
simulations without strong stellar feedback for the calculation.

Repeating now the calculation for our simulations with galactic winds, 
we estimate for \citet{Kim:09}-like observations a number of sources
of $N_{\mathrm{ACS}} \,=\, 5.8$ for A$1$+winds in an ACS-sized field. Assuming that
winds reduce the number of sources by a similar factor in all of our 
simulated average density regions, the predicted  number of galaxies in 
an ACS-size field is  $N \,=\, 3.4 \pm 0.7$. This is in good agreement with the observed 
number which is quite  sensitive to the details of the drop-out
selection. To calculate the expected number of sources in fields
containing one of our simulated bright $z \sim 6$ QSOs 
we assume that ACS observations probe a pencil beam volume
made up of one of our highly overdense regions  complemented with our average 
density regions  so as  to fill a  redshift range of $\Delta z =1$. As the excess 
of sources around the bright QSOs falls off
rapidly with radius (see again the upper right-hand panel of Figure~\ref{mockfig}) this should
be a good approximation. Based on the A$1$+winds and O$1$+winds simulations 
we estimate  $N_{\mathrm{ACS}} \,=\, 10.1$  
for the number of star-forming galaxies in an ACS-size field.
If we assume again that winds reduce the number of sources 
by a similar factor in the different simulated average density and
highly overdense regions we get  $N \,=\, 9.1 \pm 2.1$. This corresponds 
to a difference of about five sources in ACS - fields with bright  $z
\sim 6$ QSOs compared to average density fields, in good agreement
with the two observed overdense fields  in \citet{Kim:09}. 
\citet{Kim:09}, however, also observed two  underdense fields.
We have not noted any significant reduction in the number of galaxies 
due to AGN feedback in any of our simulations and the occurrence of such
underdense field would thus have to be attributed either to 
physical effects not included in our simulations or perhaps 
more likely to subtleties in the colour-selection criteria 
(physical or observational) or low-number statistics. These observational 
limitations arise primarily because of the relatively shallow depth of the 
\citet{Kim:09} observations. Since our simulations predict that the galaxy 
overdensity around the QSO halos extends to fainter luminosities with an increasing 
number of sources, deeper imaging of the underdense fields identified by \citet{Kim:09} 
would discriminate between small number statistics fluctuations and/or scatter in the 
color selection for LBGs. In addition, surveys with redshift information 
for the star-forming galaxies should -- at least in principle -- be 
able to provide stronger constraints. Obtaining redshifts at $z \sim 6$ is,
however, very challenging. Most interesting in this regard is perhaps a survey 
by \citet{Husband:13} at somewhat lower redshift  which finds
significant number of LBGs with redshifts within $\Delta z= 0.05-0.1$
of three bright $z\sim 5$ QSOs.  There have also been several attempts using $Ly\alpha$ 
emitters which, unfortunately,  have been rather inconclusive,  
not least because of the difficulty of interpreting the presence or
absence of  $Ly\alpha$  emission at these redshifts \citep{Banados:13}.
In summary, our predictions for the overdensities around bright QSOs 
appear -- apart from the reported occurrence of underdense regions in \citet{Kim:09} --
to be consistent  with observations, but there is no strong
observational evidence yet that observed overdensities 
require the bright $z\sim 6$  QSOs to be hosted in very massive haloes. 
The possibility that the host dark matter halos
have considerably lower masses and hence a lower number count of
galaxies has been recently discussed by \citet{Fanidakis:13}. In our
simulations however, we have verified that halos with viral masses in
the range $4 \-- 8 \times 10^{11} \ \mathrm{M_{\odot}}$ only host
$M_{\rm BH} \, = \, 3 \times 10^7 \-- 10^8 \, \mathrm{M_{\odot}}$
instead of the required $\sim 10^9 \, \mathrm{M_{\odot}}$. 
The formation of QSOs in lower mass halos would therefore require a picture
significantly different from the one in which supermassive black holes
grow from massive seeds by essentially Eddington limited accretion.

\begin{figure*}
\begin{subfigure}{0.5\textwidth}
\centering \includegraphics[scale = 0.5]{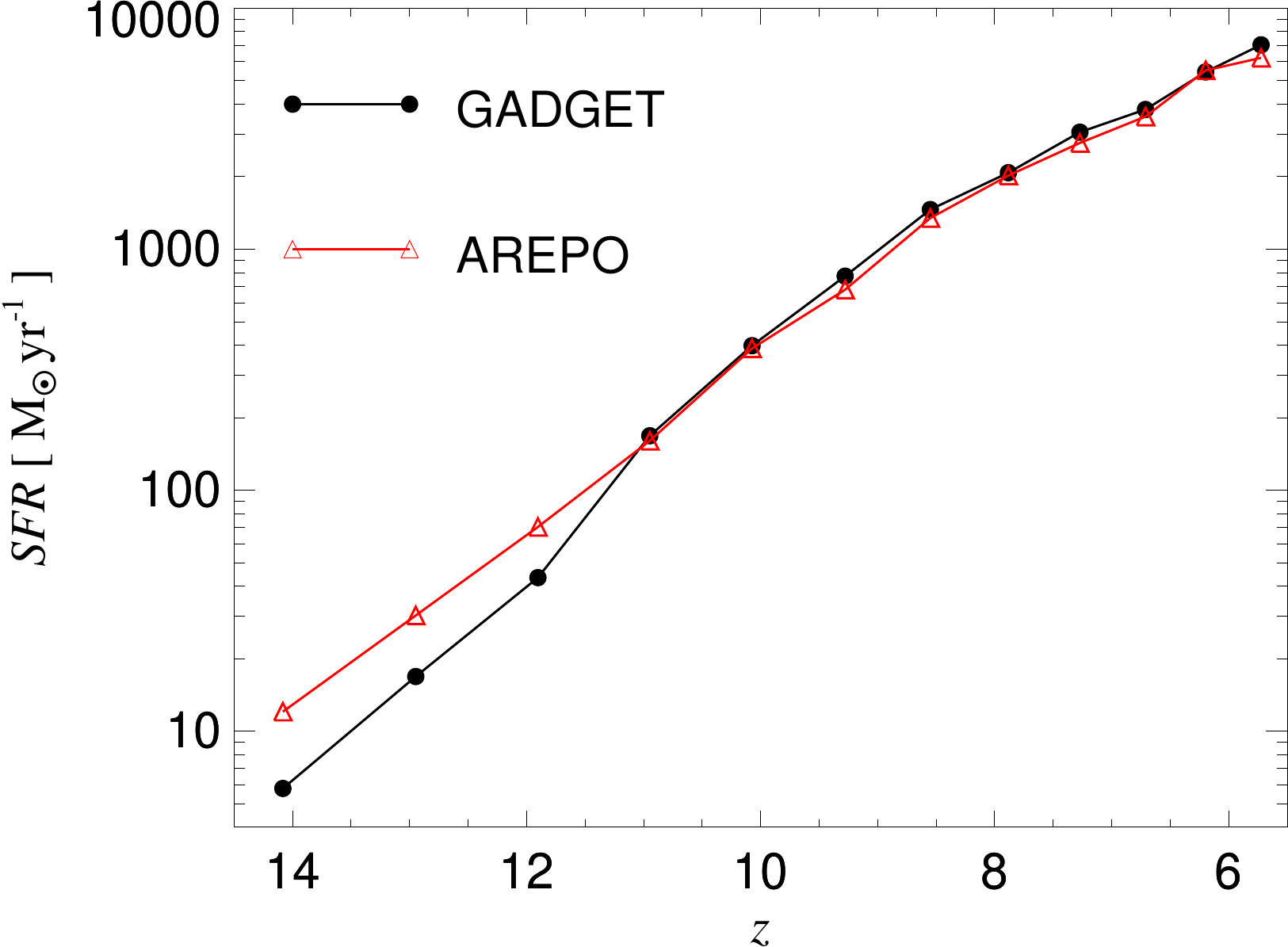}
\label{codecomp1}
\end{subfigure}%
\begin{subfigure}{0.5\textwidth}
\centering \includegraphics[scale = 0.5]{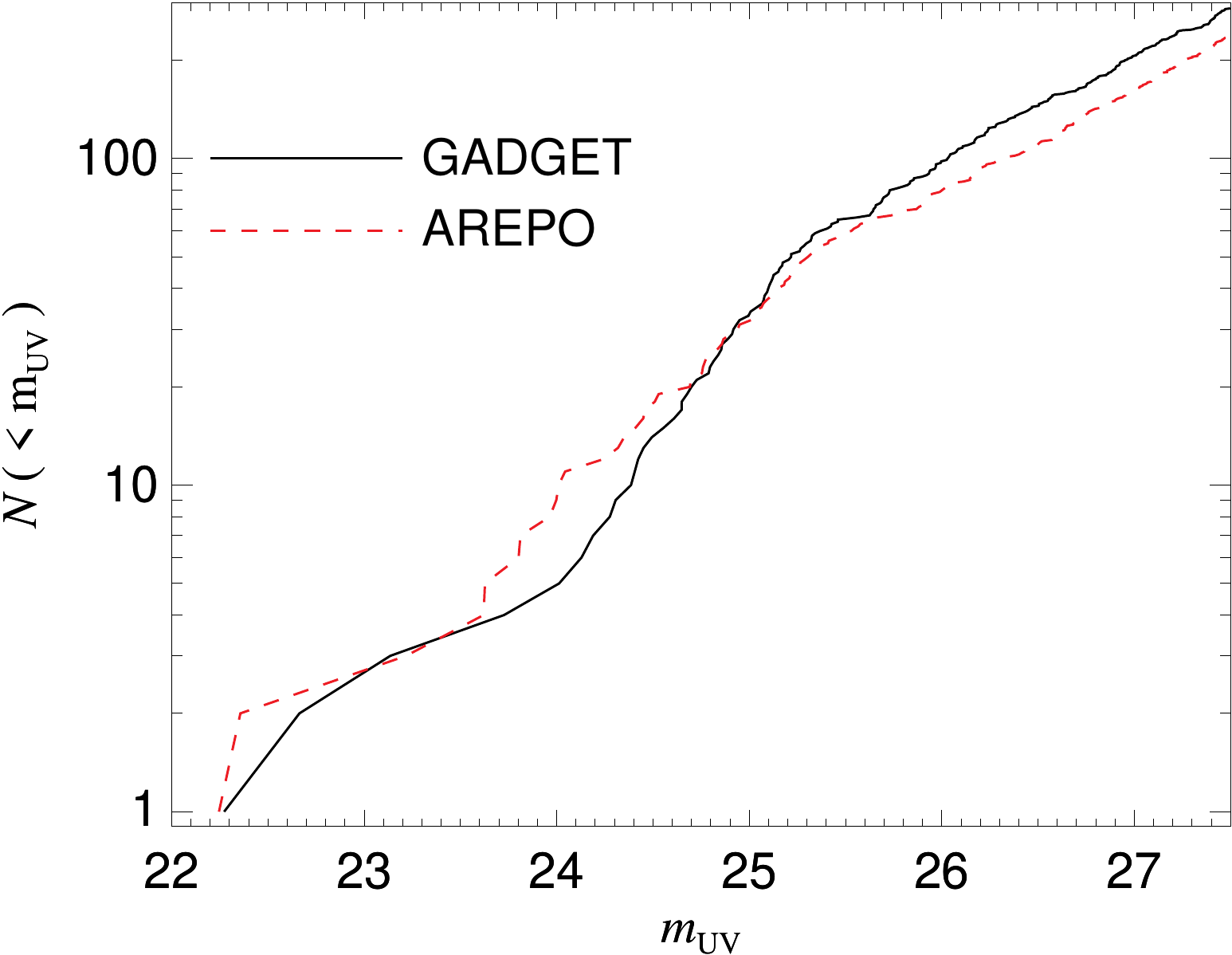}
\label{codecomp2}
\end{subfigure}%
\caption{(a) The total star formation inside a cube of side length
  $1.9 \, \mathrm{Mpc} \, (10 h^{-1} \, \mathrm{comoving \, Mpc})$
  centred on the most massive halo as a function of redshift for
  O$6$GADGET (black filled circles) and O$6$AREPO (red triangles). The
  close agreement between the predictions of both codes shows that the
  two differing hydro solvers converge on the global star formation
  rates at very high redshifts. (b) Cumulative number of sources at an
  apparent magnitude below a given value $m_{\rm UV}$ for O$6$GADGET
  (black) and O$6$AREPO (red). There is a systematic overproduction of
  galaxies at the faint end ($m_{\rm UV} \gtrsim 25.5$) in the
  simulation performed using GADGET. The excess of galaxies picked up
  in the GADGET simulation also exist in the AREPO simulation, albeit
  it is below the magnitude cut. This is a consequence of deficiencies
  in the GADGET SPH solver, which lead to more compact galaxies and
  hence higher star formation surface density, leading to brighter
  objects at the faint end.}
\label{codecomp}
\end{figure*}

\subsection{Numerical issues}

\subsubsection{Numerical resolution and comparison to semi-analytical models based on the Millennium simulation}\label{numconv}

In this section, we assess the dependence of the predicted number
counts and star formation rates of bright galaxies on the resolution
of our simulations and compare our results to semi-analytic models
based on the Millennium simulation. 
Semi-analytic models are computationally cheaper and can therefore 
be used to obtain global properties of a sample of galaxies in very large volumes. 
Cosmological hydrodynamical simulations cannot presently simulate such large volumes at sufficient resolution. However, they treat gas hydrodynamics more self-consistently together with the interaction between baryons and dark matter. They also can more realistically follow gas inflows, outflows and non-linear processes occurring in phenomena such as mergers, turbulence and fragmentation. A more detailed representation of such processes then enables closer comparison with observations and subsequent tests of 
sub-grid physics.
For this purpose, we compare our
default O$1$ simulation, with a dark matter particle resolution of
$6.75 \times 10^6 h^{-1} \, \mathrm{M_{\odot}}$ and spatial resolution
$1 h^{-1} \, \mathrm{comoving \, kpc}$ to a lower resolution
simulation of O$1$lowres with dark matter particle mass $m_{DM}
\,=\, 8.44 \times 10^8 h^{-1} \, \mathrm{M_{\odot}}$ and softening
length $\epsilon \, = \, 5 h^{-1} \, \mathrm{comoving \, kpc}$,
parameters very similar to that of the Millennium simulation. 

We ran SUBFIND \citep{Springel:01} in post-processing for snapshots at $z \, = \, 6.2$ for
the O$1$ and O$1$lowres simulations in order to identify the
gravitationally bound dark matter structures to which galaxies are
more directly related. The resulting catalogue of subhalos in each of
our simulations makes it possible to draw a fair comparison with galactic
properties from the semi-analytic models of \citet{DeLucia:07}, 
where prescriptions for galaxy formation were
applied to the dark matter subhalos identified in the same parent
Millennium simulation and are available in an on-line
database\footnote{The Millennium database is accessible on
  http://www.mpa-garching.mpg.de/millennium}. We focus here on
resolved subhalos located within a spherical volume with radius
$\approx 1.3 \, \mathrm{Mpc} \, (7 h^{-1} \, \mathrm{comoving \,
  Mpc)}$, which is well inside our high resolution region. In
the left-hand panel of Figure~\ref{milcomp}, we show the average star formation rate of
subhalos for different subhalo (dark matter) mass bins. The average
star formation rates in both our low and high resolution simulations
are in close agreement for $M_{\rm subhalo} \gtrsim 10^{11} \,
\mathrm{M_{\odot}}$, where subhalos are well resolved. The agreement
is poor for the lower mass subhalos, $M_{\rm subhalo} \lesssim 10^{11}
\, \mathrm{M_{\odot}}$. Due to low resolution, compact dense
regions where star formation is particularly efficient are not
accounted for, leading to star formation rates lower by a factor of
about four.

Note that spurious angular momentum transfer, as can arise in cosmological simulations due to low resolution, \citep{Okamoto:05, Kaufmann:07, Guedes:11} would normally lead to higher (and not lower) star formation rates as gas flows into the centre of halos more efficiently.
Moreover, as shown in Figure~\ref{milcomp}, star formation rates are converged for halos with mass
$\gtrsim 10^{11} \, \rm M_\odot$ in the low resolution simulations.
Since our high resolution simulations resolve halos with a mass $\sim 10^8 \, \rm M_\odot$, star formation rates in
halos hosting detectable galaxies, i.e. with  mass $\gtrsim 10^{10} \, \rm M_\odot$,  should also be well within numerical convergence, 
indicating that numerical angular momentum transfer does not significantly affect our results.
We have further verified that simulated galaxies have very flat rotation curves, with a ratio of peak circular velocities to circular velocity
 evaluated at the virial radius of $\lesssim 1.05$.
A detailed study of internal properties of simulated galaxies however requires even higher resolution simulations and is therefore well beyond the scope of this study.

In the same plot in Figure~\ref{milcomp}, we also show the average star formation rates of
subhalos for the corresponding galaxies according to
\citet{DeLucia:07}. There is good agreement between the semi-analytic
predictions and our numerical simulations, particularly with
O$1$lowres, where the force resolution is the same as in the Millennium
simulation on which 
\citet{DeLucia:07} is based. The small differences between the O$1$lowres and
\citet{DeLucia:07} galaxy catalogues can be explained by somewhat
different prescriptions for star formation and associated feedback
processes. We note, however, that the
semi-analytic galaxy simulations  by \citet{DeLucia:07} lack
sufficient resolution to resolve dark matter halos below $\approx
10^{10} \, \mathrm{M_{\odot}}$ and thus do not account for star
forming galaxies in these low mass halos.

In the right-hand panel of Figure~\ref{milcomp}, we show the cumulative number
count of galaxies as a function of apparent magnitude identified
following the method outlined in Section~\ref{secmethod2} for the
simulations O$1$ and O$1$lowres. A drastic reduction in galaxy numbers occurs by varying the resolution alone. 
By matching the IDs of the
stellar particles situated within the pixels associated to each galaxy
by SExtractor with their parent FoF group, we verified that the bulk
of galaxies absent in O$1$lowres simulation resides in FoF groups of
mass $\sim 10^{10} \, \mathrm{M_{\odot}}$ and in a few cases even in
as low as a few times $10^{9} \, \mathrm{M_{\odot}}$. The fact that
these halos are only marginally, if at all, resolved in our low
resolution run leads to the discrepancy of one order of magnitude in
the number of galaxies brighter than $m_{\rm UV} \,=\, 27.5$. Note also a
similar order of magnitude discrepancy for galaxies with magnitude
$m_{\rm UV} \,\le\, 26.5$. On the other hand, since our high resolution simulations
resolve halos of mass $\approx 3 \times 10^8 \, \mathrm{M_{\odot}}$
which contain galaxies fainter than the $27.5$ magnitude limit, our
predictions for the number counts of bright galaxies are numerically
converged.

\citet{Overzier:09} match the observed surface density of i-dropouts at $z \,=\, 6$ 
using the galaxy catalogues of \citet{DeLucia:07}.
However, our results unambiguously indicate that halos with masses down to $\sim 10^{10} \, \mathrm{M_\odot}$ 
can host detectable galaxies and must therefore be well resolved. 
Thus, similar semi-analytic studies based on higher resolution dark matter simulations
 would likely require stronger feedback in order 
to still match the observed surface density of i-dropouts at $z \,=\, 6$.
Differences in galaxy numbers in overdense regions according to our simulations and \citet{Overzier:09}
may also arise from the treatment of dust in their models (ignored in our simulations). 
However, the impact of dust in galaxies at $z \,=\, 6$ is thought to be low \citep{Bouwens:07, Labbe:10} 
and likely small when compared to the order of magnitude effect of varying the resolution.

\subsubsection{Hydro solver comparison}\label{seccodecomparison}

Different choices of hydro solvers used in cosmological simulation
codes can give rise to significant differences in the properties of
simulated galaxies and the intergalactic medium they are embedded
in. Recent cosmological simulations using the moving mesh code AREPO
have highlighted a number of systematic differences in gas cooling and
star formation rates, as well as in galaxy sizes and morphologies with
respect to the identical simulations performed with the GADGET code
\citep{Sijacki:12, Keres:12, Vogelsberger:12, Torrey:12, Nelson:13,
  Bird:13}. It is therefore important to verify that our predictions
for the number counts of galaxies are robust against changes in the adopted
hydro-solver scheme.

We have resimulated our O$6$ region with the moving mesh
code AREPO using the same setup as outlined in
Section~\ref{secmethod1}. Since our focus here is on comparing the
global star formation rate and the number of observable galaxies, we
have excluded black holes and AGN feedback from these
simulations. Also by dropping the black hole module we can more
straightforwardly compare the two codes, given that GADGET and AREPO
share the same implementation for gas cooling and sub-grid star
formation, allowing us thus to isolate the possible differences that
arise only due to the hydro solver. In the left-hand panel of
Figure~\ref{codecomp}, we show the total star formation rate within a
cube of $1.9 \, \mathrm{Mpc} \, (10 h^{-1} \, \mathrm{comoving \,
  Mpc})$ on a side as a function of redshift for AREPO and GADGET runs
(which we shall refer to as O$6$AREPO and O$6$GADGET
respectively). Even though at high redshifts star formation rates are
somewhat higher in AREPO, from $z \lesssim 11$ they are in very close
agreement. The reason why we do not see any significant differences in
the star formation rates predicted by these two codes is due to the
fact that gas is in the rapid cooling regime and subject to the
identical UV ionizing background. In fact, a similar result can
be seen in the top panel of Figure $2$ of \citet{Nelson:13}, where
distributions of the maximum past temperature of gas accreting onto
different mass halos are shown. Only when halos are sufficiently
massive and are at a sufficiently low redshift such that the
quasi-hydrostatic hot atmosphere develops the differences between
GADGET's and AREPO's hydro solver start to result in significant and
systematic discrepancies. This finding also supports the view that at
 this very high redshift, our simulations should not significantly suffer from numerical overcooling or spurious angular momentum transfer.

\begin{figure*}
\begin{subfigure}[l]{0.5\textwidth}
\centering \includegraphics[scale = 0.5]{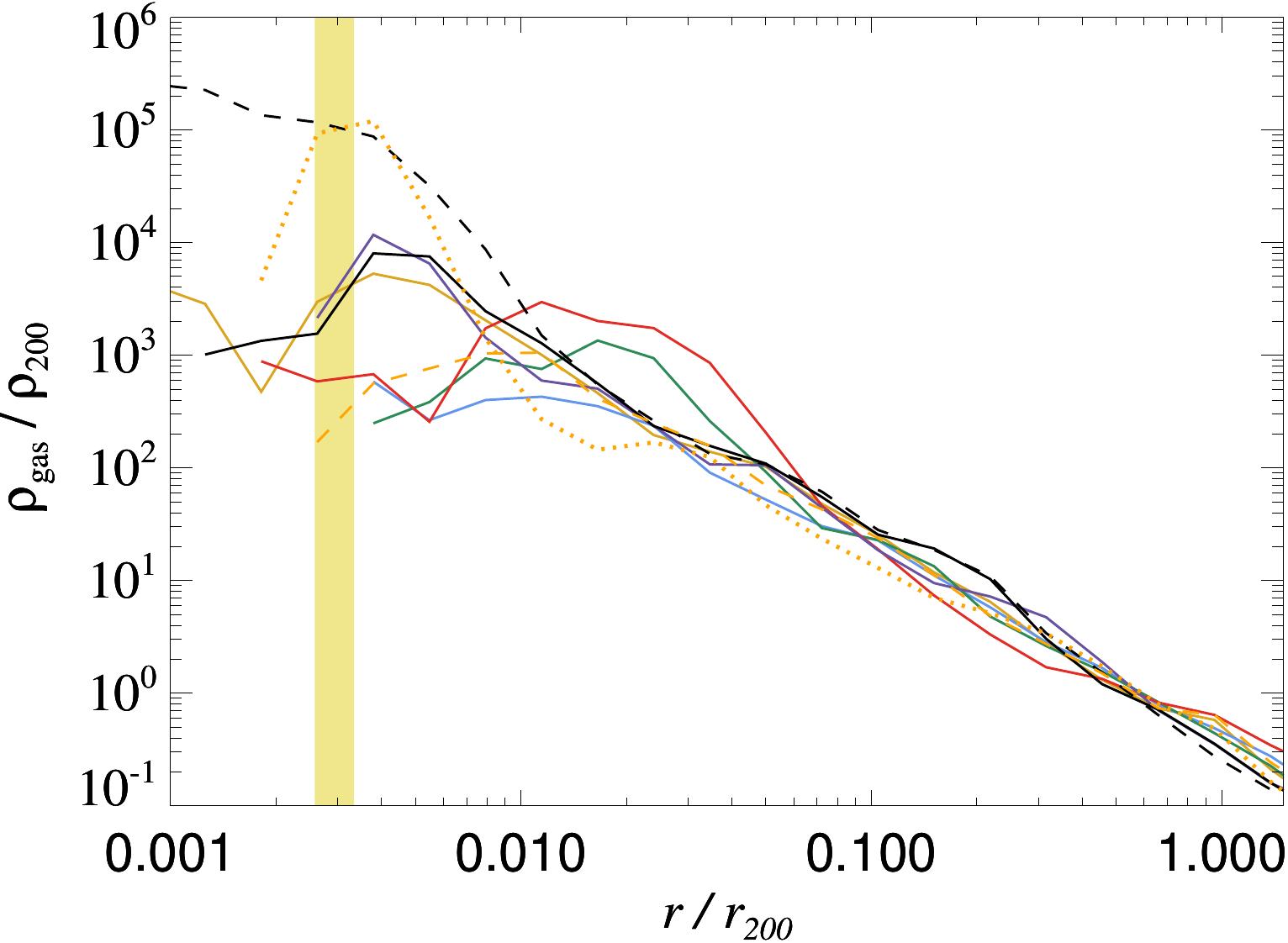}
\label{gasprofile}
\end{subfigure}%
\begin{subfigure}[r]{0.5\textwidth}
\centering \includegraphics[scale = 0.5]{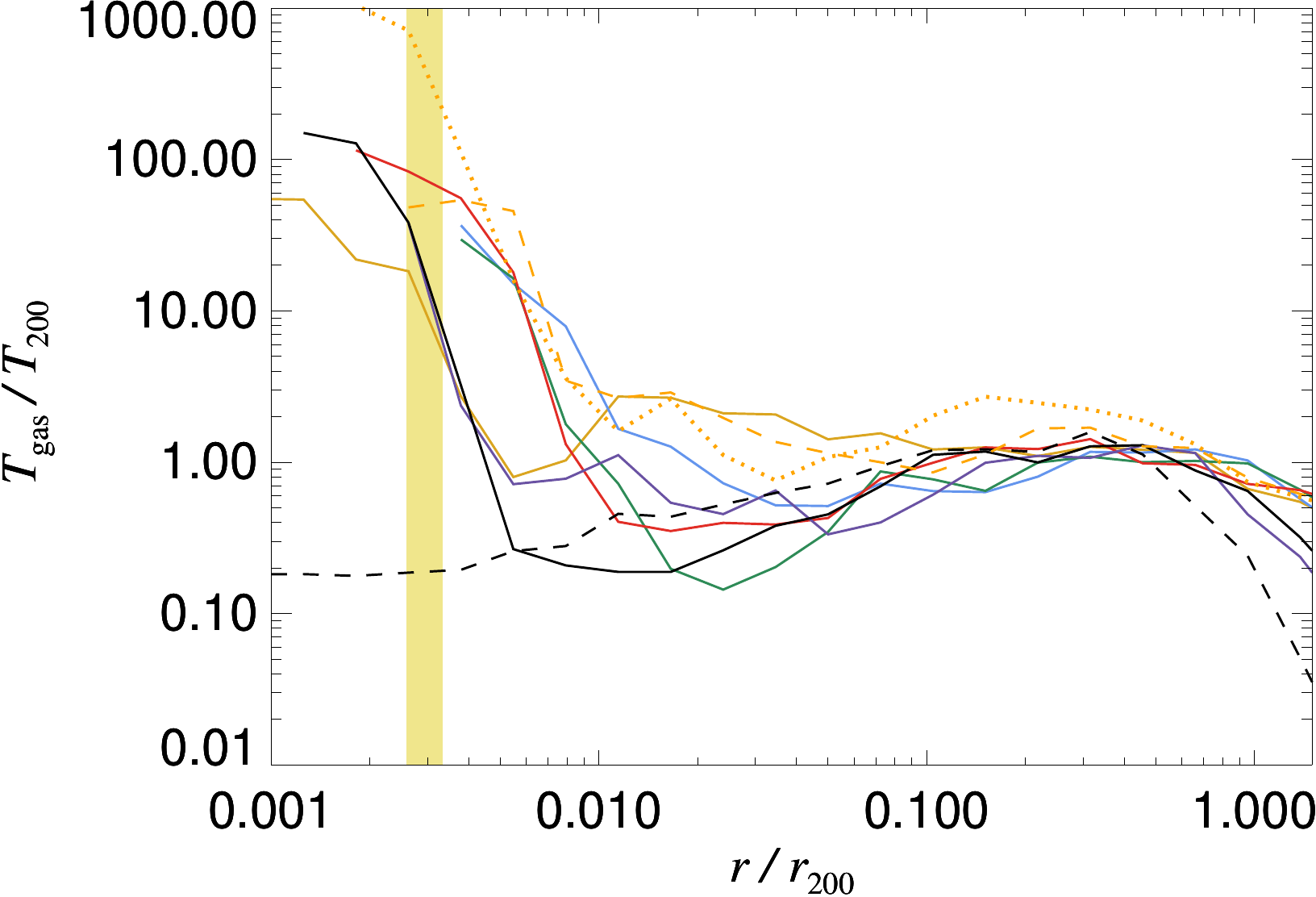}
\label{tempprofile}
\end{subfigure}
\caption{Mass-weighted radial profiles of gas (a) density and (b)
  temperature at $z \,=\, 6.2$ shown for all overdense regions
  (coloured solid lines), O$6$nobh (dashed), O$1$+winds (dashed
  yellow) and O$1$+winds(strong) (dotted yellow). Densities, temperatures and radii are scaled to the virial
  density, temperature and radius of each respective simulation. The
  vertical yellow band marks the location of the gravitational softening length
  which has a fixed physical scale and thus varies somewhat when
  scaled to the virial radius  of the different halos. The central
  density is highest for the simulations O$5$ and O$6$, where the
  central supermassive black holes are correspondingly accreting at
  the highest rate in our sample of overdense regions (see fourth
  column in Table~\ref{table3}). Comparing O$6$ with O$6$nobh clearly
  shows that AGN feedback reduces the central density by one to two
  orders of magnitude. The decrease in density corresponds to the
  increase in temperature for all regions. For the case without AGN heating, gas remains cold in the
  central regions.}
\label{profiles}
\end{figure*}

We then proceeded with the construction of the
mock images for these two simulations at $z \, = \, 6.2$. We employed
the same method as in Section~\ref{secmethod1}, except that here for
both runs we assumed the star formation rate and the UV luminosity to
be related as in \citet{Labbe:10}. The right-hand panel of
Figure~\ref{codecomp} shows the cumulative number of galaxies below a
given apparent magnitude in the UV band. While for the magnitude range
from $23.6$ to $25$, O$6$AREPO contains  about five more bright
galaxies, there are in total $\approx 1.2$ more sources in O$6$GADGET 
due to the differences at the faint end. The excess of
bright galaxies in AREPO is caused by more efficient gas cooling into
halos that already at $z \sim 6$ are starting to become sufficiently
massive so that the differences between the two hydro solvers start to
show up. To understand why there are differences at the faint end of
the luminosity function we matched galaxies between our GADGET and
AREPO simulations. Considering the coordinates of a given object in
one simulation we checked whether a matching object exists within a
radius of five pixels of the same position in the other simulation. If
such a source exists, we then take it to be the same
galaxy. Otherwise, we assume the given source to be missing if it is
selected in one simulation but not in the other. This analysis
revealed that there are no missing galaxies in AREPO, but that they
simply fall below the magnitude threshold of $27.5$. Visually
inspecting the galaxies just above this magnitude threshold in GADGET
and below in AREPO reveals that these galaxies are generally less
compact and more extended in AREPO and thus appear fainter. While we
attribute this difference to the deficiencies of the standard SPH
solver, we note that all detected galaxies in GADGET have a dark
matter halo associated with them and are not artificial SPH `blobs'
\citep{Sijacki:12, Torrey:12}. While the discrepancies in the results
produced by GADGET and AREPO as discussed above are of a numerical
nature, we note that they are small compared to previous
findings at lower redshifts. In particular, focusing on the total
number of detectable galaxies around the bright QSO in our O$6$
region, our AREPO simulation highlights the same problem we identified
with GADGET runs already -- there is a need for very strong feedback
processes to significantly suppress star formation rates in a few
times $10^{9} \mathrm{M_{\odot}}$ to $10^{11} \mathrm{M_{\odot}}$
halos to alleviate the discrepancy between cosmological simulations
and current observational findings.

\subsubsection{Uncertainties in the sub-grid physics}

Despite the high resolution of our simulations, we are still far from
reaching the resolution necessary for resolving crucial baryonic
processes relevant for galaxy formation \emph{ab initio}.  
Low resolution means that we are
forced to employ simplified sub-grid models to follow black hole
seeding, accretion and associated AGN feedback as well as star
formation and consequent outflows driven by supernova explosions. We
address each of these uncertainties in this section. 

In their cosmological simulations, \citet{Sijacki:09} investigated the
effect of varying both the halo seeding threshold mass as well as the
mass of the black hole seed (within the direct collapse model). They
verified that raising the halo threshold mass for seeding by a factor of ten 
over the value assumed in this paper leads to a black
hole mass higher by a factor of $\approx 1.5$ by $z \sim 6$. This
uncertainty however is smaller than variations in the black hole mass
due to galactic winds, which can also increase black hole mass as
shown in Section~\ref{secresults2}. Varying the black hole seed mass
from $10^5 h^{-1} \, \mathrm{M_{\odot}}$ (our adopted value) to $10^6
h^{-1} \, \mathrm{M_{\odot}}$ was also shown to lead to only a slight
increase in final black hole mass.

In a comparison of five popular black hole growth and feedback
sub-grid models, \citet{Wurster:13} use simulations of major galaxy
mergers to highlight the differences in the predicted black hole mass
assembly history. Besides leading to differing accretion histories,
the five models compared in \citet{Wurster:13} result in
post-merger black hole masses that vary by a factor up to $\sim
7$. The predicted black hole growth differs due to differences in the
evolution of the density and temperature of the surrounding gas 
caused by the different AGN feedback prescriptions. In our simulations of
overdense regions, gas inflows transport gas close to the black
holes in sufficient quantities to power accretion at the Eddington
limit until it becomes limited by feedback at $z \approx 8 \--9$. 
Note, however, that the redshift where this is expected to happen 
will depend somewhat on the mass of the assumed black hole seeds 
as well as  the mass of the halos in which these are placed
\citep{Sijacki:09}. Suggestions that the large gas supply reaches below our resolution limit 
during the exponential growth phase \citep{Dubois:12, Dubois:13} 
imply that Eddington accretion rates (as is the case in our simulations) should be robust
and nearly independent on the underlying black hole model assumed. 
However, there remains a possibility that black hole growth progresses more slowly than suggested by our simulations, where the scales at which angular momentum of the inflows may slow down accretion are not accurately resolved. 
More physical recipes of black hole accretion applied to galaxy mergers, e.g. \citet{Hopkins:10} indeed suggest that accretion can progress less efficiently at scales below $\sim 10 \, \rm pc$. 
These simulations however do not capture the cosmological environment of quasars, where large amounts of gas are brought to the vicinity of the accreting black hole, thus this question can only be settled using cosmological simulations with very high resolution. Finally, since black hole accretion rates are numerically converged \citep{Sijacki:09}, we expect spurious angular momentum transfer not to significantly affect black hole growth in our simulations.

The evolution after AGN feedback  starts to shut-off the accretion depends more sensitively
on the details of the AGN feedback implementation. For example, 
\citet{Choi:12, Wurster:13, Choib:13} give examples of how varying 
both input physics (such as
X-ray heating, or winds driven by momentum rather than energy injection)
and numerical implementation can lead to significant differences in accretion rate \citep[see also][]{Dubois:13}
 outside the regime in which gas is supplied very efficiently. We
note that an important property of our feedback model is that it leads
to self-regulated black hole mass growth. Thus, once the black hole
growth becomes feedback limited, significant changes in the feedback
prescription will still lead only to moderate changes in the black hole mass
growth (for further details see e.g. \citet{Sijacki:09}). None the
less, the properties of AGN-driven outflows can be very sensitive to
the specific feedback implementation, see e.g.  \citet{Choi:12, Choib:13} and 
\citet{Debuhr:10}, and we will discuss this point in detail in Section
3.5.

Our cooling prescription may also be too simplistic as it relies on a
cooling function that is only applicable to the primordial gas. As
stellar mass builds up and heavy elements are produced and released
into the surrounding interstellar medium, further cooling can occur
via metal lines. Efficient metal line cooling would only exacerbate
the discrepancy in galaxy counts between overdense and average regions
discussed in Section~\ref{secresults3}.  \citet{Vogelsberger:13} have
recently performed cosmological simulations using the moving mesh code
AREPO, where the effects of metal-line cooling are investigated
together with various feedback prescriptions \citep[see also][]{Schaye:10}.  
Metal-line cooling and
gas recycling are found to have little effect at high
redshift, when gas is still relatively unpolluted by metals. We
therefore expect metal line cooling not to affect any, but the most
massive galaxies in our simulations of overdense regions at $z \sim
6$.

Interestingly, \citet{Vogelsberger:13} however found gas cooling in
lower mass halos to be strongly limited by stellar feedback, which we
have shown to play a major role in our simulations at high redshift as
well. In this paper, such outflows have been shown to be a very
promising means to prevent overcooling in galaxies already at $z \sim
6$. In
the adopted sub-grid model, we have assumed a very simple prescription
in which the mass loading factor $\eta$ is constant for every
halo. There is, however, some evidence for a scaling between the mass
loading and the depth of the potential well of the galaxy the wind is
launched from. If the wind velocity is proportional to the galaxy's
escape velocity $v_{\rm w}$, then $\eta \propto v_{\rm w}^{-1}$ for
momentum-driven and $\eta \propto v_{\rm w}^{-2}$ for energy-driven
winds.  The implementation of winds with such galaxy dependent mass
loading and wind speed has been recently tested in \citet{Puchwein:13}
(see also \citet{Vogelsberger:13}), where good agreement with the low
mass end of the stellar mass function at $z \,=\,0$ (and up to $z
\,=\,2$) is obtained using energy-driven outflows. \citet{Puchwein:13}
report star formation suppression due to galactic winds in halos with
masses up to $\sim 10^{12} \, \mathrm{M_{\odot}}$, which at $z \,=\,
6$ completely encompasses the mass range of halos of all the galaxies
in the QSO fields as well as in average and intermediate regions. We
therefore expect galactic winds to remain a viable mechanism to
suppress the number counts of bright galaxies in both overdense and
average regions at $z \sim 6$. Different implementations may, however,
affect our results quantitatively, by predicting a different
shape particularly at the faint end of the luminosity function. 
Another feature of our sub-grid model for stellar feedback is that the emitted wind particles
 are temporarily hydrodynamically decoupled while they are still within the dense star forming region. Thereafter, wind particles are hydrodynamically coupled again with the rest of the gas.
\citet{Puchwein:13} and \citet{Vogelsberger:13} show that such implementation of SN-driven winds leads to luminosity functions in very good agreement with observations. 
It is nevertheless possible that this implementation can affect the internal properties of individual galaxies such as their rotational velocity and morphology.
 Capturing these properties reliably would, however, require considerably higher resolution simulations.

\begin{figure}
\includegraphics[scale = 0.5]{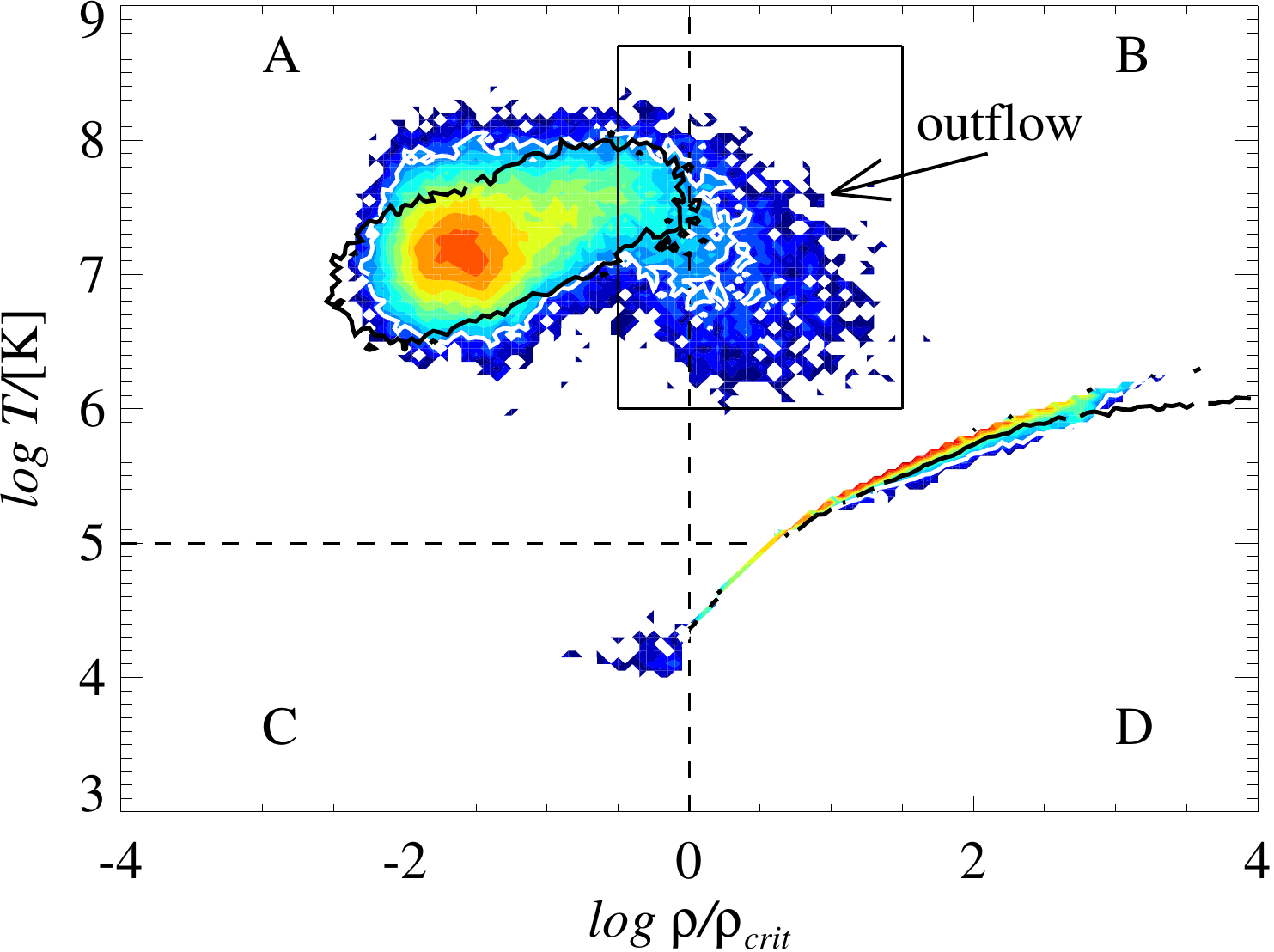}
\caption{Phase diagram of gas particles with positive radial velocity
  (in the central halo frame) within the virial radius of the QSO host
  halo at $z \,=\, 6.2$ for the O$6$ simulation (in colour). The colour
  encodes the number density of particles, with blue representing  low
  and red  high number densities.  Contours enclose the region of the
  phase diagram with  ten or more particles for the O$6$ simulation
  (white) and O$6$nobh simulation (black).  After the onset of strong
  AGN feedback, the crossover region between dense hot and diffuse hot
  gas (between   regions A and B) becomes populated. This is
  consistent with a  picture where the very dense cold gas in the
  vicinity of the QSO is  heated by the thermal AGN feedback and moved
  away  from the effective equation of state. As thermal energy is
  converted to kinetic energy, this gas escapes to regions of lower
  density. This region (marked with a black rectangle) therefore likely
  contains the bulk of the AGN-driven outflow.}
\label{phasediagram}
\end{figure}

\subsection{The effects of (thermal) AGN feedback}\label{secoutflow}

\subsubsection{AGN-driven outflows}
In our implementation of AGN feedback, a fraction $\epsilon_{\rm f} =
0.05$ of the radiated bolometric luminosity is assumed to couple with
the surrounding gas. This energy is distributed as thermal energy
over gas particles located within a smoothing length of the black hole
particle. For accretion at the  Eddington limit, the energy is
injected at a rate $L_{\rm f} \approx 1.26 \epsilon_{\rm f} \times
10^{38} \frac{M_{\rm BH}}{M_{\odot}} \, \mathrm{erg s^{-1}} \,=\, 6.3
\times 10^{45} \frac{M_{\rm BH}}{10^9 M_{\odot}} \, \mathrm{erg
  s^{-1}}$.  For one e-folding of a $10^9 \mathrm{M_{\odot}}$  black
hole, the total amount of energy injected  corresponds to $\approx 9
\times 10^{60} \, \mathrm{erg}$.   For comparison, the binding energy
of gas within the most massive halos  in our overdense sample at $z
\,=\, 6.2$ is typically only $\sim 10^{59} \, \mathrm{erg}$.

\begin{figure*}
\begin{subfigure}[l]{0.5\textwidth}
\centering \includegraphics[scale = 0.49]{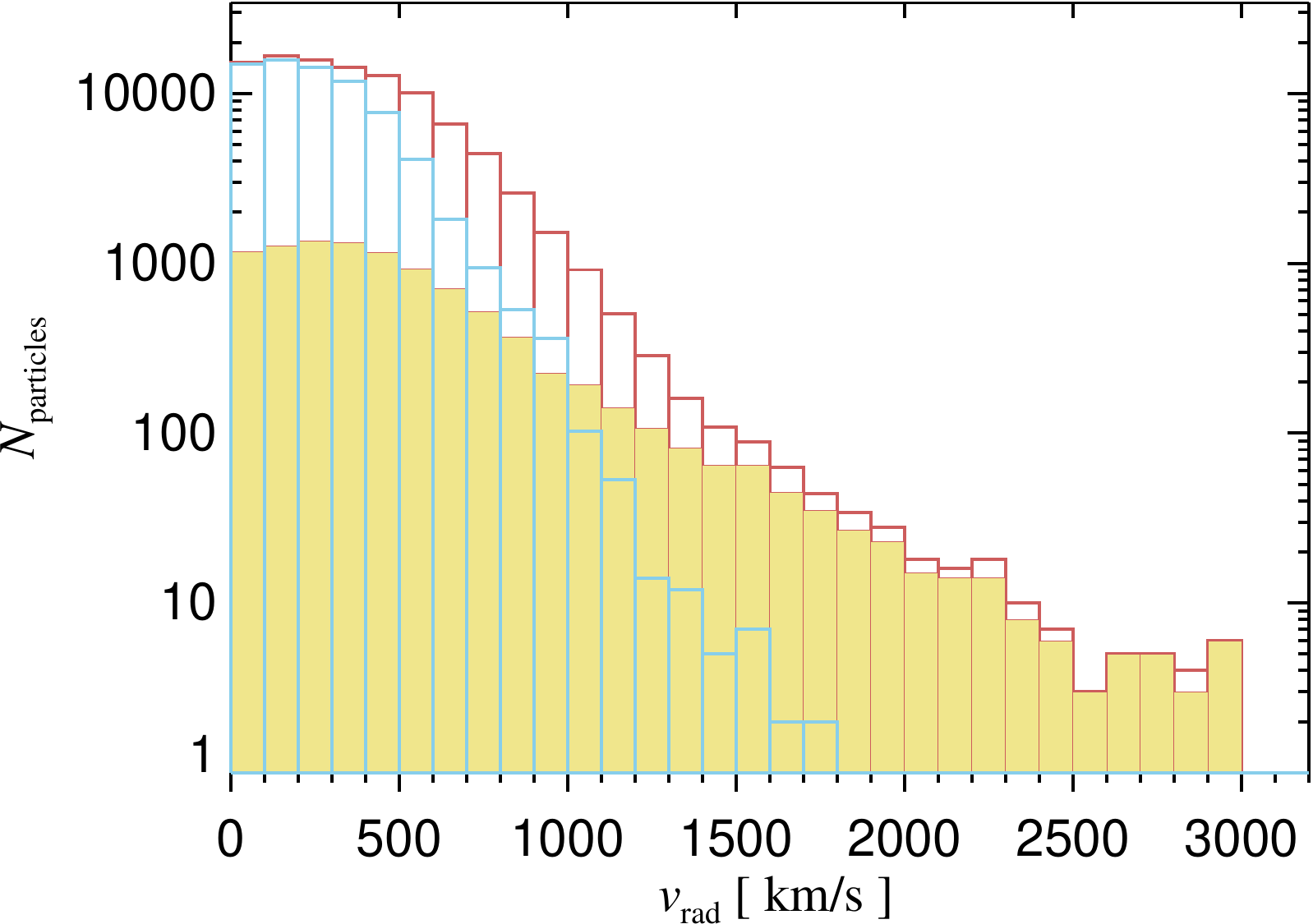}
\end{subfigure}%
\begin{subfigure}[r]{0.5\textwidth}
\centering \includegraphics[scale = 0.5]{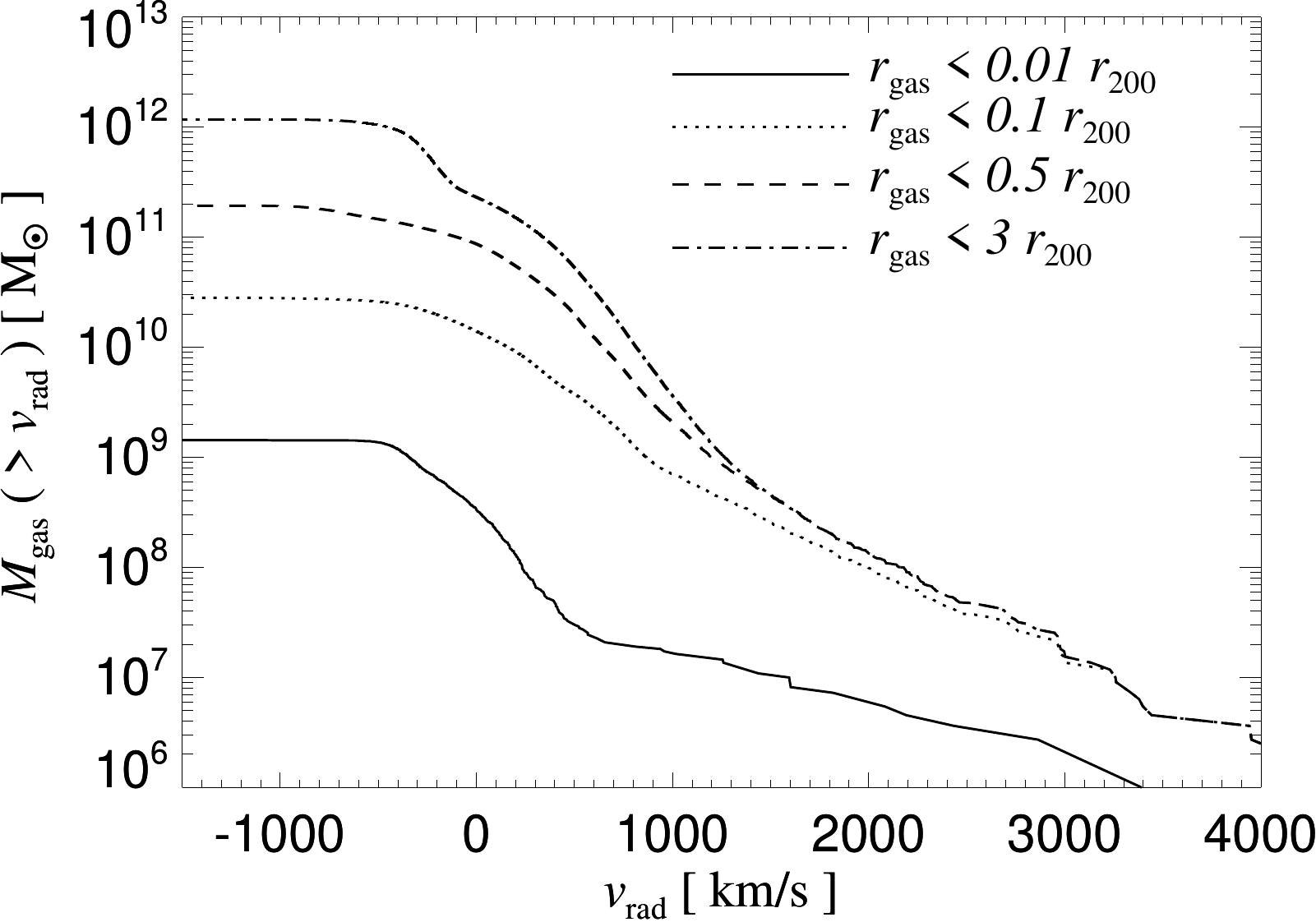}
\end{subfigure}
\caption{(a) Histograms showing the positive radial
  velocities of all gas particles within the virial radius of the most
  massive halo in the O$6$ simulation at $z \,=\, 6.2$.  The red
  histogram shows the results for  O$6$, while the blue histogram
  shows results for O$6$nobh. In yellow is the histogram for all
  particles contained in the box drawn in Figure
  \ref{phasediagram}. The yellow histogram clearly occupies the tail
  of the O$6$ histograms (red). Comparison with
  Figure~\ref{phasediagram} shows that the outflow consists of hot and
  relatively dense gas.  (b) Cumulative plot of mass of all gas particles in simulation O$6$ 
  moving with radial speed $> v_{\mathrm{rad}}$ enclosed by spheres
  with different radii as shown in the legend. The bulk of the outflow is contained within $10 \%$ of
  the virial radius of the QSO host halo.}
\label{outflow}
\end{figure*}

\begin{figure*}
\begin{subfigure}{0.5\textwidth}
\centering \includegraphics[scale = 0.5]{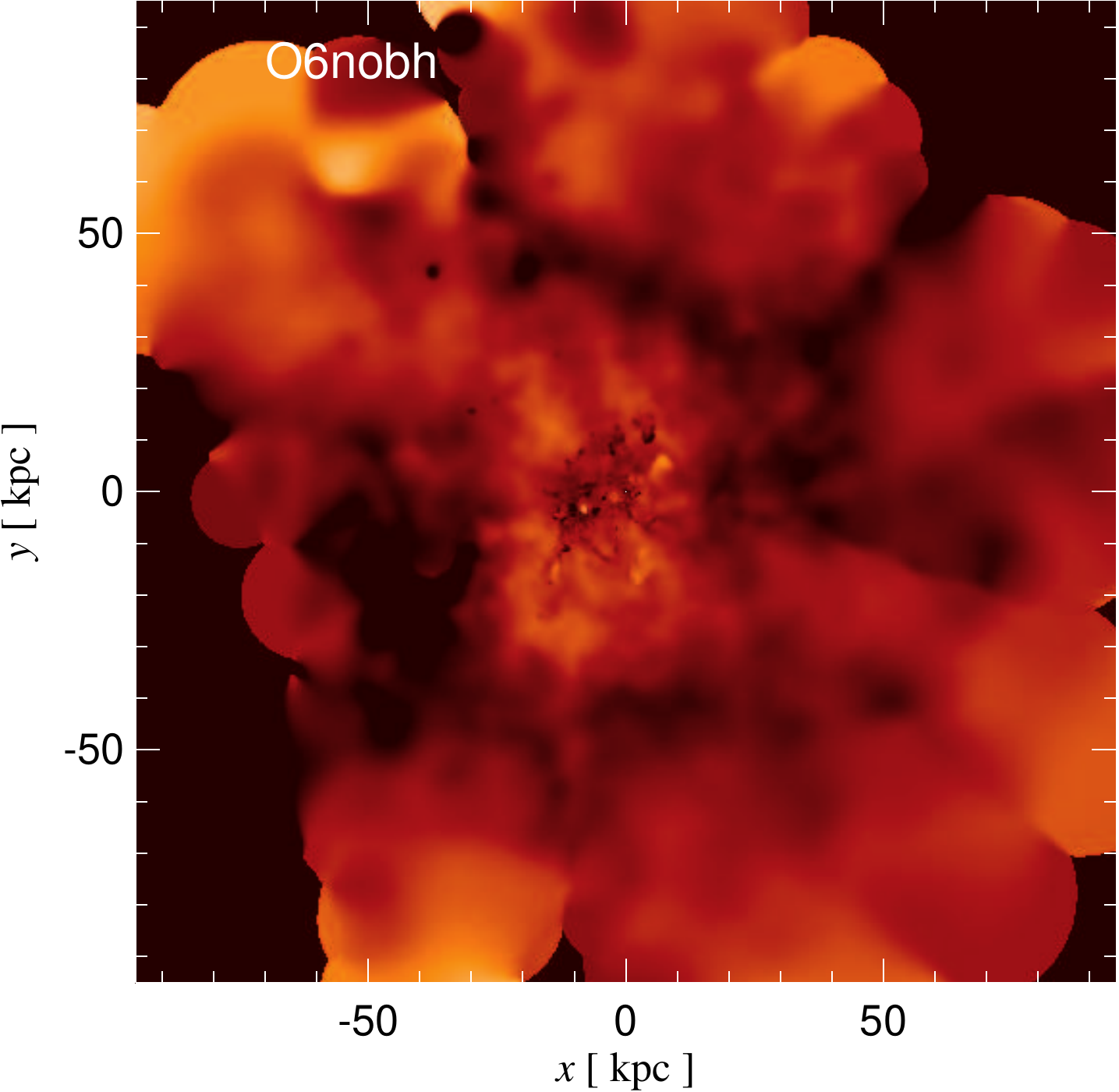}
\end{subfigure}%
\begin{subfigure}{0.5\textwidth}
\centering \includegraphics[scale = 0.5]{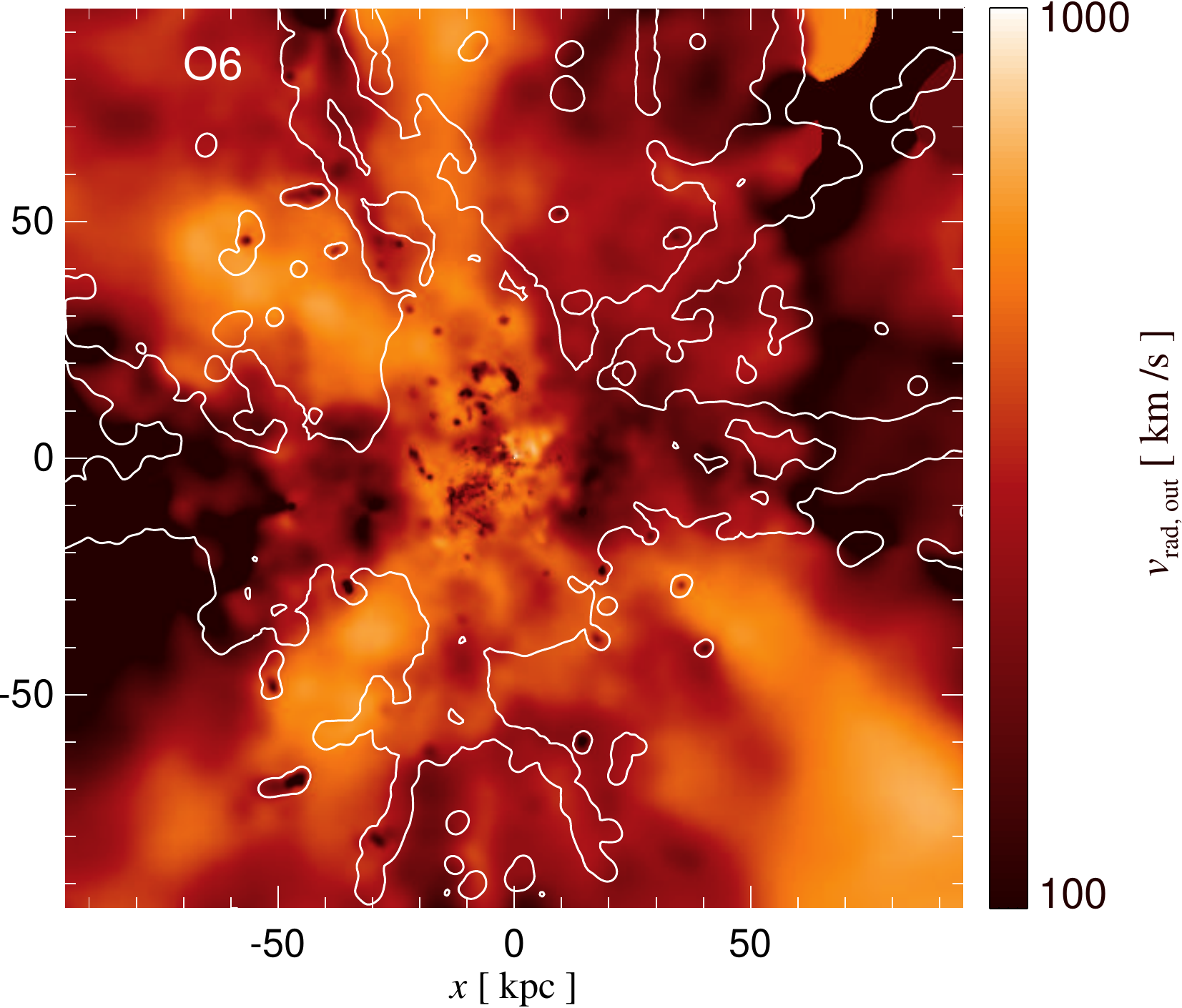}
\end{subfigure}%
\caption{Mass-weighted maps of positive radial velocity projected
  along a thin slice of thickness $57 \, \mathrm{kpc} \, (0.3 h^{-1}
  \, \mathrm{comoving Mpc})$ for O$6$nobh (left) and O$6$ (right)
  simulations at $z \,=\, 6.2$. In O$6$ simulation
  powerful outflows with speed $\sim 1000 \, \mathrm{km/s}$    are
  driven away from the supermassive black hole at the centre.  For
  O$6$, we overplot plot a gas density contour level of $\approx 2
  \times 10^5 \, \mathrm {M_{\odot} \ kpc^{-3}}$. The contours show
  that the AGN-driven outflow escapes preferentially into underdense
  regions, where it encounters lower ram pressure.}
\label{rvel_map}
\end{figure*}

The energy injected by the AGN is therefore more than sufficient for
driving an outflow, the details of which  depend on how much of this
energy is dissipated by cooling.   As we will see later, once the
black hole mass has grown sufficiently, the energy  is injected so
efficiently that the gas in its environment is heated and
over-pressurized so rapidly that the expansion time scale is  faster
than the cooling time scale and the resulting pressure gradient leads
to the acceleration of the surrounding gas, generating an
`energy-driven outflow' \citep{King:10, Ostriker:10, King:11,
  Zubovas:12, Choi:12}. The outflow transports energy to the outer regions of the halo,
modifying the density and temperature distribution and therefore the
spatially extended (Bremsstrahlung) emission from the surrounding
gas. As we show later, the gas close to the QSO is heated by the AGN
to temperatures exceeding $10^8 \, \mathrm{K}$, well in excess  of the
virial temperature of the dark matter halos hosting  the  luminous
high-redshift QSOs in our simulations.   Note that we have not tried
to model cooling of the gas due to Compton scattering by the radiation
emitted  by the accreting black hole.  Compton cooling  due to the AGN
emission  should be  important on scales smaller or comparable to our
resolution limit and  we assume here that this is accounted for by our
choice of  $\epsilon_{\rm f}$.

In this section, we investigate the physical characteristics of the
AGN outflow and explore the properties of the spatially extended
X-ray emission in our simulations.   We mainly investigate the X-ray
properties for all default simulations of overdense regions including the O$1$+winds and O$1$+winds(strong), but also 
compare to an additional simulation, in which region O$6$ is
simulated without black holes.  We shall refer
to this simulation without AGN feedback as O$6$nobh\footnote{This
  simulation is in fact the same as O$6$GADGET, but we shall refer to
  it in this section as O$6$nobh for clarity.}.

\subsubsection{Temperature and density distribution}

In Figure~\ref{profiles}, we show mass-weighted radial profiles of gas
density and temperature of the most massive halo at the centre of the
O$6$ and O$6$nobh simulations (black solid and black dashed lines,
respectively) at  $z \,=\, 6.2$ together with the corresponding
profiles of the QSO host halos from all other overdense regions.  The
coloured band indicates the  softening length of the
simulations. Densities, temperatures and radii are  scaled to the
virial gas density, the virial temperature and the virial radius of the QSO host
halo of the corresponding simulation. The central gas
density drops towards the centre in all simulations, except O$6$nobh and O$1$+wind(strong),
where the density at the centre rises by about two orders of
magnitude. Since the only difference between O$6$ and O$6$nobh is the
presence of an accreting black hole and the corresponding  thermal AGN
feedback, this decrement is obviously attributable to the AGN heating
which occurs in episodic Eddington limited bursts (see Section
\ref{secresults2}).  The reduction in density mirrors a corresponding
increase in temperature within the same radius, also of about two
orders of magnitude.  As mentioned before, the AGN heats the gas to
temperatures well in excess of  $10^8 \, \mathrm{K}$, while the
typical virial temperature of these halos is about $\sim 10^7 \,
\mathrm{K}$.

Outside the very central regions, the temperature profiles are pretty
much flat, showing that the underlying gravitational potential is
approximately isothermal.  The only simulation for which the
temperature  decreases in the centre is O$6$nobh.  Without AGN heating
the gas at the centre of this halo cools efficiently due to its high
density, leading to a core with radius $\sim 0.006 r_{\mathrm{200}}
\approx 400 \, \mathrm{pc}$ of cold gas. Note that in several cases, particularly for the models shown by the
green (O$3$) and blue (O$2$) lines, the profiles do not extend all the
way to the centre of the halo.  This is also a consequence of AGN
feedback which drives outflows and clears most of the gas away from
the central regions of these halos.

In order to further illustrate the differences due to thermal AGN
feedback,  we show a phase diagram of all gas particles with positive
radial velocity in the halo rest frame, located within the virial
radius of O$6$ in Figure \ref{phasediagram}.  We overplot
contours enclosing the region of the phase diagram  populated by more
than ten particles  for the O$6$nobh (black) and O$6$ (white) simulations for
comparison.   The phase diagram can be divided into four regions,
separated by dashed lines in Figure~\ref{phasediagram} and each labelled
with a different letter:

\begin{description}
 \item [A)] This region of the phase diagram consists of shock-heated
   hot and diffuse gas. In O$6$, the amount of gas in this region is
   considerably larger and extends to both higher and lower temperatures
   than in O$6$nobh.
 \item [B)] The gas here is above the critical density to form stars,
   but is too hot to do so. It is above region D containing the star
   forming gas lying on the effective equation of state. Most gas in this
   region was once lying along the equation of state but was heated by
   the central AGN. The gas in region B, and a portion of that in
   region A, is more abundant in the phase diagram of the O$6$ simulation
   than that of O$6$nobh as revealed by the contours.  We expect the
   gas in this area of the phase diagram to constitute the outflow and
   we mark this region with a black rectangle.
 \item [C)] The gas in this region has cooled to $T \approx 10^4 \,
   \mathrm{K}$, which is the lowest temperature to which gas can cool
   radiatively in our simulation. 
 \item [D)] Gas in this region is star forming and lies along
   the effective equation of state.
\end{description}

\begin{figure*}
\begin{subfigure}{0.5\textwidth}
\centering \includegraphics[scale = 0.5]{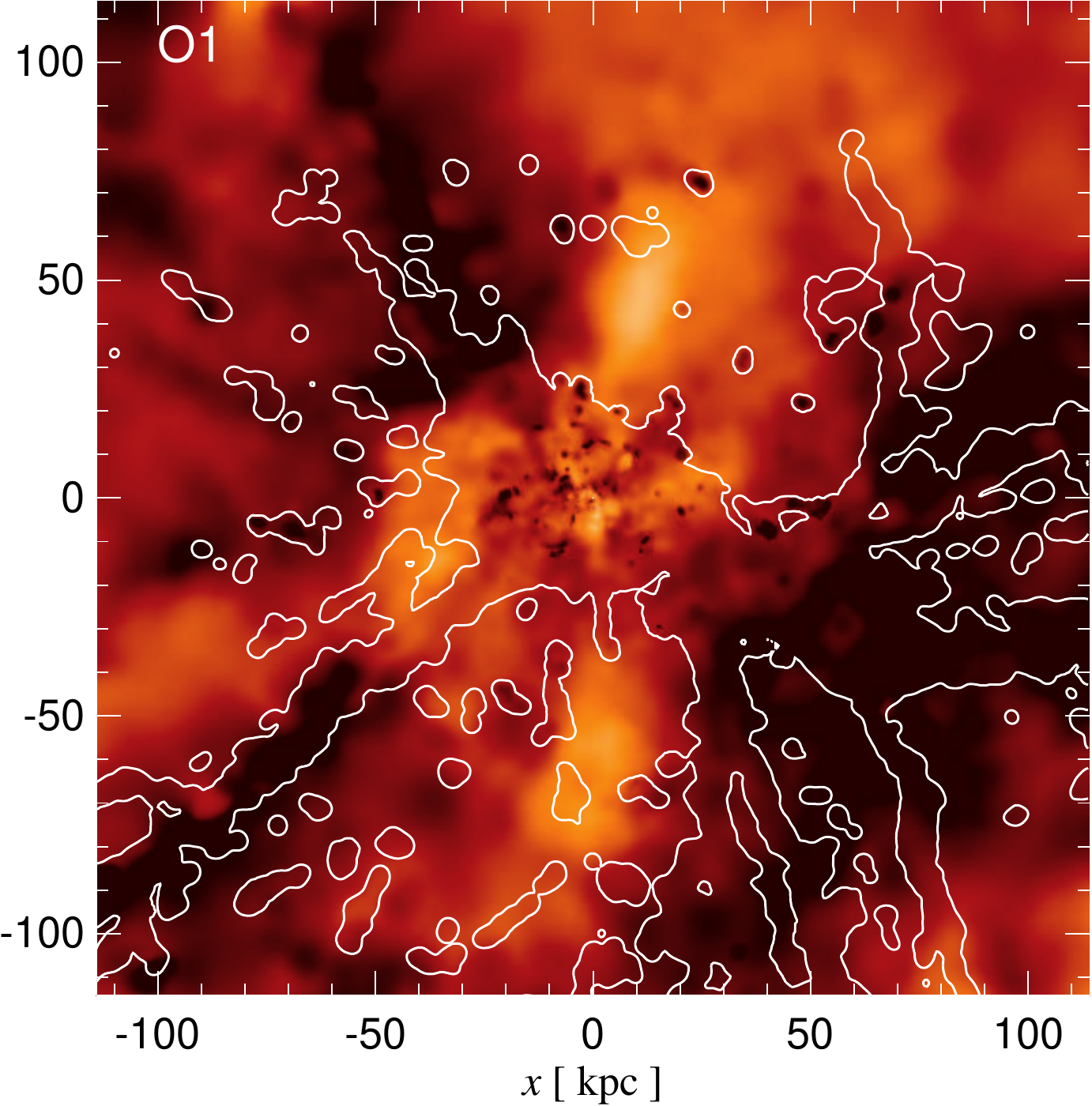}
\end{subfigure}%
\begin{subfigure}{0.5\textwidth}
\centering \includegraphics[scale = 0.5]{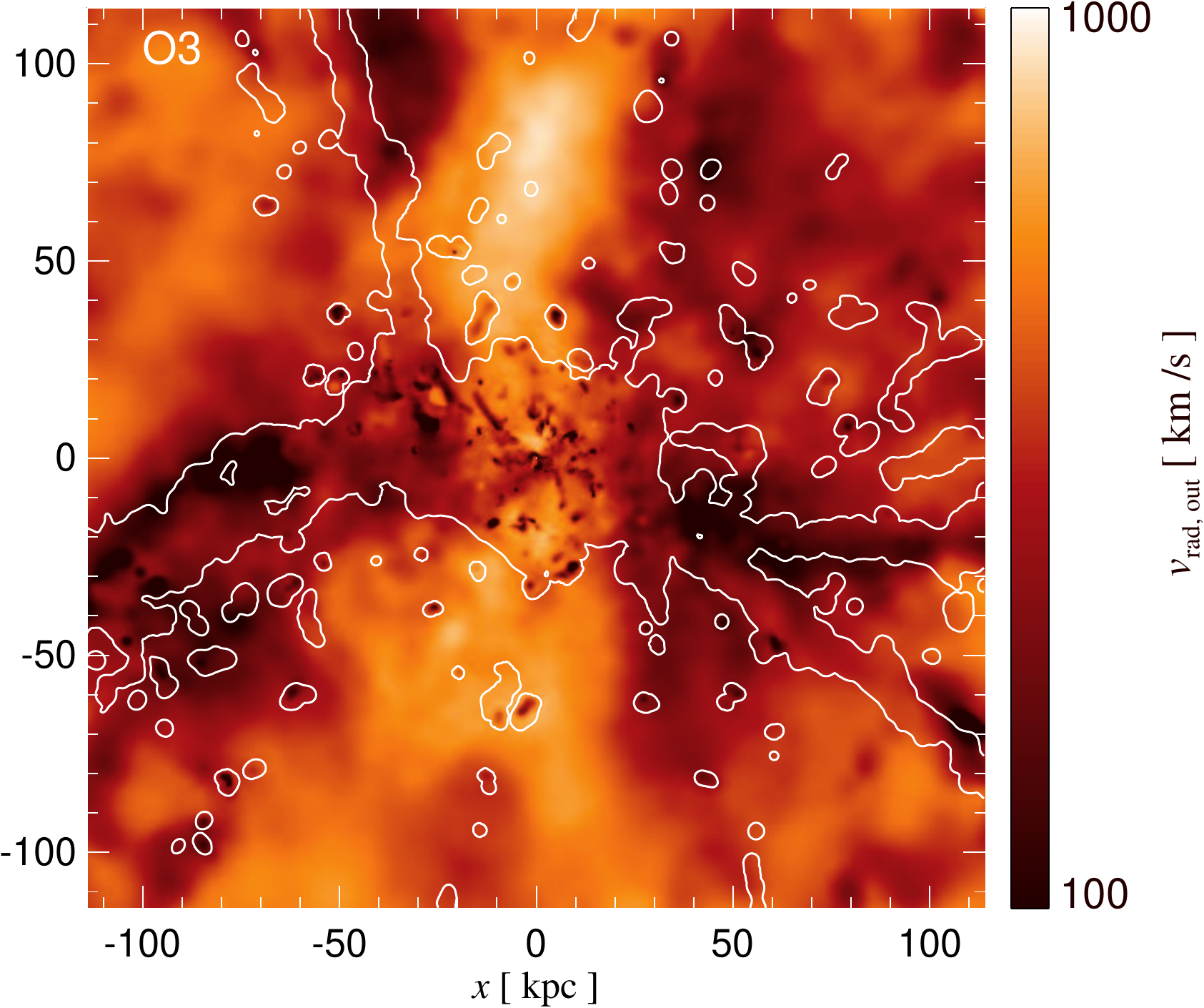}
\end{subfigure}\\

\begin{subfigure}{0.5\textwidth}
\centering \includegraphics[scale = 0.5]{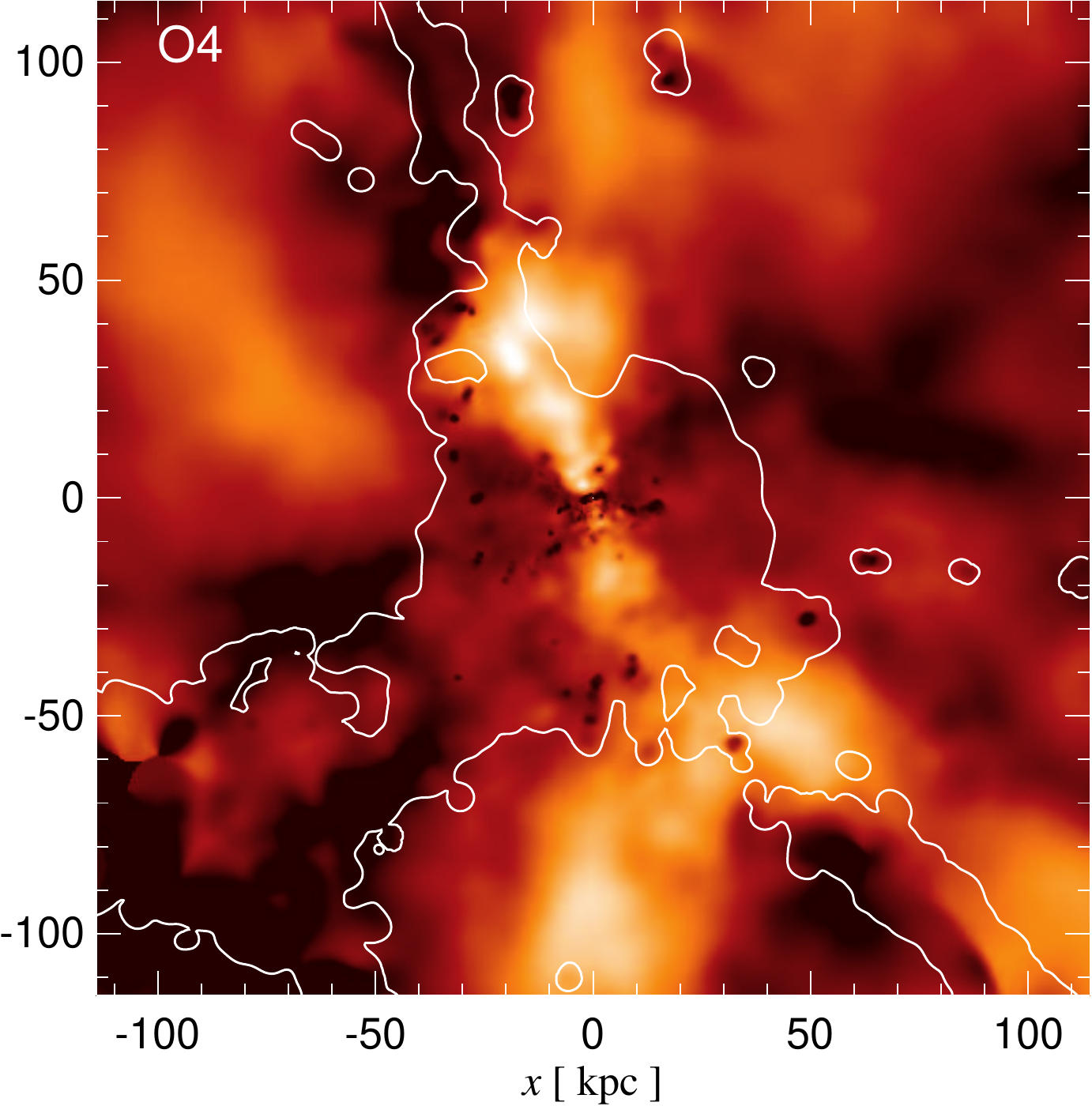}
\end{subfigure}%
\begin{subfigure}{0.5\textwidth}
\centering \includegraphics[scale = 0.5]{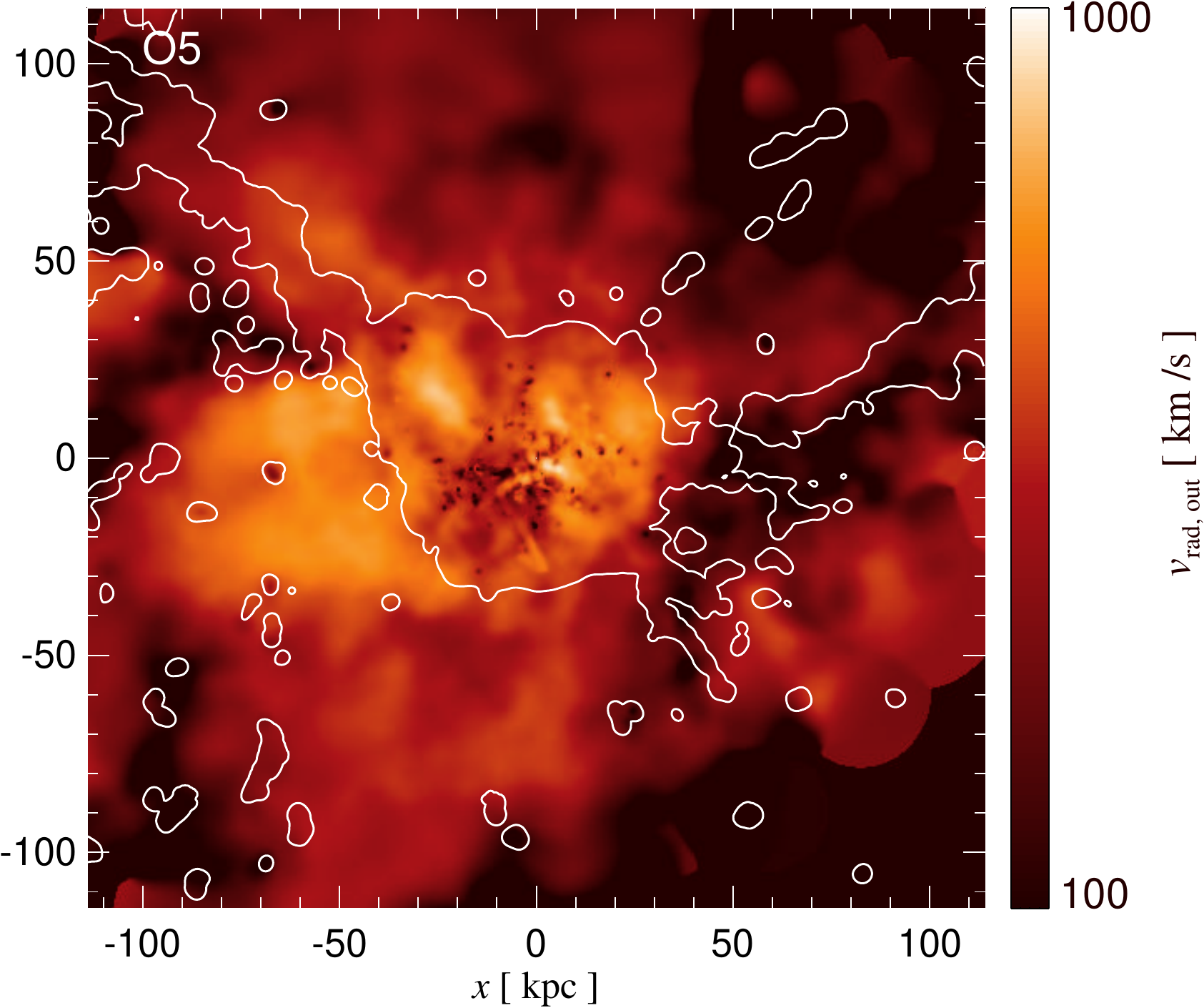}
\end{subfigure}%
\caption{Mass-weighted maps of positive radial velocity projected
  along a thin slice of thickness $\approx 57 \, \mathrm{kpc}$ in four
  of our overdense regions at $z \,=\, 6.2$. White contours enclose
  regions where the gas density is higher than $10^3 \,
  \mathrm{M_{\odot} \, kpc^{-3}}$. Powerful outflows with speeds $\sim
  1000 \-- 1500 \, \mathrm{km \ s^{-1}}$ are driven by AGN feedback out 
of the centre of the most massive halos in our
  overdense regions to distances up to $\sim 100 \,\mathrm{kpc}$ from
  the QSO. The outflows are anisotropic because they preferentially
  propagate into lower density regions.}
\label{rvel_map2}
\end{figure*}

\subsubsection{Properties of the outflow}

In Figure~\ref{outflow} we show a histogram of the radial velocities of
all outflowing particles contained within the virial radius of O$6$ (red
histogram) and O$6$nobh (blue histogram).   For O$6$nobh, the maximum
outward (positive) radial velocity is about $1700 \, \mathrm{km/s}$.
In contrast, thermal AGN feedback in O$6$ simulation gives rise to
outflowing gas with outward radial velocities reaching values as high
as $3000 \, \mathrm{km/s}$.   The yellow histogram shown in
Figure~\ref{outflow} represents the SPH particles located within the
black rectangle drawn on the phase diagram in Figure~\ref{phasediagram},
confirming that the bulk of the outflow is contained within this
region.  The AGN-driven wind thus mainly consists of hot gas that
transports  energy injected by the AGN  away from the centre with
velocities $\gtrsim 1000 \, \mathrm{km/s}$.

\begin{figure*}
\begin{subfigure}{0.5\textwidth}
\centering  \includegraphics[scale = 0.5]{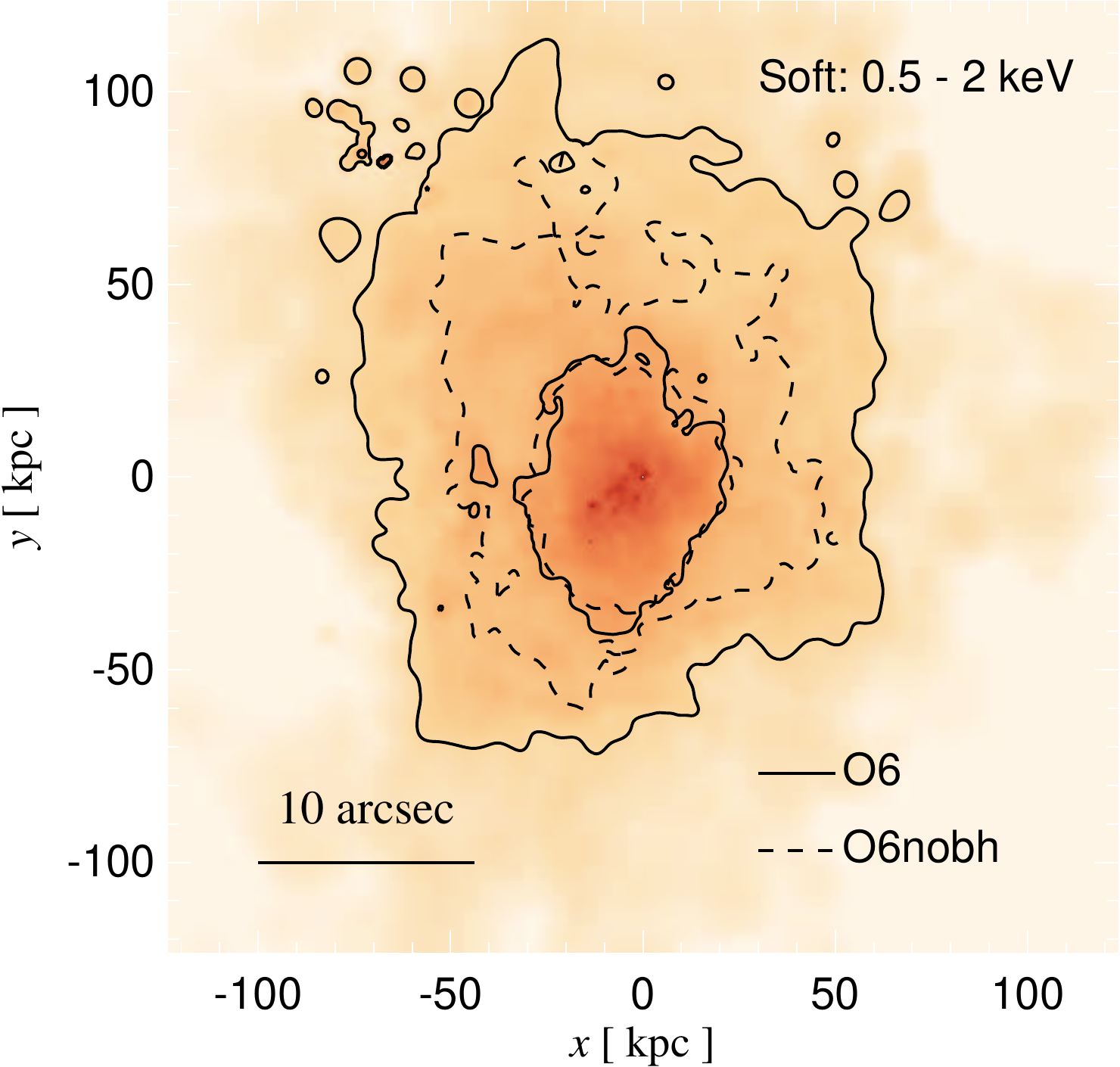}
\end{subfigure}%
\begin{subfigure}{0.5\textwidth}
\centering  \includegraphics[scale = 0.5]{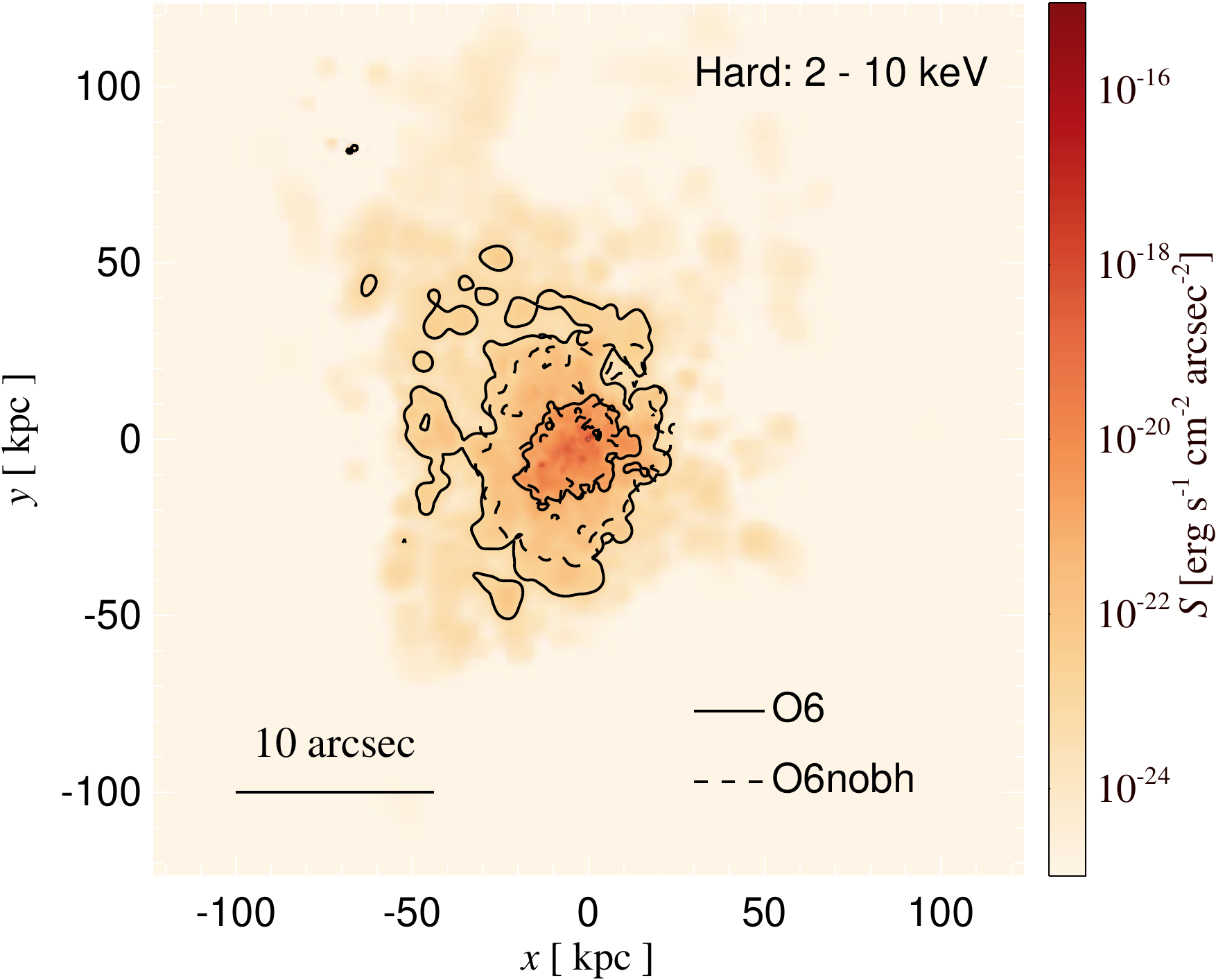}
\end{subfigure}%
\caption{The X-Ray surface brightness around the QSO host halo at $z
  \,=\, 6.2$ in O$6$ simulation for a soft energy band $0.5 \-- 2 \,
  \mathrm{keV}$ (left) and a hard band $2 \-- 10 \, \mathrm{keV}$
  (both in the observer's frame). Surface brightness contours of
  levels $10^{-21}$ and $10^{-23} \, \mathrm{erg s^{-1}
    cm^{-2} arcsec^{-2}}$ are shown as solid lines for O$6$ and dashed
  lines for O$6$nobh. AGN feedback changes the distribution of density
  and temperature of gas as the AGN-driven wind flows to the outer
  regions of the QSO host halo, leading to more extended
  Bremsstrahlung emission. In the soft X-ray band, the
  emission is more spatially extended than it would be without AGN
  heating.  In the hard band, the  X-ray emission  is dominated by
  rarer regions where the gas is extremely hot.  The profile therefore
  appears more compact and its spatial extent does not vary
  significantly between O$6$ and O$6$nobh.} 
\label{surfacebrightness}
\end{figure*}

The right-hand panel of Figure~\ref{outflow} shows a plot of the cumulative mass of
outflowing gas contained within a sphere with four different radii
surrounding the centre of the halo.  The curves overlap at large
velocities indicating that  the fastest outflowing gas is located at
$r \lesssim 0.1 \, \mathrm{r_{200}}$.     At the halo's virial radius,
the mass in gas outflowing with  $v \geq 1000 \, \mathrm{km/s}$ is
$\approx 5 \times 10^9 \, \mathrm{M_{\odot}}$. 

Recently, \citet{Maiolino:12} have detected an outflow consisting of
about $7 \times 10^9 \, \mathrm{M_{\odot}}$ of atomic gas from a QSO
at $z \approx 6.4$.  The outflow is resolved at scales of $\approx 16
\, \mathrm{kpc}$ and has a speed of $\approx 1300 \, \mathrm{km
  \ s^{-1}}$ as suggested by the width of the observed CII transition
line.  The estimate of the gas mass is likely only a lower bound on
the total mass of the outflow since some molecular gas is also
expected to be present in the wind,  suggesting that our simulated
outflow may be somewhat weaker than the strongest AGN outflows
observed so far.  The feedback luminosity in our
simulation O$6$ is, however,  not particularly strong.  As discussed in
Section~\ref{secresults2}, observed luminous $z \sim 6$ QSOs in the
current flux limited  samples  are generally caught at luminosities
and accretion rates higher by a factor of about 30 or  more than that
predicted for the QSO in our  $z \,=\, 6.2$ snapshot of the O$6$
simulations. We therefore note here  that we expect observed luminous
$z \sim 6$ QSOs  to have significantly stronger AGN-driven
winds. Our simulation O$1$+winds(strong) which contains the most
massive black holes at $z \,=\, 6.2$ with a mass of $\approx 1.3 \times
10^{10} \, \rm M_\odot$ drives such a powerful wind with velocities of
several thousand $\rm km/s$ and outflowing gas reaching to hundreds of
$\rm kpc$. Inspection of Figure~\ref{profiles} shows that the increased supply of
gas driven out of smaller galaxies in O$1$+wind(strong) leads to a very high
central gas density which efficiently radiates away the thermal AGN feedback
energy resulting in  a momentum- rather than energy-driven wind. This together
with the increased depth of the central potential is responsible for the
factor ten higher black hole mass required for the AGN feedback to limit
further accretion. While we have argued that this exceptional mass growth may
have been mediated by possibly unrealistically strong galactic winds, this may
be an example of what kind of AGN-driven  wind may be realized in even deeper
and rarer potential wells than we have considered here.

In Figure~\ref{rvel_map}, we show mass-weighted  maps of the projected
radially outward velocity of outflowing gas  for thin slices of
thickness $\approx 57 \, \mathrm{kpc} \, (0.3 h^{-1} \,
\mathrm{comoving \, Mpc})$ with origin at the centre of the halo for
O$6$nobh (left) and O$6$ (right) simulations.  The outflow in
simulation O$6$ is clearly visible and strikingly non-spherical (see also Figure \ref{rvel_map2}), with
gas flowing out with maximum speeds exceeding $\sim 1000 \, \mathrm{km/s}$.  The white
density contours show that the anisotropic outflow escapes along
directions of least resistance into lower density regions.  The
outflow thereby avoids the dense filaments and sweeps across the
voids to large distances.  For O$6$nobh, shown on the left-hand side,
positive radial velocities resulting from random motion of gas
particles as well as pressure forces in the fluid are distinctly
smaller than in O$6$.

\begin{figure*}
\begin{subfigure}{0.5\textwidth}
\centering \includegraphics[scale = 0.5]{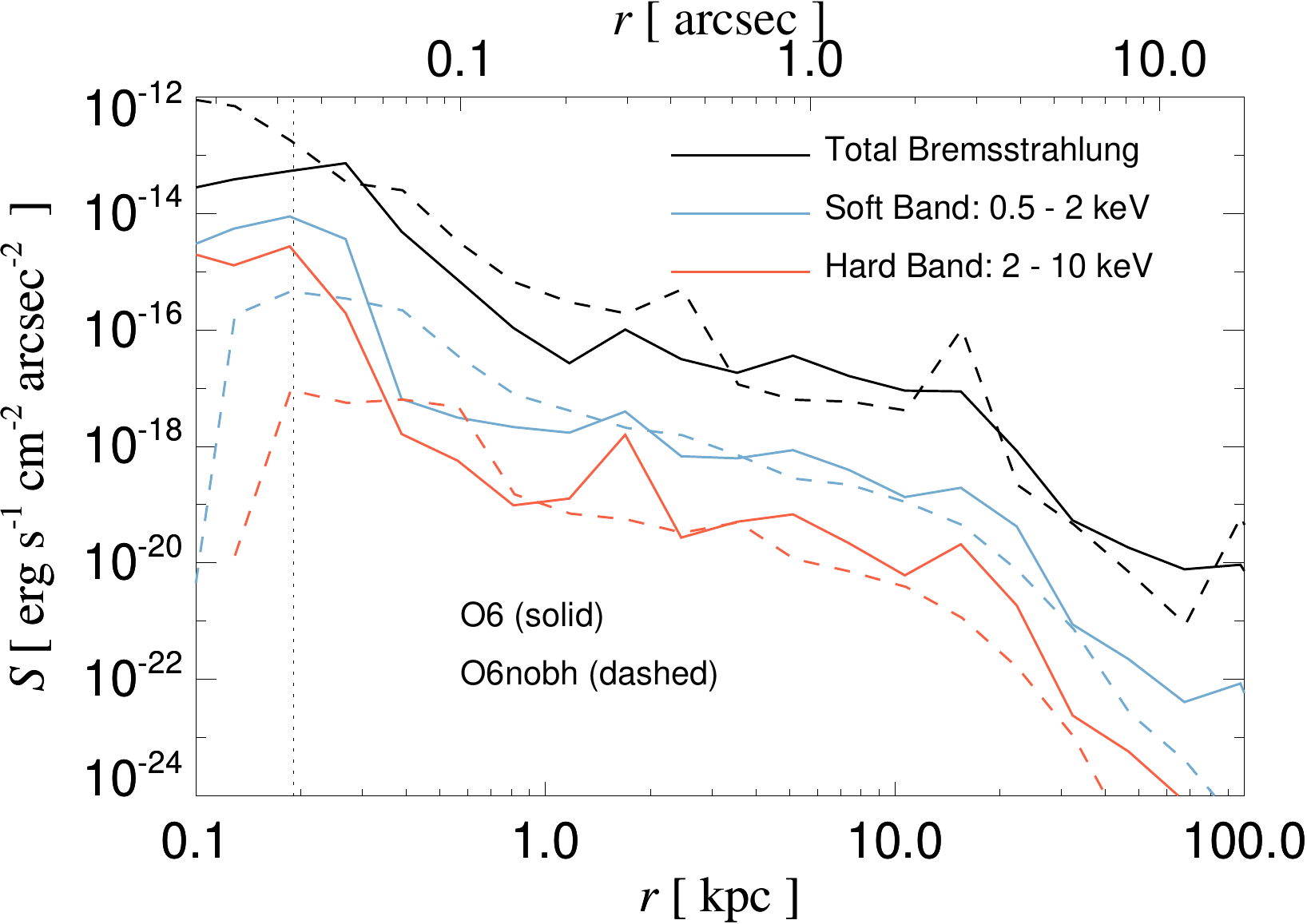}
\end{subfigure}%
\begin{subfigure}{0.5\textwidth}
\centering \includegraphics[scale = 0.5]{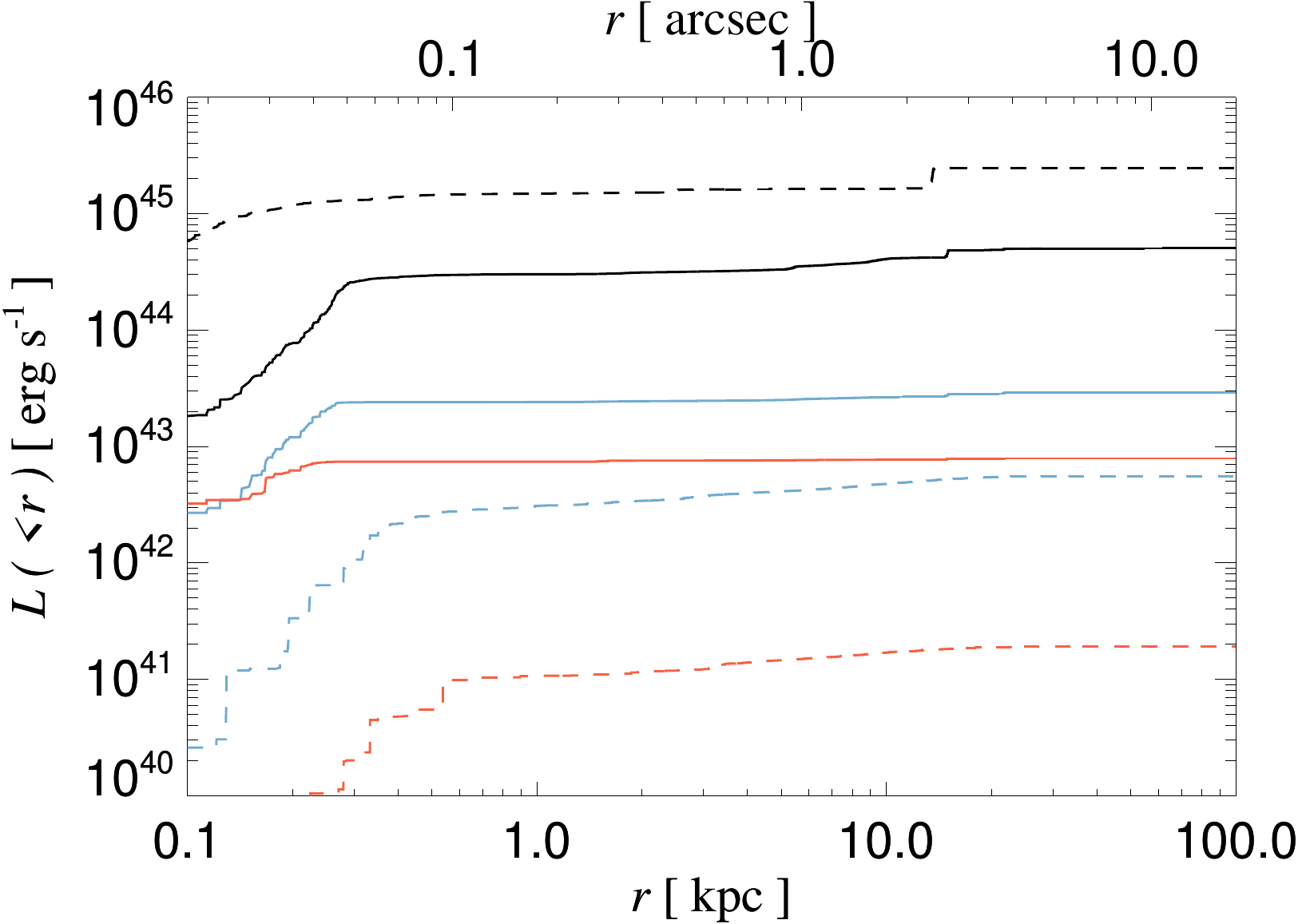}
\end{subfigure}\\
\vspace{0.5cm}
\begin{subfigure}{0.5\textwidth}
\centering \includegraphics[scale = 0.5]{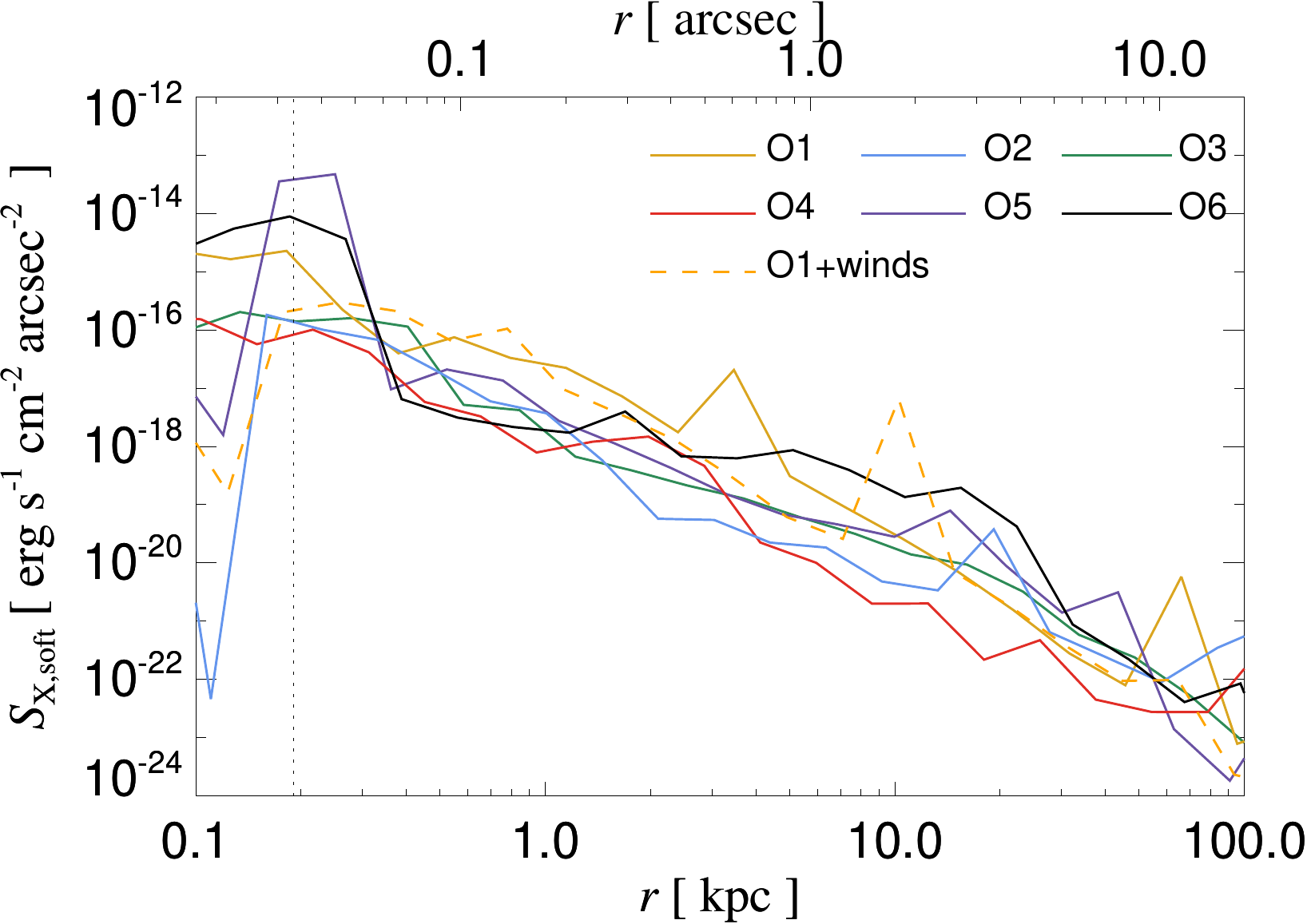}
\end{subfigure}%
\begin{subfigure}{0.5\textwidth}
\centering \includegraphics[scale = 0.5]{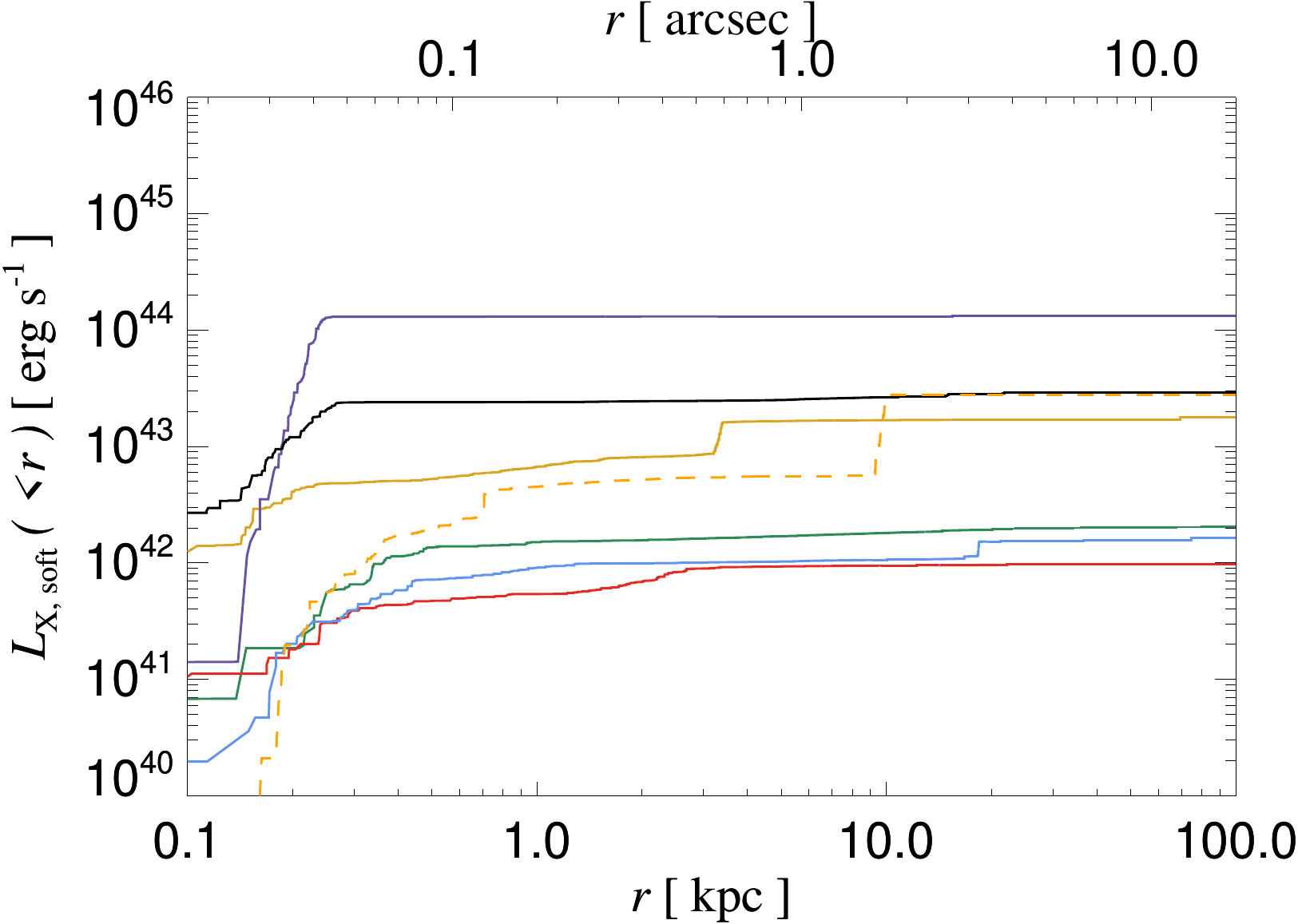}
\end{subfigure}%
\caption{Various properties of the X-ray emission in the QSO host
  halos of our sample of overdense regions. The first row compares
  Bremsstrahlung surface brightness profiles (left-hand side) and
  cumulative Bremsstrahlung luminosity for different energy bands
  (observer's frame) for O$6$ (solid lines) and O$6$nobh (dashed
  lines) at $z \,=\, 6.2$. In the X-ray bands, O$6$ shows
  systematically (moderately) higher surface brightness than O$6$nobh
  at  $r \gtrsim 2 \, \mathrm{kpc}$. Due to the presence of a hot
  bubble at the centre, a large amount of radiation in the X-rays is
  emitted in the very central regions, in stark contrast with
  O$6$nobh, where no such feature is present. This central peak
  dominates the X-ray emission within  the halo, as apparent from the
  cumulative luminosity  on the right hand side. The total
  Bremsstrahlung emission instead  is higher for O$6$nobh due to the
  presence of large quantities of very dense gas at the centre. Due to
  the high redshift most of the  energy released  will be observed  in
  the optical and UV bands.  In the bottom row, we compare the X-ray
  surface brightness profile and the cumulative X-ray luminosity in
  the observed soft X-ray band for all different QSO host halos,
  including our simulation of O$1$ with moderate galactic winds. The
  cumulative plot on the right illustrates a variation of about two
  orders of magnitude in total X-ray emission at any radii among the
  overdense sample. The brightest sources, O$1$, O$5$ and O$6$ are
  those where the QSOs are injecting the highest amounts of energy
  into their surroundings. For these regions, the central peak
  completely dominates the X-ray emission.}
\label{xrayprof}
\end{figure*}

\begin{table*}
  \centering \setlength{\tabcolsep}{3pt}
\begin{tabular}{lccccccccccccc}
\toprule Run   & $M_{\rm BH}$ & $L^{\rm bol}_{\rm QSO}$  &
$\lambda_{\rm Edd}$ & $L_{\rm QSO}^{\rm soft}$ &$L_{\rm QSO}^{\rm
  hard}$ & \multicolumn{2}{c}{$L_{\rm X}^{\rm soft}$} &
\multicolumn{2}{c}{$L_{\rm X}^{\rm hard}$} &
\multicolumn{2}{c}{$L_{\rm Brems}$} & $ \lambda_{\rm f} (<1'')$\\
 
 & $(\times 10^8 \, \mathrm{M_{\odot}})$ & $ (\times 10^{45}
\mathrm{erg \, s^{-1}})$ & & \multicolumn{2}{c}{$(\times 10^{43} \,
  \mathrm{erg \, s^{-1}})$} & \multicolumn{6}{c}{($\times 10^{43}
  \mathrm{erg \, s^{-1}})$}\\
 
 & & & & & & $(<1'')$ & $(<10'')$ & $(<1'')$ & $(<10'')$ & $(<1'')$ &
$(<10'')$ & & \\
\midrule 
O$1$  & $5.65$ & $6.48$ & $9 \%$     & $13.1$  & $33.2$  & $1.66$ & $1.70$ & $0.25$ & $0.25$ & $26.0$ & $27.4$ & $0.80$\\
O$2$  & $6.50$ & $2.32$ & $3 \%$     & $5.97$  & $15.2$  & $0.10$ & $0.16$ & $0.02$ & $0.02$ & $1.26$ & $2.86$ & $0.25$\\
O$3$  & $11.5$ & $3.21$ & $2 \%$     & $7.65$  & $19.4$  & $0.17$ & $0.20$ & $0.02$ & $0.03$ & $5.49$ & $7.11$ & $0.34$\\
O$4$  & $4.04$ & $2.40$ & $5 \% $    & $6.13$  & $15.6$  & $0.09$ & $0.10$ & $0.03$ & $0.03$ & $2.82$ & $3.63$ &  $0.46$\\
O$5$  & $3.23$ & $31.6$ & $78 \%$    & $42.8$  & $109$   & $13.1$ & $13.2$ & $1.41$ & $1.41$ & $127$  & $133$  & $0.80$\\
O$6$  & $4.68$ & $12.7$ & $22 \%$    & $21.7$  & $55.0$  & $2.54$ & $2.90$ & $0.76$ & $0.79$ & $35.3$ & $50.2$ & $0.55$\\
\midrule
O$6$+nobh&  &  &  &  &  & $0.42$ & $0.56$ & $0.01$ & $0.02$ & $162$ & $247$ & \\
O$1$+winds & $13.0$ & $5.83$ & $4 \%$     & $12.1$  & $30.6$  & $0.56$ & $2.79$ & $0.52$ & $2.64$ & $6.12$ & $18.2$ & $0.21$\\
O$1$+winds(strong) & $154$ & $1938$ & $100 \%$     & $884$  & $2245$  & $2689$ & $2689$ & $7022$ & $7022$ & $31987$ & $31987$ & $3.30$\\
\bottomrule
\end{tabular}
\caption{Properties of X-ray emission from the  six QSO hosts in our
  sample of overdense regions at $z\, = \, 6.2$. In
  the first four columns, we list the name of the different
  simulations, the mass of the black hole powering the QSO, the
  corresponding QSO bolometric luminosity and the Eddington ratio of
  the accreting black hole, respectively. In the fifth and sixth
  columns, we provide estimates for the X-ray luminosity emitted by
  the AGN in soft and hard X-ray bands i.e. $0.5 \-- 2 \, \mathrm{keV}$
  and $2 \-- 10 \, \mathrm{keV}$ (observer's frame) respectively,
  where the luminosities were obtained by integrating a mock QSO
  spectrum modelled following  \citet{Hopkins:07}. In
  the seventh and eighth columns we provide values for an X-ray
  soft band $0.5 \-- 2 \, \mathrm{keV}$ (observer's frame) from hot gas within projected radii of $1''$ and $10''$. In the
  ninth and tenth columns these values are given for an X-ray hard band $2 \-- 10 \,
  \mathrm{keV}$ (observer's frame) and in columns eleven and twelve,
  we give the total Bremsstrahlung luminosity within the same projected radii. In the last column, we
  provide a measure of how much feedback energy is radiated away by
  calculating the ratio of the total Bremsstrahlung within $1''$ to
  the instantaneous feedback power.}
\label{table3}
\end{table*}

\subsubsection{X-ray emission}

We now turn our attention to the effect of the thermal  AGN feedback
on the X-ray emission due to the cooling of the gas surrounding the
QSO.  Our results have so far suggested that the main effect of AGN
feedback is to heat the gas at the centre of the halo, converting cold
star forming gas into less dense, hot gas with temperatures in the
range $10^6 \-- 10^8 \, \mathrm{K}$.  Depending on where this gas
cools, we can expect  anything from an excess of central X-ray
emission if it cools rapidly, to enhanced X-ray emission in the
outskirts of the halo if cooling is inefficient in the vicinity of the
QSO.   In the latter case, the X-ray emission could at least in
principle be recognized as spatially extended and be separable from the
unresolved X-ray emission originating from the accretion disc. In order to compute the X-ray luminosity from the hot gas, we use the \citet{Katz:96} estimate for the Bremsstrahlung luminosity produced by each fully ionised non star-forming particle given by,

\begin{equation}\label{bremslum}
L_{\rm X,j} \,=\, 1.42 \times 10^{-27} g(T) n_{\rm j, e} n_{\rm j, i} T_{\rm j}^{1/2} \, ,
\end{equation}
 
where $g(T) \,=\, 1.1 + 0.34 \, e^{-(5.5 - \log{T})^2/3}$ is the frequency integrated Gaunt factor, $n_{\rm e}$ and $n_{\rm i}$ are the number
densities in electrons and ions respectively, $T$ is the temperature
of the emitting gas, $h$ Planck's constant and $k_{\rm B}$ the Boltzmann
constant. This expression is obtained by integrating the thermal
Bremsstrahlung emissivity:

\begin{equation}
 \epsilon_{\rm ff} \ = \ 1.42 \times 10^{-27} n_{\rm e} n_{\rm i} g(E, T)\, T^{-1/2}
 e^{-E/ k_{\rm B} T}  \, ,
\end{equation}

over all energies. We adopt the tabulated frequency dependent Gaunt factors given in \citet{Sutherland:98}. For consistency with the cooling function adopted in the code, we normalise the \citet{Sutherland:98} Gaunt factor to $g(T)$ as in Eq.~\ref{bremslum}. We have projected the contribution to the X-ray emission from all
contributing particles along the z-axis over a distance equal to three
times the virial radius of the central halo.
Figure~\ref{surfacebrightness} shows the predicted observed X-ray
surface brightness in a soft band i.e. $0.5 \-- 2 \, \mathrm{keV}$
(left-hand panel) and a hard band i.e. $2 \-- 10 \, \mathrm{keV}$ (right-hand panel) for the O$6$
region. Surface brightness contours of levels
$10^{-21}$ and $10^{-23} \, \mathrm{erg s^{-1} \, cm^{-2}\,
  arcsec^{-2}}$ are also shown for both O$6$ (solid curves) and
O$6$nobh (dashed curves) simulations.   In the case of O$6$, the X-ray
emission is more extended, particularly for the lower surface
brightness contour levels expected at large radii.  Note that at this
high redshift the bulk of the Bremsstrahlung emission has shifted to
the (far-)UV and the emission in the soft band would have been
observed at energies of the hard band and beyond in nearby and
low-redshift objects.   The (observed) hard band therefore probes rare
regions, where gas is extremely hot with temperatures $T \sim 5 \times
10^8 \, \mathrm{K}$.  The spatial extent of X-ray emission in
simulations with AGN feedback and without nearly overlap at these high
energies.

In the top row of Figure~\ref{xrayprof} we compare radial profiles of
the X-ray surface brightness and cumulative luminosities for O$6$
(solid lines) and O$6$nobh (dashed lines) for the soft and hard bands
as well as the bolometric emission as shown in the legend.   The
bolometric luminosity $L_{\rm Brems}$ is higher in O$6$nobh than in
O$6$ at almost all radii and particularly in the centre, where the gas
density is higher for O$6$nobh due to the absence of strong outflows.
In O$6$, almost all of the feedback energy is used to accelerate the
energy-driven outflow and the strongly reduced density at the centre
of the halo leads to an overall reduction of the Bremsstrahlung
emission.  In the observed X-ray bands, however, the thermal AGN
feedback leads to a clear enhancement of the X-ray emission at small
radii $\sim 400 \, \mathrm{pc}$ due to the much hotter gas found in
this region.  The central peak in fact dominates the X-ray
emission from gas in the proximity of the QSO as the cumulative plot
shown at the right-hand side of the top row indicates.  With an
angular scale considerably smaller than $1 \, \mathrm{arcsec}$, this
spatially extended X-ray emission  would still appear as a point
source even with the resolution of the Chandra X-ray observatory.  In the bottom row of
Figure~\ref{xrayprof}, we show the X-ray surface brightness profile
and the cumulative X-ray luminosity  in the observed soft band for all
QSO host halos in overdense regions (solid coloured lines), including
results for simulation O$1$+winds (dashed yellow line).  The central
X-ray emission in the soft band is  strongest for the QSO  halos in O$1$, O$5$ and
O$6$ simulations.  A look at Table~\ref{table3}, where we summarise
properties of the supermassive black holes including QSO bolometric
luminosity as well as properties of X-ray emission, reveals that these
halos are those where the largest amount of energy is being injected
by the QSO.  Comparison with the profiles shown in
Figure~\ref{profiles} shows these halos also to have  the highest
central densities, which explains why the corresponding  Eddington
ratios are higher.  Note that the higher central densities also
promote cooling, raising the ratio of cooling losses to injected
power, as shown in the last column of Table~\ref{table3}.  
In the cumulative plot at the right-hand side of the bottom row in
Figure~\ref{xrayprof}, the QSO host halos in the O$1$, O$5$ and O$6$
simulations therefore  show the strongest central X-ray
emission in the soft band and the corresponding AGN-driven winds loose significant 
amounts of the injected thermal energy. 
The emission is almost completely dominated by the central
peak,  especially in O$5$ and O$6$, where the Eddington ratio is
$\gtrsim 20 \%$.  In other halos, QSOs are experiencing comparatively
quiescent accretion since gas was moved outwards due to AGN outflows,
explaining the lower central densities found in these halos (see
Figure~\ref{profiles}).  The redistribution of the hot central gas
leads to moderately more extended emission. However, emission levels
are up to two of orders of magnitude lower due to the absence of a
dominant central peak of X-ray emission.  Thus, either central densities are high,
promoting high black hole accretion rates and hence high AGN energy
injection, and cooling via Bremsstrahlung is boosted in the central
regions or the central regions have been evacuated by an outflow and
the central emission  is less dominant and emission more extended.
Including galactic winds further alters the X-ray emission profile,
reinforcing the notion that X-ray emission is critically dependent on
feedback physics. In O$1$+winds(strong), where we are probing the effect of
extremely efficient accretion and outflows, dissipation of kinetic
energy heats the dense gas at the centre  so much that X-ray emission
temporarily exceeds the injected power from the AGN. We note, however, that 
the results of O$1$+winds(strong) should be 
considered as a (generous) upper limit for the effect of thermal AGN
feedback  on the surrounding gas and it's X-ray emission.

Inspection of Table~\ref{table3} suggests that predicting observable
effects of thermal AGN feedback on the  X-ray emission from the host
halos of luminous $z \sim 6$ QSOs and predicting prospects of
discriminating this emission from the direct emission  released during
the accretion onto the central supermassive black hole will be
difficult.  AGN-driven winds generally lead to X-ray emission which is
more extended than expected from hot halo gas alone.  This is a
non-trivial result, because it depends sensitively on where injected
energy is radiated away and on the input physics of our simulations.
At low accretion rates an appreciable fraction  of the total emission
in the soft band (0.5 -2 keV) should be spatially extended,
but our simulations predict  flux levels too low for a detection with
Chandra. With increasing accretion rate/feedback luminosity the X-ray
emission due to AGN feedback becomes more compact in our simulations
and is probably detectable but not resolvable with Chandra. In this
case the spectral signature of thermal emission from hot gas with
temperatures and luminosities significantly higher than expected from
the cooling  of gas in a halo without AGN feedback would be the
diagnostic of choice. Our simulation
O$1$+winds(strong) thereby suggests that an ultra-hard spectrum due to
very  high temperatures may be such a signature to look out for.  The
predicted X-ray emission will, however, undoubtedly,    depend also
strongly on the details of the AGN feedback implementation.  Indeed,
\citet{Choi:12} show that injection of momentum instead of thermal
energy can have a strong effect on both the predicted wind  velocities
and the X-ray emission.  We therefore leave more definite statements
to a future detailed study.

\section{Conclusions}\label{secconclusions}

We have used a comprehensive suite of high-resolution hydrodynamical
resimulations  of the Millennium simulations to study the environment
of luminous  high-redshift ($z \sim 6$) QSOs harbouring supermassive
black holes with (several) billion solar masses. In our
simulations these black holes grow in the most massive dark matter
halos by Eddington limited  accretion from massive seed black holes
and their growth is self-regulated by strong thermal AGN feedback. We
have  produced realistic mock images of star forming galaxies and  the
spatially extended X-ray emission to investigate whether these probes of the
environment of the observed luminous high-redshift QSOs  can verify
our picture for the growth of these early forming  supermassive black
holes. 

Our simulation suite consists of cosmological hydrodynamical
simulations of six highly overdense regions, six regions of
moderate/intermediate overdensity and six regions of average density.
 We find that the central dark matter halos in the
highly  overdense regions are not only more massive, but their
environment is also markedly different with the presence of well
defined massive filaments  intersecting the central dark matter
halo. This position at the intersection of prominent filaments is
instrumental  for the continuous Eddington limited fuelling of the
central supermassive black hole which in turn is responsible for their
very efficient growth. From $z \sim 8$ black hole growth becomes AGN feedback limited
and  continues in short bursts of accretion at the Eddington limit
which result  in strong  mostly energy-driven AGN winds
interdispersed by longer  phases of more moderate accretion.  
In contrast, in our simulations of average regions of the Universe, 
the massive seed black holes generally only grow to a few
times  $10^6 \, M_\odot$.

Our simulations predict that at the  luminosities of the observed $z
\sim 6$ QSOs the luminosity function
is very steep. The peak accretion luminosities of the supermassive black holes are in
good agreement   with the luminosities of observed bright
high-redshift QSOs  which are expected to be strongly biased  towards
large Eddington ratios.  If AGN-driven  winds indeed severely limit the gas 
supply already  at $z\sim 8$ then there should be large numbers of  somewhat
fainter QSOs with luminosities in the range $10^{45-46}$ erg/s which
are powered by similarly massive black holes accreting well 
below the Eddington limit. The masses of our supermassive black holes 
 appear thereby to be somewhat lower than the highest values measured
 in observed bright QSOs at this redshift, albeit introducing (strong)
 galactic winds due to stellar feedback may raise the predicted
 masses of the most massive black holes. It is thus not clear 
 whether the most massive black holes may have to be hosted 
 in halos more massive than the most massive haloes present in the
 Millennium simulation with its about a factor hundred smaller volume than probed by 
 current surveys. 

We found that our predictions of the number of star-forming galaxies
in the field as well as in the environment of luminous high-redshift QSOs  at currently
achievable flux limits are only weakly affected  by
photo-(re-)ionization feedback,  but are very sensitive  to the
presence and strength of supernova-driven galactic winds. We have
further shown that previous studies of this kind based both on
numerical simulations and semi-analytic modelling  have suffered
significantly  from insufficient resolution. For efficient galactic
wind implementations consistent with observation of galaxy properties
at lower redshift, the predicted number of star-forming galaxies
with small angular separation from luminous high-redshift QSOs is  
moderate and consistent with observations and the apparent failure to detect the large 
excess of  star-forming galaxies which had been predicted by simple analytical models of QSOs
that are hosted by the most massive dark matter halos. The predicted number of excess  
sources  increases, however,  strongly for fainter flux limits. Deeper 
observations should thus be key for
making progress with constraining the host halo masses of bright $z \sim 6$ QSOs.

Once the black hole masses in the most massive halos in our simulation
have grown to a few billion solar masses the  (thermal) AGN feedback
in our simulations strongly affects the overall  spatial distribution
and the thermal state of the gas   in the halos. The injection of
$5\%$ of the bolometric luminosity of the AGN powered by the accreting
supermassive black hole  as thermal energy leads to ``blast-waves''
which efficiently launch (mostly) energy-driven winds which move several
billion solar masses of gas at speeds $\gtrsim 1000 \, \rm km \,
s^{-1}$ to distances of $\gtrsim 10$ kpc similar to what has been
recently observed for AGN outflows, albeit mostly  at lower redshift.  We have created
mock images of the predicted spatially extended X-ray emission from
the environment of the luminous high-redshift  QSOs.  At low accretion
rates/luminosities, our simulations predict  a significant fraction of
the X-ray emission to be spatially extended,  but unfortunately at
flux levels too low for a detection with Chandra.  Spectral signatures
of hot gas in the more compact and probably unresolvable  emission at
higher accretion rates/luminosities are therefore the most  promising
diagnostic of the effect of AGN feedback on the gas surrounding
luminous high-redshift QSOs. 

In summary, our simulations show that current observations of the
environment of bright $z \sim 6$ QSOs are fully  consistent with the
picture of Eddington limited early growth of supermassive black holes
from massive seed black holes with masses in the range $10^5 \-- 10^6
\mathrm{M_{\odot}}$.  If massive seed black holes can indeed form at
$z\sim 15$ and bright $z \sim 6$ QSOs are hosted in
halos as massive as in our simulations there appears to be no need for super-Eddington
accretion.  The observed number of star-forming
galaxies suggests  that efficient
galactic winds are already operating at this high redshift.  There is
also the exciting prospect to detect not  only the direct X-ray
emission from the QSOs powered by accretion onto   these supermassive
black holes but also the effect of the thermal AGN feedback on the
spatial extent and the spectrum of the X-ray emission.    At these
high redshifts, the detection of resolved spatially extended
X-ray emission will be very challenging. At lower redshifts however,
spatially extended X-ray emission around bright QSOs  should be an
important tool to study the mechanism of how the feedback  from the
energy released during the accretion onto supermassive black holes
drives observed AGN outflows.  

\section{Acknowledgements}

We thank George Becker for useful discussions and suggestions and Roberto 
Maiolino, Roderik Overzier and the anonymous referee for comments on the manuscript. We are also
grateful to Simon White for granting access to Millennium Simulation
data and Stefan Hilbert for assistance.  All simulations presented in
this paper were run using the DiRAC Complexity cluster based at the
University of Leicester and the COSMOS Supercomputer based at
 the University of Cambridge funded by STFC. TC acknowledges an STFC
studentship. MH acknowledges support by the FP7 ERC Grant
Emergence-320596.  
\bibliographystyle{mn2e} 
\bibliography{references}

\end{document}